\documentclass[fleqn,usenatbib,useAMS]{mnras}

\usepackage{mathptmx}

\usepackage[T1]{fontenc}
\usepackage{ae,aecompl}

\usepackage[pdftex]{graphicx}   
\usepackage[intlimits]{amsmath} 
\usepackage{amssymb}            
\usepackage{tikz}               
\usepackage{multirow}           

\usepackage{verbatim}        
\usepackage{color}           

\usepackage{enumitem}

\definecolor{black}{rgb}{0.0,0.0,0.0}  
\definecolor{grey}{rgb}{0.5,0.5,0.5}   
\definecolor{gold}{rgb}{0.85,.66,0}    
\definecolor{red}{rgb}{1,0,0}          
\definecolor{orange}{RGB}{255,140,0}   
\definecolor{green}{RGB}{0,200,0}      
\definecolor{edcolour}{RGB}{10,128,5}
\definecolor{edremcolour}{RGB}{150,150,150}

\PassOptionsToPackage{pdfpagelabels=false}{hyperref} 
\usepackage{hyperref}           

\newcommand{\xmm}{\textit{XMM-Newton}}

\newcommand{\chandra}{\textit{Chandra}}
\newcommand{\rosat}{\textit{ROSAT}}
\newcommand{\spitzer}{\textit{Spitzer}}
\newcommand{\iras}{\textit{IRAS}}

\newcommand{\swift}{\textit{Swift}}
\newcommand{\swiftx}{\textit{Swift-XRT}}
\newcommand{\xrt}{\textit{Swift-XRT}}

\newcommand{\wise}{\textit{WISE}}
\newcommand{\neowise}{\textit{NEOWISE}}
\newcommand{\ero}{\textit{eROSITA}}
\newcommand{\erosita}{\textit{eROSITA}}

\newcommand{\allw}{{AllWISE}}
\newcommand{\sdss}{{SDSS}}
\newcommand{\sdssiv}{{SDSS-IV}}

\newcommand{\boss}{{BOSS}}
\newcommand{\eboss}{eBOSS}

\newcommand{\ebossts}{eBOSS/TDSS/SPIDERS}
\newcommand{\rass}{{RASS}}
\newcommand{\RASS}{{RASS}}
\newcommand{\irxs}{{1RXS}}
\newcommand{\iirxs}{{2RXS}}
\newcommand{\rassb}{{RASS-BSC}}
\newcommand{\rassf}{{RASS-FSC}}
\newcommand{\rassbf}{{RASS~(BSC+FSC)}}
\newcommand{\rassw}{{RASS+AllWISE}}
\newcommand{\xrayw}{{X-ray+AllWISE}}
\newcommand{\xrayws}{{X-ray+AllWISE+SDSS}}
\newcommand{\rassws}{{RASS+AllWISE+SDSS}}
\newcommand{\rasswsz}{{RASS+AllWISE+SDSS+z}}
\newcommand{\rassagn}{\texttt{SPIDERS\_RASS\_AGN}}
\newcommand{\xmmslagn}{\texttt{SPIDERS\_XMMSL\_AGN}}
\newcommand{\rassclus}{\texttt{SPIDERS\_RASS\_CLUS}}
\newcommand{\xclassclus}{\texttt{SPIDERS\_XCLASS\_CLUS}}

\newcommand{\xmmsl}{{XMMSL}}
\newcommand{\xmmslver}{{XMMSL~(release 1.6)}}
\newcommand{\xmmslw}{{XMMSL+AllWISE}}
\newcommand{\xmmslws}{{XMMSL+AllWISE+SDSS}}

\newcommand{\tdss}{TDSS}

\newcommand{\spiders}{SPIDERS}
\newcommand{\spidersagn}{SPIDERS-AGN}
\newcommand{\sequels}{SEQUELS}
\newcommand{\CSC}{CSC}
\newcommand{\xmmiii}{3XMM}
\newcommand{\xmmb}{3XMM-Bright}
\newcommand{\xmmbw}{3XMM-Bright+AllWISE}

\newcommand{\sxps}{1SXPS}

\newcommand{\sqdeg}{deg$^{2}$}
\newcommand{\sqdegpix}{deg$^{2}$\,pix$^{-1}$}
\newcommand{\psqdeg}{deg$^{-2}$}
\newcommand{\psqcm}{cm$^{-2}$}
\newcommand{\ergs}{erg\,s$^{-1}$}
\newcommand{\logergs}{log$_{10}[\mathrm{erg\, s}^{-1}]$}
\newcommand{\logcm}{log$_{10}[\mathrm{cm}^{-2}]$}
\newcommand{\cps}{ct\,s$^{-1}$}
\newcommand{\cpspa}{ct\,s$^{-1}$\,arcmin$^{-2}$}
\newcommand{\cgs}{erg\,cm$^{-2}$\,s$^{-1}$}

\newcommand{\ecfunitsb}{[erg\,cm$^{-2}$\,s$^{-1}$]/[ct\,s$^{-1}$]}

\newcommand{\um}{$\mu$m}
\newcommand{\si}{$\sim$}

\newcommand{\xspec}{\textsc{xspec}}

\newcommand{\speccy}{\textsc{speccy}}
\newcommand{\nway}{\textsc{Nway}}

\newcommand{\wo}{$[W1]$}
\newcommand{\wt}{$[W2]$}
\newcommand{\wthree}{$[W3]$}
\newcommand{\wfour}{$[W4]$}
\newcommand{\wowt}{$[W1$-$W2]$}
\newcommand{\nh}{N$_\mathrm{H}$}
\newcommand{\post}{P$_\mathrm{post}$}
\newcommand{\lyalpha}{Ly\,$\alpha$}
\newcommand{\halpha}{H\,$\alpha$}
\newcommand{\hbeta}{H\,$\beta$}
\newcommand{\oiii}{[O\,\textsc{III}]}
\newcommand{\mgii}{Mg\,\textsc{II}}

\newcommand{\sub}[1]{\textsubscript{#1}}
\newcommand{\salvato}{(Salvato et~al., in prep.)}

\newcommand{\salvatoalt}{Salvato et~al., in prep.}

\newcommand{\eng}[2]{$#1 \times 10^{#2}$}

\usetikzlibrary{shapes,arrows,positioning}
\tikzstyle{block}  = [rectangle, draw, fill=blue!20, 
    text width=6em, text centered, rounded corners, minimum height=2em,node distance=3.5em]
\tikzstyle{result}  = [rectangle, draw, fill=green!20, 
    text width=11em, text centered, rounded corners, minimum height=3em,node distance=4.5em,thick]
\tikzstyle{blockd} = [rectangle, draw, fill=orange!20, 
    text width=3.4em, text centered, rounded corners, minimum height=2em, node distance=6em]
\tikzstyle{decision} = [diamond, draw, fill=yellow!20,
    text width=1.0em, text badly centered, node distance=3.5em, inner sep=0pt]
\tikzstyle{line} = [draw, -latex']
\tikzstyle{discard} = [circle, draw, fill=orange!20, 
    text width=1.0em, text badly centered, node distance=3.5em, inner sep=0pt]

\title[SPIDERS: AGN Target Selection]{SPIDERS: Selection of spectroscopic targets using AGN candidates detected in all-sky X-ray surveys}
\author[Dwelly et al.]{T.~Dwelly$^{1}$\thanks{E-mail:\href{mailto:dwelly@mpe.mpg.de}{dwelly@mpe.mpg.de}}, 
M.~Salvato$^{1}$, 
A.~Merloni$^{1}$, 
M.~Brusa$^{1,2,3}$, 
J.~Buchner$^{4,5}$, 
\newauthor S.~F.~Anderson$^{6}$,
Th.~Boller$^{1}$, 
W.~N.~Brandt$^{7,8,9}$,
T.~Budav{\'a}ri$^{10,11,12}$, 
\newauthor N.~Clerc$^{1,13,14}$, 
D.~Coffey$^{1}$, 
A.~Del~Moro$^{1}$, 
A.~Georgakakis$^{1}$, 
P.~J.~Green$^{15}$,
\newauthor C.~Jin$^{1}$, 
M.-L.~Menzel$^{1}$,  
A.~D.~Myers$^{16}$,
K.~Nandra$^{1}$, 
R.~C.~Nichol$^{17}$,
\newauthor J.~Ridl$^{1}$, 
A.~D.~Schwope$^{18}$, 
T.~Simm$^{1}$ \\
$^{1}$Max-Planck-Institut f{\"u}r extraterrestrische Physik, Giessenbachstr., D-85748, Garching, Germany\\
$^{2}$Dipartimento di Fisica e Astronomia, Universit{\`a} di Bologna, viale Berti Pichat 6/2, I-40127 Bologna, Italy\\
$^{3}$INAF-Osservatorio Astronomico di Bologna, via Ranzani 1, I-40127 Bologna, Italy\\
$^{4}$Millenium Institute of Astrophysics, Vicu{\~n}a, MacKenna 4860, 7820436 Macul, Santiago, Chile\\
$^{5}$Pontificia Universidad Cat{\'o}lica de Chile, Instituto de Astrof{\'i}sica, Casilla 306, Santiago 22, Chile\\
$^{6}$Department of Astronomy, University of Washington, Box 351580, Seattle, WA 98195, USA\\
$^{7}$Department of Astronomy and Astrophysics, 525 Davey Lab, The Pennsylvania State University, University Park, PA 16802, USA\\
$^{8}$Institute for Gravitation and the Cosmos, The Pennsylvania State University, University Park, PA 16802, USA\\
$^{9}$Department of Physics, 104 Davey Lab, The Pennsylvania State University, University Park, PA 16802, USA\\
$^{10}$Department of Applied Mathematics and Statistics, Johns Hopkins University, 3400 N. Charles St., MD 21218, USA\\
$^{11}$Department of Computer Science, Johns Hopkins University, 3400 N. Charles St., MD 21218, USA\\
$^{12}$Department of Physics and Astronomy, Johns Hopkins University, 3400 N. Charles St., Baltimore, MD 21218, USA\\
$^{13}$CNRS, IRAP, 9 Av. colonel Roche, BP 44346, F-31028 Toulouse cedex 4, France\\
$^{14}$Universit{\'e} de Toulouse; UPS-OMP; IRAP; Toulouse, France\\
$^{15}$Harvard-Smithsonian Center for Astrophysics, 60 Garden St., MS \#20, Cambridge, MA 02138, USA\\
$^{16}$Department of Physics and Astronomy, University of Wyoming, Laramie, WY 82071, USA\\
$^{17}$Institute of Cosmology and Gravitation, Dennis Sciama Building, University of Portsmouth, Portsmouth PO1 3FX, UK\\
$^{18}$Leibniz-Institut f{\"u}r Astrophysik Potsdam, An der Sternwarte 16, 14482 Potsdam, Germany}

\date{Accepted 2017 April 05. Received 2017 March 31; in original form 2017 January 31; latest version \today}

\pubyear{2017}

\begin{document}
\label{firstpage}
\pagerange{\pageref{firstpage}--\pageref{lastpage}}
\maketitle

\begin{abstract}
  \spiders\ (\textbf{SP}ectroscopic \textbf{ID}entification of \textbf{eR}OSITA
  \textbf{S}ources) is an \sdss-IV survey running in parallel to the
  \eboss\ cosmology project. \spiders\ will obtain optical
  spectroscopy for large numbers of X-ray-selected AGN and galaxy
  cluster members detected in wide area \ero, \xmm\ and \rosat\
  surveys.  We describe the methods used to
  choose spectroscopic targets for two sub-programmes of \spiders\
  targets: X-ray selected AGN candidates detected in the \rosat\ All
  Sky and the \xmm\ Slew surveys.  We have exploited a Bayesian
  cross-matching algorithm, guided by priors based on mid-IR
  colour-magnitude information from the \wise\ survey, to
  select the most probable optical counterpart to each X-ray
  detection.  We empirically demonstrate the high fidelity of our
  counterpart selection method using a reference sample of bright
  well-localised X-ray sources collated from \xmm, \chandra\ and \xrt\
  serendipitous catalogues, and also by examining blank-sky locations.
  We describe the down-selection steps which resulted in the
  final set of \spiders-AGN targets put forward for
  spectroscopy within the \ebossts\ survey, and present catalogues of
  these targets.  We also present catalogues of \si12\,000 \rosat\ and
  \si1500 \xmm\ Slew survey sources which have existing optical
  spectroscopy from \sdss-DR12, including the results of our visual
  inspections. On completion of the \spiders\ program, we expect to
  have collected homogeneous spectroscopic redshift information over a
  footprint of \si7500\,\sqdeg\ for $>$85~percent of the \rosat\ 
  and \xmm\ Slew survey sources having optical counterparts in the
  magnitude range 17$<$$r$$<$22.5, producing a large and highly
  complete sample of bright X-ray-selected AGN suitable for
  statistical studies of AGN evolution and clustering.
\end{abstract}


\defcitealias{Anderson07}{A07}

\section{Introduction}
\label{sec:intro}

  X-ray emission is a signpost of accretion of matter onto the
  super-massive black holes that seed the whole population of massive
  galaxies and may strongly influence their formation and subsequent
  evolution.  X-ray selected samples of active galactic nuclei (AGN)
  are particularly powerful because, compared to UV/optical/mid-IR
  selection methods, X-ray selection is much less susceptible to (but
  not completely immune from) the deleterious effects of obscuration by
  intervening material and the dilution of AGN light by the host
  galaxy \citep[e.g. see the recent review by][and references
  therein]{Brandt05}.

  Samples of X-ray selected AGN have been relatively small compared to
  the purely optically selected AGN available from large area optical
  surveys such as the Sloan Digital Sky Survey \cite[\sdss;][]{York00}.
  Although there are many X-ray survey fields currently
  under active study, the spectroscopic completeness is typically low,
  or the survey extents are typically small \citep[][]{Brandt05}. As a
  result, even amongst the most intensively studied fields, the number
  of X-ray AGN with spectroscopic identifications does not exceed a
  few thousand sources per field, as for example, in the
  XBo{\"o}tes/AGES survey \citep[][]{Kochanek12}, and in the Baryon
  Oscillation Spectroscopic Survey (\boss) ancillary project within
  the Northern XMM-XXL field \citep[][]{Menzel16}.
  It is therefore not surprising that our understanding of black hole
  growth across cosmic time lags significantly behind investigations
  of galaxy evolution. Indeed, the physical conditions under which AGN
  are fuelled likely depend on a number of parameters, such as host
  galaxy stellar mass or position within the cosmic web. Disentangling
  the relative significance of those factors requires large samples to
  account for the potentially large intrinsic scatter of relations and
  co-variances between parameters of interest.

  Efforts to obtain complete redshift information for all-sky X-ray
  samples, the exemplar being the R{\"o}ntgen Satellite
  All-Sky Survey, \citep[\rass;][]{Voges99,Voges00,Boller16}, have
  been hampered by the typically rather poor positional accuracy of
  the X-ray detections.  For example, the mean and 95th percentile of
  the 1\,$\sigma$ error radii are $\sim$20\,arcsec and
  $\sim$35\,arcsec respectively for sources in the \rass\ catalogue.
  Historically, this has made selection of the correct optical
  counterparts difficult because, even at the relatively shallow
  depths of currently available wide area imaging
  (e.g. the $r$\si22.5\,mag limit reached by \sdss\ imaging;
  \citealt[][]{Aihara11}), there are already many possible optical
  counterparts found within the error circle of each all-sky X-ray
  source.  Despite these difficulties, several groups have reported
  the results of cross-matching the \rass\ catalogues to counterparts
  found in wide-area optical and near-IR surveys
  \citep[e.g.][]{Veron-Cetty04,Mickaelian06,Parejko08,Haakonsen09,Greiner15}.
  However, these studies have often relied on some degree of human
  interaction in the cross-matching process, or have been limited to
  only the bright end of the population, both of which are undesirable
  features when compiling well-understood and complete samples.

  Even where the cross-matching hurdle has been overcome, the
  follow up of tens of thousands of X-ray sources with single object
  spectrographs requires a prohibitively large telescope time
  allocation. The largest \rass\ follow-up
  programs to date have, by necessity, focused on the optically bright
  part of the X-ray source population, and have been carried out as a
  small component of of large scale galaxy redshift surveys exploiting
  wide-field highly-multiplexed fiber-fed spectrographs.
   For example, \citet[][hereafter \citetalias{Anderson07}]{Anderson07} report
  spectroscopically identified counterparts for $\sim$7000 \rass\
  sources covering 5740~\sqdeg\ (i.e. the spectroscopic footprint of
  the \sdss\ 7th Data Release, DR7, \citealt{Abazajian09}).  Their sample comprises 6224
  broad line AGN (BLAGN), 515 emission line galaxies (ELGs) and 266
  BL~Lacs, based on visual inspection of the spectra. Another large
  sample was presented by \citet{Mahony10}, who report reliable
  spectroscopic identifications for 1715 \rass\ bright catalogue sources covering
  17\,046~\sqdeg\ obtained as part of the 6-degree-Field Galaxy Survey
  \citep[6dFGS;][]{Jones04,Jones09}. The \rass-6dFGS sample has a
  90~percent redshift success rate at b\sub{j}=17.5\,(Vega) but has a
  rapidly declining success rate towards fainter fluxes, and so is
  dominated by optically bright objects.  Unfortunately, these
  existing large spectroscopic samples are still incomplete, since
  a large fraction of the X-ray detections still lack a
  spectroscopically measured counterpart, and inhomogeneous, since a
  variety of criteria or supporting data have been used to select the
  counterparts to X-ray sources and to determine which of those
  counterparts receive spectroscopic follow up.

  \spiders\ (\textbf{SP}ectroscopic \textbf{ID}entification of
  \textbf{eR}OSITA \textbf{S}ources) is an observational programme
  within the \sdssiv\ project \citep{Blanton17} 
  which seeks to improve upon the aforementioned situation.  \spiders\ will
  run for up to 6~years (2014--2020) alongside the Extended Baryon
  Oscillation Spectroscopic Survey \citep[\eboss;][]{Dawson16} and
  Time Domain Spectroscopic Survey \citep[\tdss;][]{Morganson15}
  projects. The primary goal of \spiders\ is to obtain extensive,
  homogeneous and complete spectroscopic follow-up of extragalactic
  X-ray sources, both point-like and extended, using data from X-ray
  satellites and over the \sdss\ extragalactic imaging footprint.
  \spiders\ naturally splits into two main components; an AGN
  programme and a galaxy clusters programme \citep[the latter is
  described by][]{Clerc16}.  The \spiders-AGN programme has been
  designed to collect $\sim$40\,000 spectra of X-ray AGN, and to bring
  population studies of accreting super-massive black holes to a new
  level of accuracy. First demonstrations of the \spiders\ science
  applications, based on the \boss\ follow-up of X-ray selected AGN in
  the Northern XMM-XXL field, were presented by
  \citet{Menzel16,Liu16}. \spiders\ will target X-ray sources detected by the
  forthcoming all-sky X-ray survey to be carried out by \erosita\
  (\textbf{e}xtended \textbf{RO}entgen \textbf{S}urvey with an
  \textbf{I}maging \textbf{T}elescope \textbf{A}rray;
  \citealt[][]{Merloni12,Predehl16}).  However, we have started the
  \spiders\ project in advance of the \erosita\ launch (scheduled for
  2018).  We present in this paper the first phase of the \spiders-AGN
  programme (a.k.a. `Tier-0') which exploits existing (pre-\erosita)
  all-sky X-ray source catalogues to explore the bright end of the
  X-ray AGN population.

  An initial goal of the \spiders\ project is to obtain highly
  complete and reliable identifications for the optical counterparts
  to all \rass\ sources \citep[from both the bright and faint
  catalogues;][]{Voges99,Voges00}, that fall within the
  \eboss\ survey footprint and that have possible counterparts with
  magnitudes within the accessible range ($17<i<22.5$). In addition,
  \spiders\ will obtain redshifts for sources detected in the \xmm\
  Slew Survey \citep[\xmmsl;][]{Saxton08}, which covers a very wide
  sky area (around 2/3 of the full sky). The \xmmsl\ is a factor of a
  few shallower than the \rass\ but has the advantage of being
  sensitive over a broader and harder energy range (0.2--12 versus
  0.1--2.4\,keV). 

  As we discuss in detail later, the mid-IR, specifically the all-sky survey performed
  by the \textit{Wide-field Infrared Survey Explorer} \citep[\wise;][]{Wright10}, 
  is the vital stepping-stone that allows us to correctly select optical  
  counterparts to the bright X-ray selected AGN in the \rass\ and \xmmsl\ surveys. 
  Studies made using the
  \textit{Infrared Astronomical Satellite} \citep[\iras;][]{Neugebauer84},
  \spitzer\ \citep{Werner04} and \wise\ observatories have shown that
  AGN activity is almost always associated with mid-IR emission
  \citep[e.g.][]{Elvis94,Stern05,Stern12,Assef13}.  Indeed, the
  spectral energy distribution of X-ray selected AGN is characterised
  by a tight correlation between near-IR and X-ray flux,
  \citep[e.g.][]{Mainieri02,Brusa05,Civano12,Marchesi16}, particularly when
  high-spatial resolution mid-IR and hard-X-ray measurements are
  available \citep{Gandhi09,Asmus14}. The tendency of luminous AGN to
  stand out from other astronomical populations in the mid-IR has been
  extensively exploited to separate them from the field galaxy
  population \citep[e.g.][]{Lacy04,Stern05,Hickox07,Donley12,Assef13,Mateos13}.  In this
  work we show how the combination of \wise\ imaging data and a
  Bayesian cross-matching algorithm (full details of which will be
  presented by \salvatoalt), can be used effectively to overcome most
  of the issues listed above, which have so far hampered the
  realization of highly complete follow-up programs for the \rass\ and
  \xmmsl\ surveys.

  The paper is laid out as follows.  In section~\ref{sec:data} we
  describe the data sets used in this study. In
  section~\ref{sec:method} we describe the details of the
  cross-matching process, the selection of targets for spectroscopy
  within \spiders, and the process by which we have visually inspected
  \sdss-DR12 spectra associated with our X-ray samples.  In
  section~\ref{sec:verify} we present several independent tests of the
  fidelity of our cross-matching and target selection schemes.  In
  section~\ref{sec:results}, we discuss the properties of the $>$13\,000
  \rass\ and \xmmsl\ sources with existing \sdss-DR12
  spectral identifications. In section \ref{sec:discussion}
  we compare our sample to that of \citetalias{Anderson07} and describe our
  expectations for the completed \spiders-AGN program.

Throughout this paper we express magnitudes in their native systems:
AB magnitudes for \sdss\ \citep{Fukugita96}, and Vega magnitudes for
\wise\ \citep{Assef13}.  In order to allow direct comparison with
existing works from the X-ray survey literature, we adopt a flat
$\Lambda$CDM cosmology with 
$h=H_0$/[100~km~s$^{-1}$~Mpc$^{-1}$]=0.7; 
$\Omega_M$=0.3; $\Omega_{\Lambda}$=0.7, broadly
consistent with the most recent \textit{Wilkinson Microwave Anisotropy
  Probe} and \textit{Planck} determinations
\citep[e.g.][]{Hinshaw13,Ade15}. The data products released in this work can be obtained from the 
MPE X-ray surveys website (\url{http://www.mpe.mpg.de/XraySurveys}).

\section{Preparation of input data sets}
\label{sec:data}

In this section we detail the steps taken to collate and prepare the
input data sets that have been used to produce the final lists of
\spiders\ targets.  In all cases we have only considered targets that
lie within the area defined by the 10\,778~\sqdeg\ of the \sdss-\boss\
imaging footprint\footnote{\url{http://www.sdss3.org/dr9/algorithms/boss_tiling.php}}. This
footprint is a superset of the area that will be considered for the
\eboss\ observations. The \boss\ imaging footprint consists of two
large contiguous regions; 70~percent of the total area is in the
North Galactic Cap (NGC), and the remainder is in the South Galactic Cap (SGC), see
e.g. Fig.~\ref{fig:RASS_XMMSL_maps}. It is expected that, after six
years of operations, \eboss\ will have observed approximately
7500\,\sqdeg\ within this footprint.  The spatial filtering was
carried out using the \textsc{polyid} tool from the
{\textsc{mangle}}\footnote{\url{http://space.mit.edu/~molly/mangle/}}
software suite \citep{Hamilton93,Hamilton04,Swanson08}.

\subsection{\rosat\ All Sky Survey catalogue (\rass)}
\label{sec:data:rass}

The R{\"o}ntgen Satellite \citep[\rosat;][]{Truemper82} was used to
carry out a 6-month-long scanning sky survey (\rass) in 1990--91,
covering around 99.7~percent of the entire sky.  Despite the many
advanced X-ray observatories that have been launched since \rosat,
none has had a larger survey `grasp' (used here to mean the product of
telescope collecting area and field-of-view) in the soft X-rays, and
so the \rass\ remains the most sensitive all-sky survey in the soft
X-ray band (0.1--2.4~keV).

\begin{figure}
\begin{center}
\includegraphics[angle=0,width=84mm]{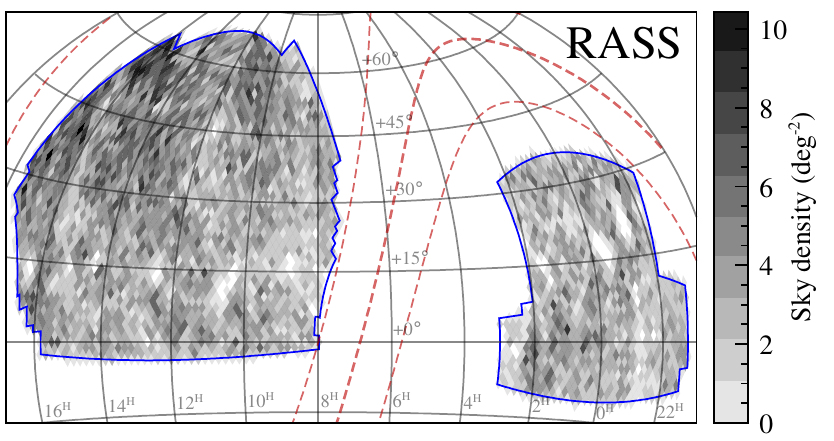}
\end{center}
\vspace{-0.5cm}
\begin{center}
\includegraphics[angle=0,width=84mm]{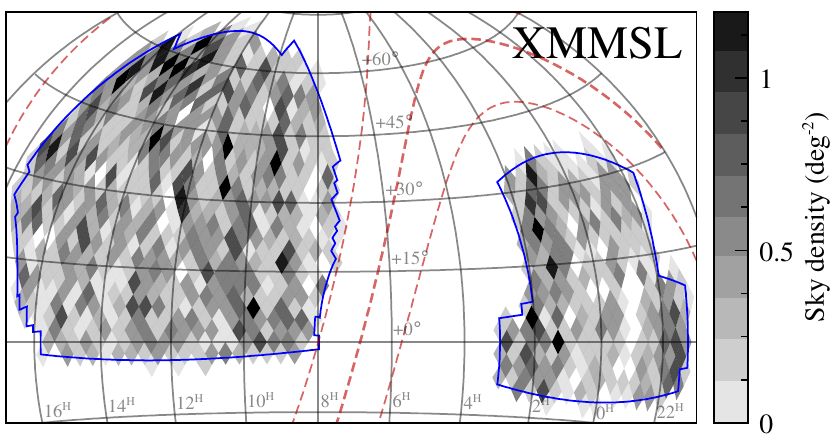}
\end{center}
\caption{Sky distribution of the X-ray source samples considered in this paper.
  {\em Top panel:} \rass\ source density
  map gridded with a HEALPix pixelisation (NSIDE=32, 3.36\,\sqdegpix)
  and displayed with an Equatorial Hammer-Aitoff projection.  The
  density variations (and holes) are primarily due to the variations
  in effective exposure time and background count rate during the
  \rosat\ survey.  
  {\em Bottom panel:} Same for the \xmmsl\ sources,
  but shown with a coarser pixel scale (NSIDE=16, 13.4\,\sqdegpix).
  Note the differences in the greyscale ranges. The (solid blue) line 
  indicates the perimeter of the \boss\ imaging footprint. 
  The Galactic plane is indicated (dashed red lines at $b=0,\pm15$\,deg). }
\label{fig:RASS_XMMSL_maps}
\end{figure}

Two first-generation \rass\ source catalogues were produced, the
Bright Source Catalogue \citep[BSC;][]{Voges99}, containing 18\,806
X-ray sources (detected with $>0.05$\,\cps, at least 15 X-ray counts
and a minimum detection likelihood of 15), and the Faint Source
Catalogue \citep[FSC;][]{Voges00} which contains 105\,924 sources
(detected with at least 6 X-ray counts and a minimum detection
likelihood of 6.5).  We constructed a parent sample of 32\,408 X-ray
sources from the concatenation of all \rassb\ and \rassf\ sources
located within the \boss\ imaging footprint.  A small number (17) of
\rass\ detections with undefined positional errors (likely to be
detection algorithm artefacts) were then removed, leaving 32\,391
sources.  The median positional uncertainty (1$\sigma$ radius,
including a 6\,arcsec systematic error) of the remaining \rass\
sources is 17\,arcsec, and 95~percent have uncertainties smaller than
34\,arcsec.  No attempt was made at this stage to filter the \rass\
catalogue any further, for example, by detection likelihood.  We
discuss the frequency and impact of spurious X-ray detections later in
section~\ref{sec:verify:RASS_spurious}.  The sky distribution of the
\rass\ sample is shown in Fig.~\ref{fig:RASS_XMMSL_maps}.  The mean
sky density of sources is 3.0\,deg$^{-2}$, but their distribution is
far from uniform, due primarily to the uneven sensitivity limit of the
\rosat\ all-sky survey.

\begin{figure}
\begin{center}
\includegraphics[angle=0,width=84mm]{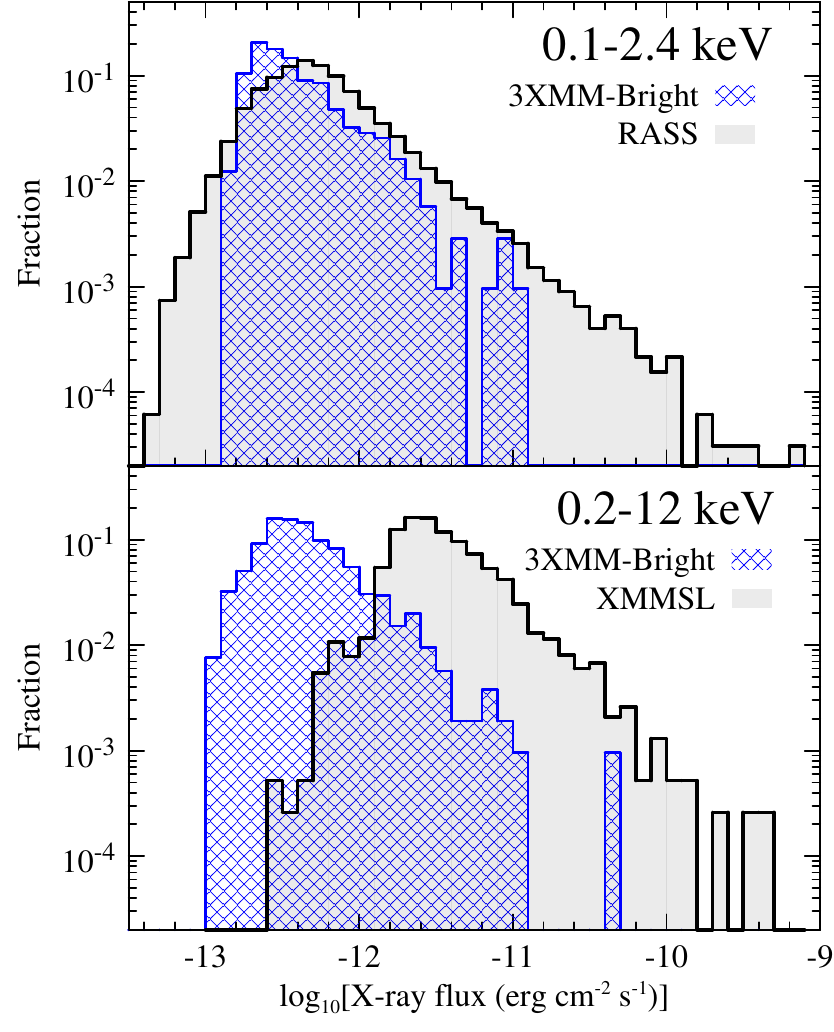}
\end{center}
\caption{The X-ray flux distributions (corrected for Galactic
  absorption) of the \rass\ sample (0.1--2.4\,keV band, upper panel),
  and the \xmmsl\ sample (0.2--12\,keV band, lower panel).  The flux
  distributions of the \xmmb\ 
  reference sample 
  in the same bands are shown for comparison 
  (see section~\ref{sec:data:3xmmb}).  
  The \xmmb\ fluxes have been converted from their native
  system and corrected for Galactic absorption, as described in 
  Appendix~\ref{sec:3XMM_ECF_etc}.}
\label{fig:xflux_histo_vs_xmmb}
\end{figure}

We describe in Appendix~\ref{sec:RASS_ECF_etc} our method to estimate
unabsorbed X-ray fluxes from the \rass\ count rates (i.e. correcting
for the photoelectric absorption due to the Galactic column density in
the direction of the source).  The unabsorbed 0.1--2.4~keV flux
distribution of the \rass\ sources is presented in
Fig.~\ref{fig:xflux_histo_vs_xmmb}.  The distribution is strongly
peaked with a median of $5.2\times 10^{-13}$\,\cgs, and with
86~percent of the sources lying within $\pm$0.5\,dex of this value.

We note that \citet{Boller16} have recently presented the second
revision of the \rosat\ X-ray Survey catalogue (\iirxs), using
additional \rosat\ survey data, improved source detection algorithms,
and improved source characterisations.  Unfortunately the \iirxs\
catalogue was released after \spiders-AGN targets had been submitted
for observation, and so we do not consider it further here.  An
associated work by several of us \salvato\ will present \wise\
associations for \iirxs\ and \xmmsl\ sources covering the entire extragalactic sky,
using similar cross-matching techniques to those presented here.

\subsection{\xmm\ Slew survey catalogue (\xmmsl)}
\label{sec:data:xmmsl}

As part of the normal operations of \xmm, data are accumulated by the
European Photon Imaging Camera pn detector during slews between
pointed observations \citep{Saxton08}.  As of early 2014, the \xmm\
slew observations have covered 65~percent of the sky at least
once\footnote{{\url{http://www.cosmos.esa.int/web/xmm-newton/xsa\#download}}}.
The \xmmsl\ catalogue is built from sources detected in at least one of the 
following energy bands: soft (0.2--2\,keV), hard (2--12\,keV) and full (0.2--12\,keV).

Our starting point is the `Clean' version of the \xmmsl\
catalogue release~1.6
(2014~Feb.~26)\footnote{{\url{http://nxsa.esac.esa.int/catalogues/xmmsl1D_clean.fits.gz}}}, from which we select the 4325 X-ray
detections which fall inside the \boss\ imaging footprint. However,
some of the catalogue entries are multiple detections of the same
X-ray source, whereas we require a list of unique sources.  Therefore,
where groups of \xmmsl\ detections have the same
\texttt{UNIQUE\_SRCNAME} we choose only the best spatially constrained
(i.e. having the smallest value of \texttt{RADEC\_ERR}). Additionally,
we choose only the best spatially constrained detection from any pairs
of sources that lie within a 30\,arcsec radius circle. These filtering
steps leave a catalogue of 3843 unique \xmmsl\ sources\footnote{We
  note here that an earlier version of our sample selection algorithm
  accidentally excluded all \xmmsl\ sources with multiple X-ray
  detections.  This affects $\sim$10~percent of the \xmmsl\ sources
  lying in the first 4010\,\sqdeg\ of \eboss\ plates (i.e. the tiling
  `chunks' named eboss1, 2, 3, 4, 5, 9 and 16).  We corrected the
  target selection error before submitting \spiders\ targets for the
  remainder of the \eboss\ sky.  In the following sections, unless
  otherwise noted, all quoted statistics refer to the correctly
  filtered \xmmsl\ catalogue. We indicate in the supplied catalogues
  (see Appendix~\ref{sec:catalogues}) which \xmmsl\ sources were
  affected by this error.}.
The sky distribution of the \xmmsl\ sample
is shown in Fig.~\ref{fig:RASS_XMMSL_maps}.
The rate of spurious X-ray detections in the `Clean' version of the 
\xmmsl\ catalogue is estimated to be 4~percent in the full energy 
band\footnote{\url{http://www.cosmos.esa.int/web/xmm-newton/xmmsl1d-ug}}.
The median statistical positional
uncertainty (1$\sigma$ radius, excluding systematic errors)
of the unique \xmmsl\ sources is 4.0\,arcsec, and 95~percent have
statistical uncertainties smaller than 15.4\,arcsec.

The \xmmsl\ catalogue includes flux estimates for each source,
calculated from the observed count rates using a simple linear energy
conversion factor which assumes a $\Gamma=1.7$ powerlaw spectrum
absorbed by a fixed Galactic column density of $3 \times
10^{20}$\,\psqcm.  The ratio of unabsorbed to absorbed flux over the 0.2--12~keV band for this
particular spectral model is 1.10 (calculated using
WebPIMMS\footnote{\url{http://heasarc.gsfc.nasa.gov/cgi-bin/Tools/w3pimms/w3pimms.pl}}),
and we have therefore multiplied the catalogue fluxes by this factor
to correct for Galactic absorption.  Given the large statistical
uncertainties on the fluxes of the \xmmsl\ sources, and the relatively
weak sensitivity of the \xmm\ bandpass to moderate levels of Galactic
absorption, we have applied a global correction to the \xmmsl\
catalogue fluxes rather than attempting to correct for the Galactic
column density local to each source.
The unabsorbed 0.2--12~keV flux 
distribution of the \xmmsl\
sources is presented in Fig.~\ref{fig:xflux_histo_vs_xmmb}. 
The distribution is strongly peaked with a median of $3.0\times
10^{-12}$\,\cgs, and with 89~percent of the sources lying within
$\pm$0.5\,dex of this value.

We note that 1934/3843 (50~percent) of the \xmmsl\ sources have at
least one \rass\ detection within 60\,arcsec, and so the two X-ray
samples used in our work are far from independent. However, since the
\xmmsl\ and \rass\ samples have very different characteristics and
were selected in very different ways (telescopes, energy ranges,
detection routines etc.) it is most convenient to treat these two
samples separately.

\subsection{\allw\ candidate counterpart catalogue}
\label{sec:data:allw}

The \wise\ mission \citep{Wright10} carried out an all-sky survey in
the 3.4, 4.6, 12 and 22\,\um\ bands (we denote magnitudes measured in
these bands as \wo, \wt, \wthree\ and \wfour\ respectively).  The
`\allw' catalogue was released in late 2013 \citep{Cutri13}, and
incorporates all data obtained by \wise\ in its original mission
phase, including a second season of observations in the two shorter
wavelength bands.  The survey scan pattern results in inhomogeneous
coverage, (deeper at the Ecliptic poles than at the Ecliptic
Equator). Throughout this work we only use the \texttt{w[1234]mpro}
magnitude estimates from the \allw\ catalogue, which are appropriate
for point-like sources. For 95~percent of the extragalactic sky
($|b|>15$\,deg), the 5\,$\sigma$ point-source magnitude limits in the
\wo, \wt, \wthree\ and \wfour\ bands are at least as deep as 17.6, 16.1, 11.5 and
7.9\,mag (Vega) respectively. We note that, for every source detected
above 5$\sigma$ in at least one band, the \allw\ catalogue contains a
measurement of brightness in all four \wise\ photometric bands, even
where such measurements fall well below the nominal detection
limits. At the faint limit of the \allw\ catalogue, the vast majority of
sources have their highest SNR detection in the shortest wavelength
\wise\ band.

We have restricted our search for \allw\ counterparts to within a 
60\,arcsec radius of the X-ray position of each member of the
\rass\ sample described in section~\ref{sec:data:rass}.   No
down-selection is carried out on this \allw\ sample apart from
removing the few duplicate entries which fall inside the search radii 
around two or more adjacent \rass\ sources.  This procedure
results in a catalogue of 450\,409 unique potential \allw\
counterparts (a mean of 13.9 potential counterparts per \rass\
source).

An identical procedure was carried out to build a list of potential
\allw\ counterparts to the \xmmsl\ sample described in section
\ref{sec:data:xmmsl}, resulting in a catalogue of 46\,389 unique
potential \allw\ counterparts (a mean of 13.4 potential counterparts
per \xmmsl\ source).  Note that although the positions of \xmmsl\
catalogue sources are typically much better determined than those of
the \rass\ sources, a small fraction do have large ($>$30\,arcsec)
positional uncertainties. We have therefore used a relatively large
search radius of 60\,arcsec when compiling this initial list of
potential \allw\ counterparts.

\subsection{\sdss\ photometric counterpart catalogue}
\label{sec:data:sdss_photo}
As part of the first three phases of the Sloan Digital Sky Survey (\sdss), a
wide-area multi-band ($ugriz$) imaging survey was carried out using the
2.5-metre telescope at the Apache Point Observatory, New Mexico
\citep{Gunn06,Aihara11}. The most recent photometric catalogue derived
from these imaging data was released as part of \sdss-DR13
\citep{Albareti16}. The nominal 95~percent completeness limits of the \sdss\ imaging are 
22.0, 22.2, 22.2, 21.3 and 20.5\,mag in the $ugriz$ bands, 
respectively\footnote{\url{http://www.sdss.org/dr13/scope}}.
Starting with the \sdss-DR13 catalogue (specifically we considered primary objects from the `Datasweep' 
files\footnote{\url{http://www.sdss.org/dr13/imaging/catalogs/}}), we 
down-selected a sample of potential
optical counterparts to X-ray sources by searching within 65\,arcsec
of the X-ray positions of all members of the \rass\ and \xmmsl\
samples described in sections~\ref{sec:data:rass} and
\ref{sec:data:xmmsl}. The mean sky density of \sdss\ photometric
objects lying near the X-ray sources in our sample is
\eng{3.8}{4}\,\psqdeg.  Note that we adopt a slightly larger search
radius here compared to that used to select potential \allw\
counterparts. This is in order to prevent the effects of astrometric
uncertainties scattering potential \sdss\ counterparts to \allw\
sources outside our considered search area.

\subsection{\sdss\ spectroscopic counterpart catalogue}
\label{sec:data:sdss-spec}
Since its first light in 1998, and up to its 12th data release (DR12),
the \sdss\ project has obtained over five million astronomical
spectra, with various spectrographs and within a number of different
survey programs \citep{Alam15}. 

For practical reasons we have used two versions of the spectroscopic
catalogues associated with \sdss-DR12: i) an early prototype version
of the DR12 spectroscopic catalogue, and ii) the full official DR12
spectroscopic catalogue.  At the time when we created the \spiders\
target catalogues (March--April 2014), the final version of the \boss\
spectroscopic catalogue was not yet completely finalised (and did not
include the spectra collected as part of \sequels, a few \boss\
main-survey plates, and several plates associated with ancillary
programmes).  However, the combination of 
a pre-release version of the \boss\ spectroscopic catalogue, joined with
the official \sdss-DR8 spectroscopic catalogue included more than
95\% of the spectra that would go on to form the final official DR12
spectroscopic catalogue. 
To form a `proto-DR12' catalogue, we first
filtered the DR8 and pre-release \boss\
catalogues to include only spectra with \texttt{SPECPRIMARY}=1
(i.e. for each object with multiple spectroscopic observations we only
considered the best quality one, see \citealt{Alam15} for a
description of how this choice is made), and have \texttt{ZWARNING}=0
(i.e. we discarded spectra for which the \boss\ analysis pipeline
identified potential issues, including those spectra for which the
fitting routine could not determine a reliable redshift).  The 
combination of the filtered DR8 and pre-release \boss\ catalogues
(containing 1\,674\,844 and 1\,857\,100 sources respectively) were considered when we
decided which potential \spiders\ targets already had good-quality
\sdss\ spectra available (and hence should be excluded from the list of
potential targets in \sdssiv).  Use of this incomplete reference
catalogue for target selection is acceptable, as its only impact is a
slight decrease in observing efficiency, due to a small fraction 
(estimated to be less than 1\,percent) 
of \spiders\ AGN targets with existing \boss\ spectra being re-observed
during the \ebossts\ project.

However, as part of this paper we also release samples of \rass\ and
\xmmsl\ sources with existing \sdss\ spectra. Therefore, for the sake
of clarity and repeatability, and to allow reference to fully
documented data products, we prefer in these cases to base our samples
on the officially released \sdss-DR12 spectroscopic sample, which
contains all spectra collected by the \sdss\ and \boss\ optical
spectrographs \citep{Smee13} in the MJD range
[51578:56837]\footnote{\url{http://data.sdss3.org/sas/dr12/sdss/spectro/redux/specObj-dr12.fits}}.
In cases where multiple \sdss-DR12 spectra are available for a
photometric object, we consider only the `best' spectrum (i.e. the one
flagged with \texttt{SPECPRIMARY}=1).  Since we visually inspect all
spectra matched to our X-ray source samples (see section
\ref{sec:VI}), we have not filtered the DR12 spectra on the basis of
pipeline redshift warning flags.

\subsection{\xmmb\ reference catalogue}
\label{sec:data:3xmmb}
Within the Bayesian cross-matching framework (described in section
\ref{sec:method}), the complex task of identifying reliable
counterparts of \rass\ (and, to a lesser extent, \xmmsl) sources
benefits dramatically from the availability of a sub-sample of
well-characterized X-ray sources at similar flux levels, with good
enough positional accuracy to make the cross-matching exercise
non-problematic.  To this end, we have exploited the \xmmiii\
serendipitous source catalogue (DR4)\footnote{\url{http://nxsa.esac.esa.int/catalogues/3XMM_DR4cat_slim_v1.0.fits.gz}},
which is derived from over 7427 \xmm\ pointings performed up to
December 2012, covers a total of 794\,\sqdeg, and contains 372\,728
unique detections (for a description of the \xmmiii\ program, and a more recent
revision of the catalogue, see \citealt{Rosen16}). The median
flux of \xmmiii\ sources in the soft (0.2--2~keV) energy band is $\sim
6 \times 10^{-15}$\,\cgs, around two orders of magnitude fainter than
the flux limit of the \rass. However, the \xmmiii\ catalogue does
include a significant tail of bright sources, which overlaps with the
flux ranges spanned by the \rass\ and \xmmsl\ samples.
Therefore, we have exploited the X-ray bright end of the \xmmiii\ catalogue to
provide a reference catalogue of well understood X-ray bright sources
over the \boss\ imaging footprint. This (near) `truth' sample is used to
derive the priors that inform the Bayesian cross-matching process
(section~\ref{sec:method}).  
 
From the \xmmiii\ parent sample we select sources which meet all of
the following criteria: i) lie inside the \boss\ imaging footprint,
ii) have X-ray fluxes $\ge 10^{-13}$\,\cgs\ in the 0.2--2\,keV energy
band
(where $F_{\mathrm{0.2-2keV}}$ is calculated as the unweighted sum
over the individual flux measurements in the standard \xmmiii\ energy
bands: 0.2--0.5, 0.5--1 and 1--2\,keV), iii) have a very high detection
likelihood (\texttt{SC\_DET\_ML}$>$50), iv) have low likelihood of being
extended in the X-ray (\texttt{SC\_EXT\_ML}$<$8), v) have no warning
flags set (\texttt{SC\_SUM\_FLAG}=0), vi) are not associated with
observations of Solar System objects, and vii) are not in \xmm\
observations associated with poor astrometry (this final cut was
applied retrospectively)\footnote{Our first version of the \xmmb\
  catalogue included an additional eight sources, all of which had no
  \allw\ counterparts within 3\,arcsec. These eight sources were
  located close to each other within the `XMM-XXL North' field and
  were detected within two consecutive \xmm\ mosaic-mode observations
  (targets `XXLn074' and `XXLn094').  We have visually inspected
  independently processed images derived from these \xmm\ data-sets,
  and find no bright X-ray sources at the catalogued positions of the
  eight \xmmiii\ sources (N. Clerc, private communication). 
  Furthermore, the \xmmiii\ \texttt{POSCOROK} flag was set to \texttt{FALSE} 
  for 7/8 of these detections, indicating that the pipeline 
  had been unable to correct the X-ray astrometry against external optical/IR catalogues.
  Therefore we removed these objects from the \xmmb\ sample.}.

These cuts result in a high quality reference sample of 1049 bright,
well measured, point-like X-ray sources (which we will hereafter call
the \xmmb\ sample). The mean and 95th percentile of the 1$\sigma$
error radii for the \xmmb\ sources are 0.6\,arcsec and 1.4\,arcsec
respectively, making unambiguous cross-correlation with
multi-wavelength catalogues relatively simple (see
section~\ref{sec:priors} below).  Fig.~\ref{fig:xflux_histo_vs_xmmb}
demonstrates that the \xmmb\ sources have a similar range of X-ray
fluxes to sources in the \rass\ sample, 
but are approximately ten times fainter than sources in the \xmmsl\ sample\footnote{See 
Appendix~\ref{sec:3XMM_ECF_etc} for details of the method used to estimate
the unabsorbed fluxes of the \xmmb\ sample in the \rass\ and \xmmsl\ energy bands
(0.1--2.4 and 0.2--12\,keV, respectively).}.

\begin{table*}
  \caption{Summary of input catalogues after application of 
    preparation and filtering steps described in the text. 
    Note that the \rass, \xmmsl\ and \sdss-spectroscopy catalogues have been filtered to include only 
    objects lying inside the 10\,788.3\,\sqdeg\ \boss\ imaging footprint. 
    The \allw\ and \sdss\ imaging catalogues have been filtered to only include 
    objects lying within the indicated radii of \rass\ and \xmmsl\ sources.}
\label{tab:input}
\begin{tabular}{@{}lcrclc}
\hline
Catalogue name                     &  Waveband                        & Number of   & Mean density              & Notes      &  Section \\
                                   &                                  &  objects    & (\psqdeg)                 &            &          \\
\hline
\rass\                             & 0.1--2.4\,keV                    &     32\,408 & 3.0                       & `1RXS'. Combined BSC+FSC.  & \ref{sec:data:rass} \\
\rule{0pt}{4ex}\xmmsl\             & 0.2--12\,keV                     &        3843 & 0.36                      & Release~1.6, 2014~February~26. Unique sources. & \ref{sec:data:xmmsl}\\
\rule{0pt}{4ex}\multirow{2}{*}{\allw} & \multirow{2}{*}{3.4--22\,\um} &    450\,600 & \multirow{2}{*}{\eng{1.6}{4}} & $<$60\,arcsec from \rass\ sources  & \multirow{2}{*}{\ref{sec:data:allw}} \\
                                   &                                  &     51\,725 &                           & $<$60\,arcsec from \xmmsl\ sources & \\
\rule{0pt}{4ex}\sdss\ photometry   & \multirow{2}{*}{$ugriz$}         & 1\,081\,654 & \multirow{2}{*}{\eng{3.8}{4}} & $<$65\,arcsec from \rass\ sources  & \multirow{2}{*}{\ref{sec:data:sdss_photo}} \\
(DR13)                             &                                  &    126\,606 &                           & $<$65\,arcsec from \xmmsl\ sources & \\
\rule{0pt}{4ex}\sdss\ spectroscopy & 380--920\,nm  (SDSS)             & \multirow{2}{*}{3\,316\,373} & \multirow{2}{*}{310} & All \sdss-I,II, most \sdss-III         & \multirow{2}{*}{\ref{sec:data:sdss-spec}} \\
(proto-DR12)                       & 365--1040\,nm (BOSS)             &             &                           & \texttt{SPECPRIMARY}=1, \texttt{ZWARNING}=0         & \\
\rule{0pt}{4ex}\sdss\ spectroscopy & 380--920\,nm  (SDSS)             & \multirow{2}{*}{3\,658\,581} & \multirow{2}{*}{340} & All \sdss-I,II,III in \boss\ footprint & \multirow{2}{*}{\ref{sec:data:sdss-spec}} \\
(official-DR12)                    & 365--1040\,nm (BOSS)             &             &                           & \texttt{SPECPRIMARY}=1                              & \\
\hline
\end{tabular}
\end{table*}

\section{Selecting counterparts to X-ray sources}
\label{sec:method}
We summarise in this section the Bayesian cross-matching method used
to associate \rass\ and \xmmsl\ sources with optical/IR
counterparts. This method expands upon the techniques introduced by
\citet{Budavari08,Rots11}.  We refer the reader to \salvato\ for a
full and generalised description of the Bayesian cross-matching
method, a description of `\nway', the \textsc{Python}
implementation of the algorithm that we have used within this work, and a comparison to the likelihood ratio technique
\citep{Sutherland92}. Given that the \rass\ and \xmmsl\ surveys
  are far from the confusion limit, we make the simplifying assumption
  that each X-ray detection is dominated by a single X-ray source
  having up to one counterpart at longer wavelengths.  We briefly
reiterate below (section~\ref{sec:post_recipe}) those formulae that
are pertinent to the case considered here, that is, where we just wish
to find the best counterpart to each of a sample of X-ray sources from
a catalogue of potential counterparts.

\subsection{Bayesian posterior probability of associations}
\label{sec:post_recipe}

The posterior probability for a cross-match between an X-ray source,
$i$, and a potential counterpart $j$, is given by,
\begin{equation}
  \label{eq:post}
  P_{\mathrm{post}} = \left[1 + {\Pi(\mathbf{x}_j)\,B_{ij}\,\frac{1-P}{P}} \right]^{-1}
\end{equation}
where $\Pi(\mathbf{x}_j)$ is a prior (or `bias') term dependent on the
location of counterpart $j$ in some parameter space $\mathbf{x}$ (see
below), $B_{ij}$ is the Bayes factor for the geometric association, given by,
\begin{equation}
\label{eq:b_ij}
  B_{ij} = 
  \frac{2}{\sigma_{i}^2 + \sigma_{j}^2} 
  \exp{\left[- \frac{\psi^2_{ij}}{
  2(\sigma_{i}^2 + \sigma_{j}^2)} \right] }
\end{equation}
where $\psi_{ij}$ is the angular separation (in radians), and 
$\sigma_{i}$, $\sigma_{j}$ are the respective positional uncertainties for sources $i$ and $j$.
$P$ is a normalising factor which 
takes account of the mean sky density of potential counterparts ($\rho_s$, units \psqdeg), 
and the expected fraction of X-ray sources that have a true counterpart, $\eta_{x}$,
\begin{equation}
\label{eq:post_final}
  P= \frac{\eta_{x}}{4 \pi (180/\pi)^2 \rho_s} = 2.4241\times 10^{-5}\, \frac{\eta_{x}}{\rho_s}. 
\end{equation}

The value of the prior,
  $\Pi(\mathbf{x})$, at some location in N-dimensional measurement
  space, $\mathbf{x}$, is given by the ratio
  $f_{\mathrm{xray}}(\mathbf{x})/f_{\mathrm{all}}(\mathbf{x})$, where
  $f_{\mathrm{xray}}(\mathbf{x})$ is the probability density function
  of true counterparts to X-ray sources, (normalised such that
  $\int{f_{\mathrm{xray}}(\mathbf{x})d\mathbf{x}} = 1$ over the
  parameter range of interest), and $f_{\mathrm{all}}(\mathbf{x})$ is
  the probability density function of all potential counterparts,
  similarly normalised.
In general, $\Pi(\mathbf{x})$ can be used to encode some or all of our
prior knowledge of the distribution of the measurable
 properties of true counterparts to
X-ray sources (e.g. magnitudes, colours). Setting
$\Pi \equiv 1$ reduces Eqn.~\ref{eq:post} to the standard
unweighted form \citep[e.g.][]{Rots11}.

In practice we won't know $\Pi(\mathbf{x})$ exactly (as to do so would
require that we had already measured the multi-wavelength properties
of the X-ray sample), but we can estimate it, $\Pi^\prime(\mathbf{x})
\sim \Pi(\mathbf{x})$, using a 
training sample of well measured X-ray sources which we
expect to be representative of objects in our main X-ray sample.  In
section \ref{sec:priors} we describe how we have computed the
$\Pi^\prime(\mathbf{x})$ used to select counterparts to \rass\ and
\xmmsl\ sources.

We note that the $P_{\mathrm{post}}$ statistic measures the
  probability of association of individual pairs of sources
  independently of the presence of other possible pairings.  So, for
  example, although $P_{\mathrm{post}}$ cannot tell us the probability
  that at least one of the possible counterparts to an X-ray source is
  the correct one, it can be used to choose the most probably of these
  counterparts.  A more complete treatment of such cases is
  implemented within the soon to be released version of the \nway\
  code \salvato.  See also the treatment of multiple potential
  counterparts by \citet{Pineau11}.

\subsection{Bayesian priors derived from a bright X-ray reference sample}
\label{sec:priors}

We investigated the properties of the \xmmb\ sample searching for
simple combinations of parameters in which the X-ray sources stand out
clearly from the general field population.  This is complicated by the
heterogeneous nature of the bright X-ray source population.  For
example, \citet{Zickgraf03} found that the optically bright end of the
\rassb\ catalogue is associated with a mix of AGN, galaxies, galaxy
clusters, M~stars, white~dwarfs, K~stars, F-G~stars, and cataclysmic
variables.

High luminosity AGN (i.e. QSOs) typically outshine their host galaxies
in the optical--mid-IR bands, but for lower luminosity AGN (Seyferts),
the host-galaxy emission may match or exceed the AGN emission.
In addition, extinction along the line of sight could also mask any AGN
signature in the UV/optical bands, and even into the near-IR bands as
the extinction increases. However, mid-IR colours of bright AGN are
less susceptible to these effects, and promise to provide a more
universal tool for AGN identification.

The release of the \allw\ catalogue in November 2013 \citep{Cutri13}
spurred us to examine whether mid-IR imaging information could assist
in optical counterpart selection for the \rass\ and \xmmsl\
sources. The vast majority of the X-ray bright AGN detected in \rass\
and \xmmsl\ are expected to be bright and red in the two shortest
wavelength \wise\ channels (i.e. 3.4 and 4.6\,\um). Therefore, \rass-selected
AGN should stand out from field stars and galaxies in the mid-IR. In
addition, counterparts to the non-AGN `contaminants' in the X-ray
sample (i.e. Galactic stars, and bright nearby galaxies) are also
expected to stand out (they are likely to be significantly brighter in
the \wo\ and \wt\ bands than the bulk of the \allw\ field population).
Furthermore, the sky density of field sources in the \allw\ catalogue
is less than half that in the \sdss\ imaging catalogue (see
Table~\ref{tab:input}), which dramatically reduces the rate of false
identification w.r.t. a purely optical-based counterpart selection
scheme. The useful dynamic range of the \allw\ catalogue is somewhat
larger than that of the \sdss\ catalogue (as expected, given that
\wise\ was designed as a true-all sky surveyor, whereas the \sdss\
imager was primarily designed to target faint galaxies and QSOs).  In
particular, the treatment for X-ray sources associated with very
bright stars is simpler with \allw; point sources start to saturate at
\wt$<6.7$\,mag in \wise\ \citep{Cutri13}, but saturate at
$r<14.1$\,mag in \sdss\ imaging \citep[][]{gunn98}.

We start with the \xmmb\ catalogue described in Section
\ref{sec:data:3xmmb}, and search for counterparts in the \allw\
catalogue, using a simple cone search. We find that 1000/1049 (95.2\%)
of the \xmmb\ sources had exactly one \allw\ counterpart within
3\,arcsec of the X-ray position. We denote these 1000 matches as the
`\xmmbw' sample. 
The mean sky density of \allw\ field sources local to the \xmmbw\ sources is 15\,300
  objects\,\psqdeg\ (estimated by measuring the density of \allw\ sources 
  within 1\,arcmin radius control regions placed at a distance 6\,arcmin from each X-ray source). 
  Therefore we would na{\"i}vely expect 0.033 field
  sources to fall within the 3\,arcsec radius circle we searched
  within for each \xmmb\ source.  However, the typical \allw\ counterparts to
  the \xmmb\ sources are much brighter than typical field sources
  (the median \wt\ magnitude of the \xmmbw\ matches is 13.2\,mag
  compared to 16.6\,mag for the \allw\ field population), and so an 
  average \allw\ field source would be
  overwhelmed by the average \allw\ counterpart to a \xmmb\ source
  (see also section~5 of \citealt[][]{Broos11}). We have
  empirically estimated the rate at which random associations are both
  close enough and bright enough to contaminate the \xmmbw\ sample:
  for each \xmmbw\ source we measured the probability of a field \allw\
  source brighter (in \wt) than the true \allw\ counterpart to lie 
  within any randomly placed 3\,arcsec circle. 
  We tested 100 randomized locations per \xmmbw\ source for a total of 10$^5$
  samples, and find an almost negligible overall contamination rate of just 0.12\,percent.
We note that using a  matching radius 
larger than 3\,arcsec would result in a higher
completeness (fraction of X-ray sources with \allw\ counterparts), but
would lead to more contamination from chance aligned field sources, and
would also increase the number of cases where there is more than one
potential \allw\ counterpart per \xmmb\ source (making unambiguous
associations difficult). In Appendix~\ref{sec:xmmb_ir_faint} we
discuss in more detail the nature of the 49/1049 \xmmb\ sources
which do not have \allw\ counterparts within 3\,arcsec, and conclude
that only 3/49 are likely to be associated with genuinely mid-IR faint
sources, the remaining cases are the result of various problems with
either \xmm\ or \allw\ photometry, e.g. blending/confusion, presence
of saturated bright stars, and poor astrometry at the edge of the
\xmm\ field of view. Therefore, we are confident that our reference
sample is not artificially excluding a significant part of the bright
X-ray population.

\begin{figure}
\begin{center}
\includegraphics[angle=0,height=59mm]{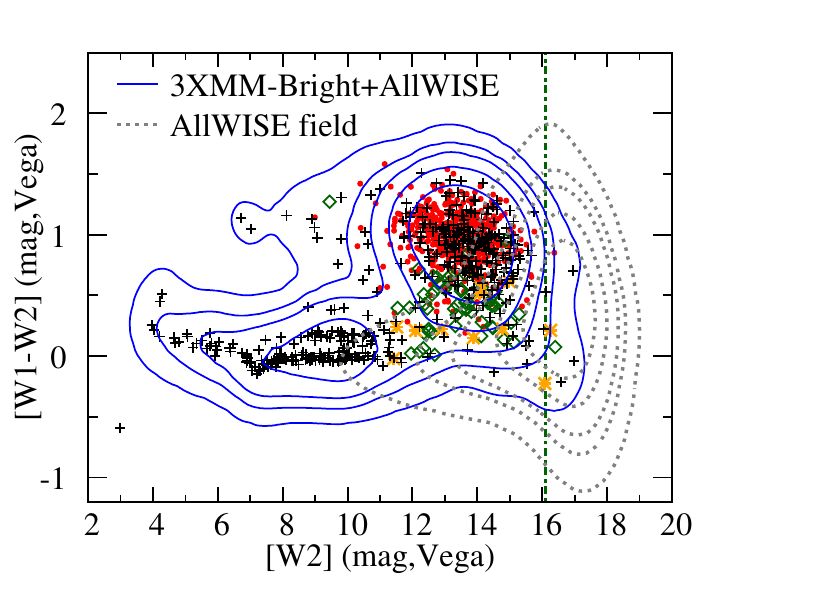}
\end{center}
\begin{center}
\vspace{-0.8cm}
\includegraphics[angle=0,height=59mm]{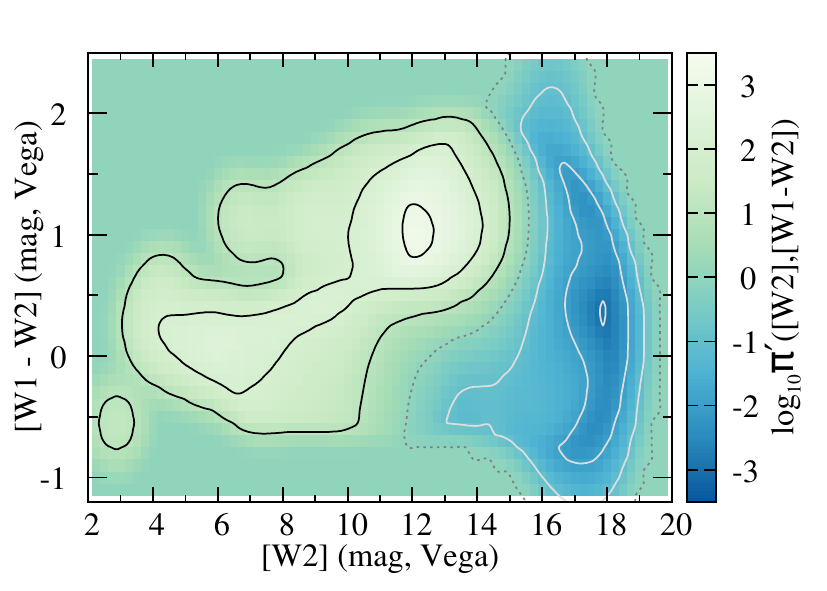}
\end{center}
\caption{{\em Top panel:} Distribution of \xmmbw\ and \allw\ field
  sources in \allw\ colour-magnitude space. \xmmbw\ sources with
  \sdss-DR12 spectroscopic classifications are shown with red filled
  points (\texttt{CLASS=QSO}), green open diamonds (\texttt{CLASS=GALAXY}), and
  orange asterisks (\texttt{CLASS=STAR}). \xmmbw\ sources lacking
  spectroscopic classifications are shown with small black + signs.
  Contours contain 99, 95, 90, 75, and 50~percent of the smoothed
  density of each catalogue.  The vertical line (green dot-dashed)
  shows the nominal 5$\sigma$ detection limit for the \wt\ band.  {\em
    Bottom panel:} Map of the prior $\Pi^\prime([W2],[W1-W2])$,
  illustrating the weighting given to potential counterparts as a
  function of location in \wt,\wowt\ colour-magnitude space (see
  section~\ref{sec:post_recipe}).  Potential counterparts lying in
  locations with light (green) colours are up-weighted, and those lying in
  regions having darker (blue) colours are down-weighted.  Contours are drawn
  at log$_{10}\Pi^\prime([W2],[W1-W2])$ = 3,2,1,0,-1,-2,-3.}
\label{fig:priors}
\end{figure}

We have used the following recipe to compute a pixelized map of
  the prior, $\Pi^\prime(\mathbf{x})$, in the \wt,\wowt\
  colour-magnitude space, using as input the training sample (\xmmbw\
  sources) and a `field' sample (consisting of \allw\ objects lying within 1\,arcmin radius control
  regions placed at a distance of 6\,arcmin from each of the sources in the
  \xmmbw\ sample): i) compute 2D histograms from each of the training
  and field samples, covering the parameter interval
  0$\le$\wt$\le$25,~-5$\le$\wowt$\le$5, with pixel steps of 0.25\,mag
  in \wt\ and 0.1\,mag in \wowt, ii) smooth each 2D histogram with a
  2D Gaussian kernel having $\sigma_{\mathrm{[W2]}}$=0.5 and
  $\sigma_{\mathrm{[W1-W2]}}$=0.2, iii) normalise each 2D histogram
  such that the sum of the pixel values over the considered range
  (0$\le$\wt$\le$25,~-5$\le$\wowt$\le$5) equals unity, iv) threshold
  each 2D histogram such that no pixel has a value smaller than one
  over the total number of objects in the field sample.  The chosen
  smoothing kernel is a fair compromise between the desire to reduce
  the shot noise of the training sample (pushing to larger kernels),
  versus the desire to retain the separation between the distributions
  of the X-ray reference sample and of field sources (pushing to
  smaller kernels). The parameter interval over which we have defined
  the prior contains more than $1-10^{-6}$ of the smoothed PDF for the
  \xmmbw\ sample.  
The distribution of the \xmmbw\ sample in \allw\
colour-magnitude space is presented in the upper panel of
Fig.~\ref{fig:priors}, illustrating clearly how these bright X-ray
selected sources stand out from the field population in this
measurement space.  The ratio of the density distribution of our
training (the \xmmbw\ sources) and field samples
constitutes our Bayesian prior
  $\Pi^\prime([W1],[W1$-$W2])$, which is also shown in the bottom
panel of Fig.~\ref{fig:priors}, illustrating which parts of
colour-magnitude space are up- and down-weighted by the cross-matching
routine.

In order to get a sense of the mix of sources that make up the \xmmbw\
sample, we have cross-matched to the \sdss~DR12 spectroscopic
catalogue, searching within a 2\,arcsec radius from the \allw\
position. Of the 557 \xmmbw\ sources with \sdss\ spectra and reliable
\boss\ pipeline automated classifications, 490 are classified as
\texttt{QSO}, 56 as \texttt{GALAXY} (13 of which have a
  sub-class indicating AGN activity) and 11 as \texttt{STAR}.  The
selection function for the sub-sample of \xmmbw\ sources with spectra
is difficult to determine, but is certainly highly incomplete at the
bright and faint ends.  Therefore, we should not make quantitative
predictions from these fractions. However, we can at least say with
some confidence that for the range of X-ray fluxes probed by the
\rass\ and \xmmsl\ samples, the spectroscopic samples collected by
\spiders\ will be dominated by QSOs, but with significant minorities
of both normal galaxies and Galactic stars.  Note that of the \xmmbw\
sources with no spectral classification in \sdss~DR12, many lie in the
part of colour-magnitude space expected to be occupied by Galactic
stars and very bright nearby galaxies (\wt$<11$, \wowt$<0.3$, see
upper panel of Fig.~\ref{fig:priors}).

\subsection{Selection of \rassagn\ targets}
\label{sec:assoc:rass}
Armed with the priors described in the previous section, we can now
proceed to the identification of the \rass\ sources.  We start with
the \allw\ catalogue described in section~\ref{sec:data:allw}.  All
except 11/32\,391 (0.03~percent) of the \rass\ sources in our sample
have at least one potential \allw\ counterpart lying within 1\,arcmin.
For each \rass\ source, we used \nway\ (version 1.0) to calculate the
posterior probability of it being associated with each of the possible
\allw\ counterparts within 1\,arcmin, taking account of the X-ray
position and its uncertainty, the \allw\ position, the number density of the sources, the \wt\ magnitude
and the \wowt\ colour.  
At the time
  of generation of the \rassagn\ target catalogue we made a decision
  to adopt a fixed radial positional uncertainty of 0.3\,arcsec
  (1$\sigma$) for all \allw\ sources.  In retrospect this was an
  unnecessary and sub-optimal choice, and it would have been much more
  correct to use the positional uncertainties tabulated in the \allw\
  catalogue. However, we do not expect that this has had a
  significant effect on our target selection given that the X-ray
  positional uncertainties are almost always much larger than those of \allw\
  sources, and so dominate the denominators of Eqn.~\ref{eq:b_ij}.

In an ideal world we would obtain spectra for all potential
counterparts for each X-ray source above some minimum \post.  However,
the combination of the \boss\ spectrograph plug-plate
fiber-collision constraint (minimum fibre separation $>$62\,arcsec),
the single-pass survey strategy over the main \ebossts\ footprint
\citep{Dawson16}, and the limited fibre-budget allocated to the
\spiders-AGN program, mean that we only attempt to target a single
potential counterpart per \rass\ source. Therefore, for each \rass\
source, only the `best' \allw\ counterpart (i.e. the one having the
highest posterior probability) was considered in the following steps.

\begin{figure}
\begin{center}
\includegraphics[angle=0,width=84mm]{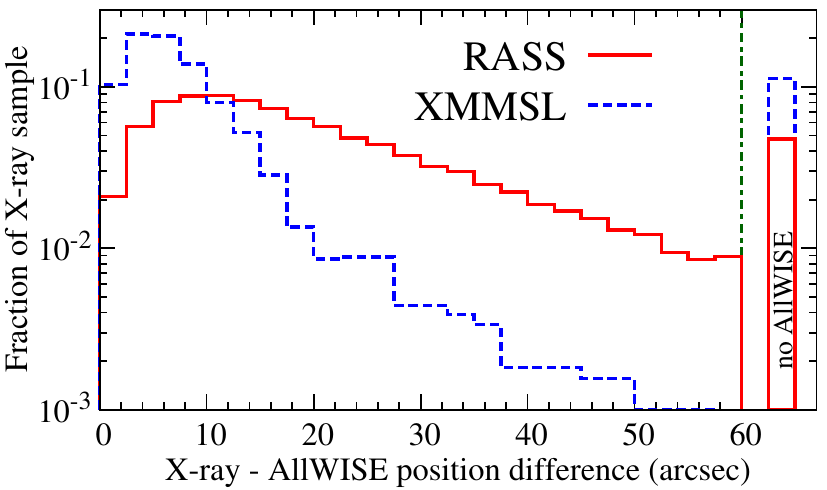}
\end{center}
\caption{Distribution of separations between X-ray positions and
  \allw\ positions for the \rassw\ (red solid line) and \xmmslw\ (blue
  dashed lines) samples.  X-ray sources without any valid \allw\
  counterpart (i.e. no counterparts with \post$\ge$0.01), are
  represented in the rightmost bin.  The vertical line (green
  dot-dashed) shows the maximum search radius considered.}
\label{fig:Xray_AllWISE_offset}
\end{figure}

\begin{figure}
\begin{center}
\includegraphics[angle=0,width=84mm]{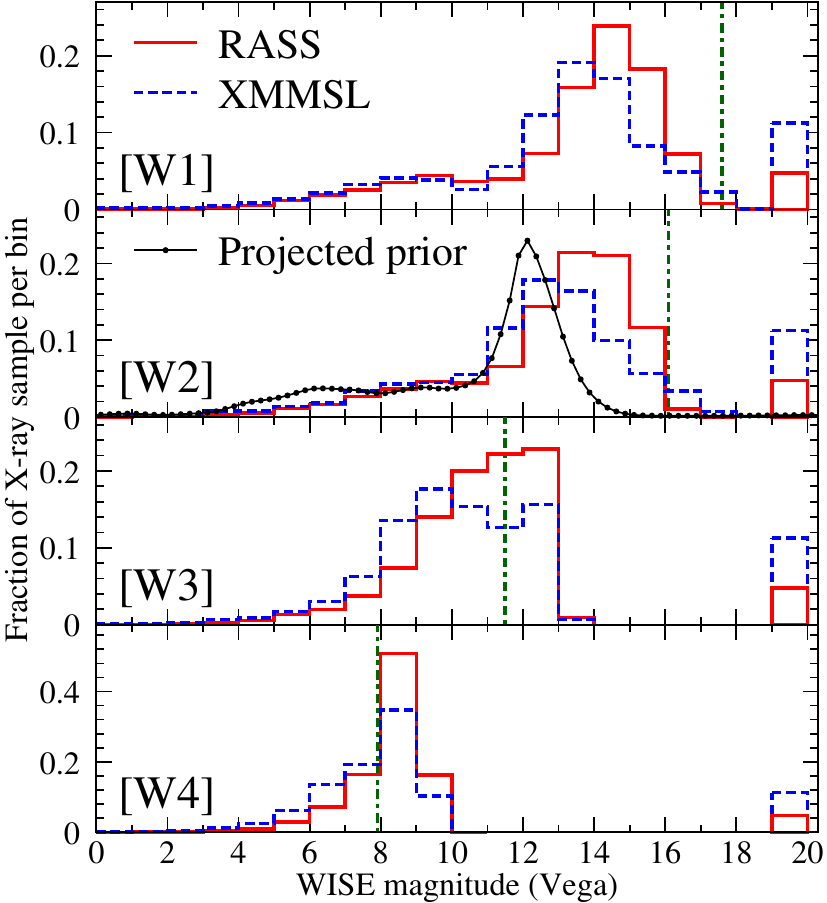}
\end{center}
\caption{Distribution of mid-IR brightness in the \wise\ (\wo, \wt,
  \wthree\ and \wfour) bands for the \rass\ (red solid line) and
  \xmmsl\ (blue dashed lines) samples, as measured with \allw\
  point-source photometry.  X-ray sources with no \allw\ counterpart
  are represented in the faintest magnitude bin. Vertical lines (green dot-dashed)
  show the nominal 5$\sigma$ point source {\em detection} limits (valid for
  $>95$~percent of the extragalactic sky) in each \wise\ band. In the second 
  (\wt) panel we also show (black line with points) an arbitrarily scaled 
  projection of $\Pi^\prime([W1],[W1$-$W2])$, (averaged over the \wowt\ axis).}
\label{fig:AllWISE_mag_histos}
\end{figure}

We find that for 30\,855/32\,391 (95.3~percent) of \rass\ sources we
have a best matching \allw\ counterpart with posterior probability
\post$\ge$0.01 (see section~\ref{sec:verify:spurious_rate} for a
discussion of the \post\ threshold below which the sample becomes
significantly contaminated by interlopers).  
\rass\ sources with best \allw\ counterparts having \post$<$0.01 were
not considered further.  
The best matching \allw\ counterparts
  have a median \post\ of 0.86 compared to 0.025 for the second best
  counterparts. In 82\,percent of cases, the \post\ for the best match
  is more than twice that for the second best match indicating a very
  secure choice. In a small fraction of cases (4\,percent), the second
  best match has a \post\ within 10\,percent of the best match, and
  for such cases we cannot differentiate significantly between the
  best and second best \allw\ counterparts.  See
section~\ref{sec:verify:RASS_spurious} for a discussion of how the
fraction of \rass\ sources lacking any \post$\ge$0.01 \allw\
counterparts depends on X-ray detection likelihood. The distribution
of position differences between the \rass\ sources and their best
matching \allw\ counterparts is shown in
Fig.~\ref{fig:Xray_AllWISE_offset}. The magnitude distributions of the
best matching \allw\ counterparts are shown in
Fig.~\ref{fig:AllWISE_mag_histos}.

\begin{figure}
\begin{center}
\includegraphics[angle=0,width=84mm]{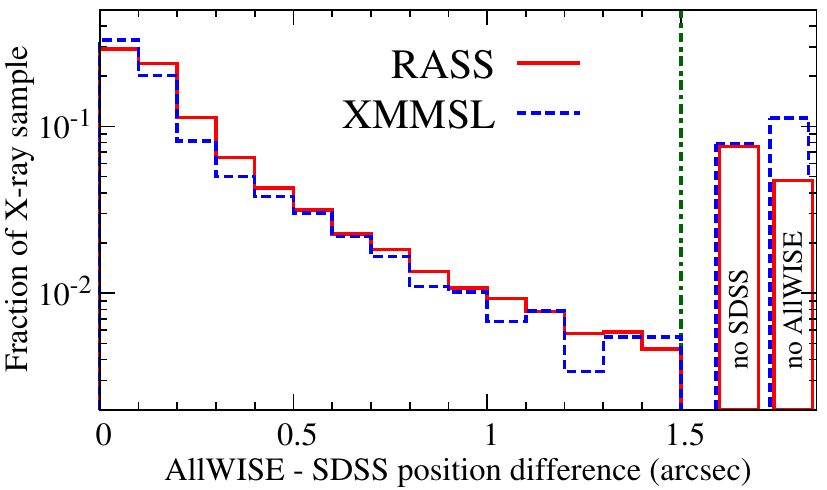}
\end{center}
\caption{Distribution of separations between \allw\ positions and
  \sdss\ photometric positions for the \rass\ and \xmmsl\ sources with
  \allw\ counterparts. For comparison, the fraction of each X-ray sample 
  without a valid \allw\ counterpart and the fraction of X-ray+\allw\ sources without an optical
  counterpart, are represented in the two rightmost bins.  The vertical line
  (green dot-dashed) shows the maximum search radius considered.}
\label{fig:AllWISE_SDSS_offset}
\end{figure}

\begin{figure}
\begin{center}
\includegraphics[angle=0,width=84mm]{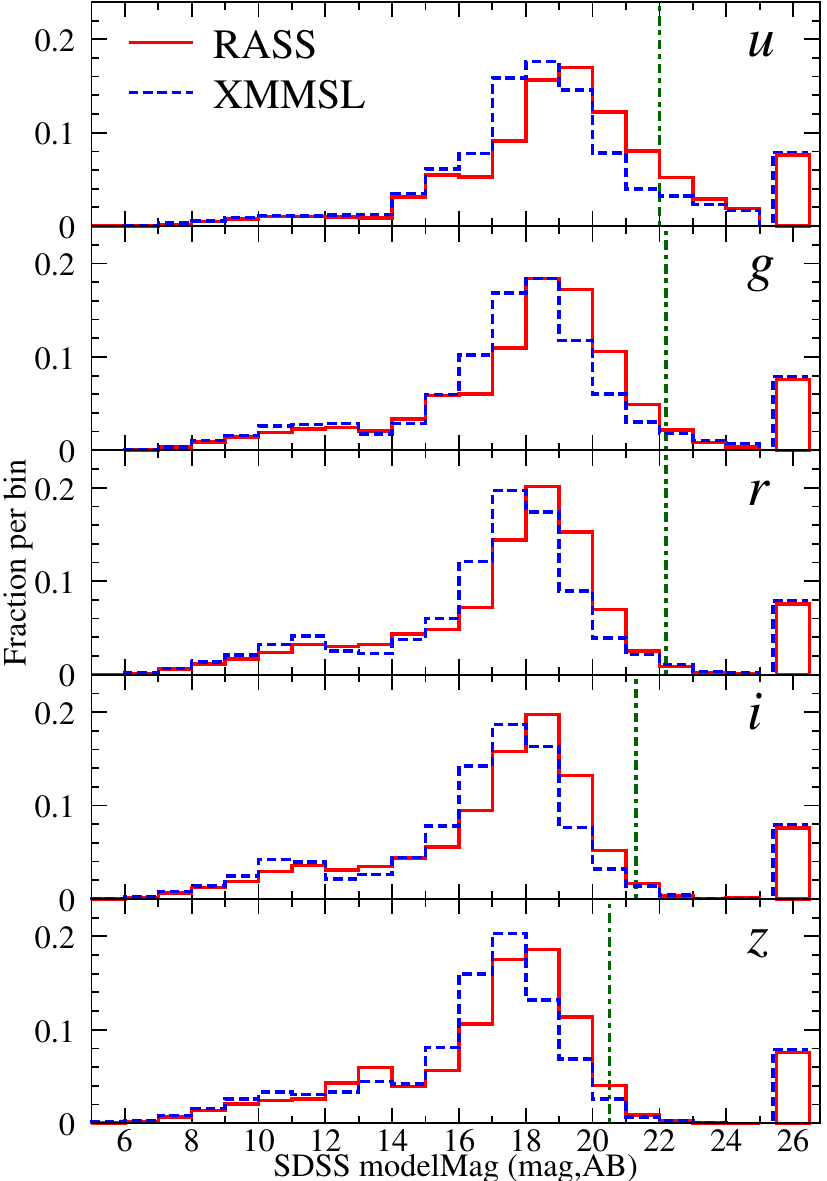}
\end{center}
\caption{Distribution of optical brightness in the \sdss\ $ugriz$
  filters for the \rass\ and \xmmsl\ sources with \allw\ counterparts,
  as measured by the \sdss-DR13 photometric catalogue
  \texttt{modelMag} parameter.  X-ray+\allw\ sources without an
  optical counterpart are represented in the faintest magnitude bin.
  Vertical lines (green dot-dashed) show the nominal 95~percent
  completeness limits in each \sdss\ filter. }
\label{fig:optical_mag_histos}
\end{figure}

The best \allw\ counterpart for each \rass\ source was then matched to
an optical counterpart from the \sdss-DR13 photometric catalogue. The
brightest (as measured by \texttt{modelMag}$\_r$) object within
1.5\,arcsec of the \allw\ position is chosen as the most probable
optical association.  At the time of generating the \spiders\
  targeting catalogues (early 2014), this choice seemed
  appropriate. However, later analysis suggests that better choices
  could have been made. In section~\ref{sec:xray_no_sdss_photom} we
  discuss how our simplistic approach to associating \allw\ sources
  with optical counterparts has impacted the target selection. We will
  make improvements to this step in future \spidersagn\ studies,
  taking into account the positional uncertainties and sky densities
  of the mid-IR and optical populations.
We find that 28\,515/30\,855 (92.4~percent) of the \rassw\ sources have
at least one \sdss-DR13 photometric optical counterpart within
1.5\,arcsec of the \allw\ position. The distribution of \allw--\sdss\
positional offsets is shown in Fig.~\ref{fig:AllWISE_SDSS_offset}, and
the magnitude distribution is shown in
Fig.~\ref{fig:optical_mag_histos}.

We then further down-select the 28\,515 remaining potential \rassws\
targets to get to our final list of \spiders\ objects to be forwarded
to the \eboss\ tiling team.  Firstly, we removed previously
spectroscopically identified sources by matching against a list of
good-quality primary \sdss-proto-DR12 spectra (see
section~\ref{sec:data:sdss-spec}).  There are 11\,643 \rassws\ sources
with at least one reliable \sdss-proto-DR12 spectrum (within 1\,arcsec
of the optical photometric catalogue position), and thus, which are
not considered for targeting.  The majority of these (77~percent) have
a pipeline classification of \texttt{CLASS}=\texttt{QSO}.  The
properties of the spectroscopically identified sources are discussed
in more detail in section~\ref{sec:RASS_with_spec}.

We then remove \rassws\ objects brighter than the nominal magnitude
limit of \eboss\ (i.e. \texttt{fiber2Mag}\_i$<$17.0, where \texttt{
  fiber2Mag}\_i is a measure of the expected flux from the object that
would be enclosed within a 2\,arcsec diameter fiber under average
seeing conditions). This bright source cut removes 7092 objects.  To
improve our robustness against imperfect modelling of very bright (and
possibly saturated) objects, we also remove a small number (319) of
objects which escape the \texttt{fiber2Mag}\_i cut but which have
\texttt{modelMag}\_i$<$16.0.  These bright limits are motivated by the
desire to avoid the on-chip spectra of very bright stars overwhelming
the spectra of their neighbours, which can be many magnitudes fainter.
We also remove the 283 very faint targets
(\texttt{fiber2Mag}\_i$>$22.5, for which we do not expect to be able
to obtain useful spectra with a 2.5\,m-class telescope), and the 150
targets which fall within the \boss-DR10 bright star
mask\footnote{{\url{http://data.sdss3.org/sas/dr10/boss/lss/reject_mask/bright_star_mask_pix.ply}}}.
The final \rassagn\ target list contains 9028 candidate targets over
the full \boss\ targeting footprint (area 10\,778\,\sqdeg), a density
of 0.84\,targets\,\psqdeg.

\begin{figure}
\begin{center}
\includegraphics[angle=0,width=84mm]{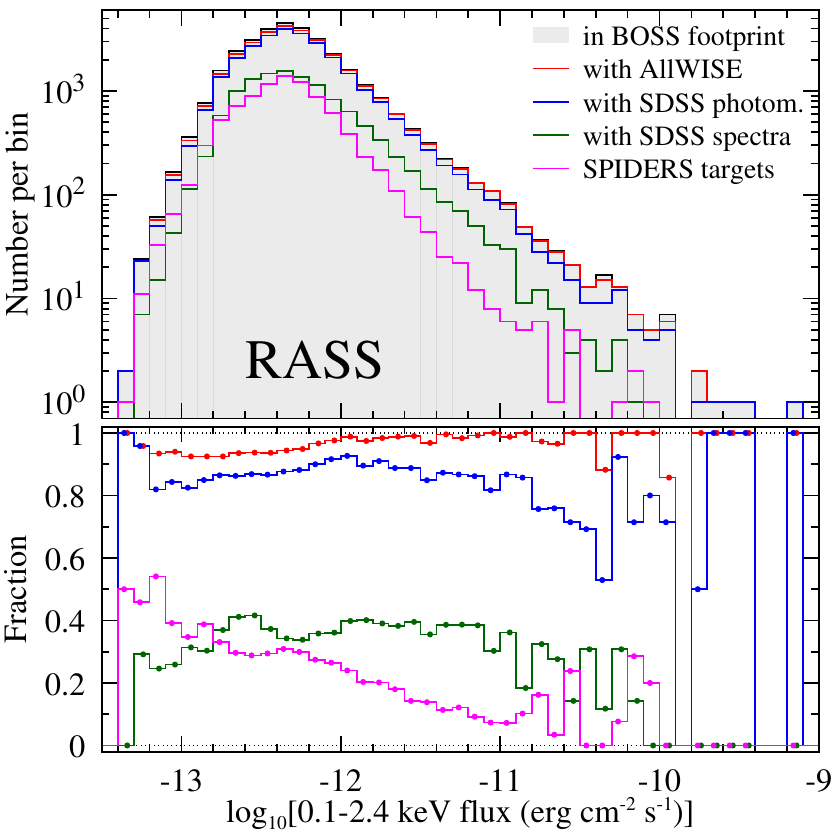}
\includegraphics[angle=0,width=84mm]{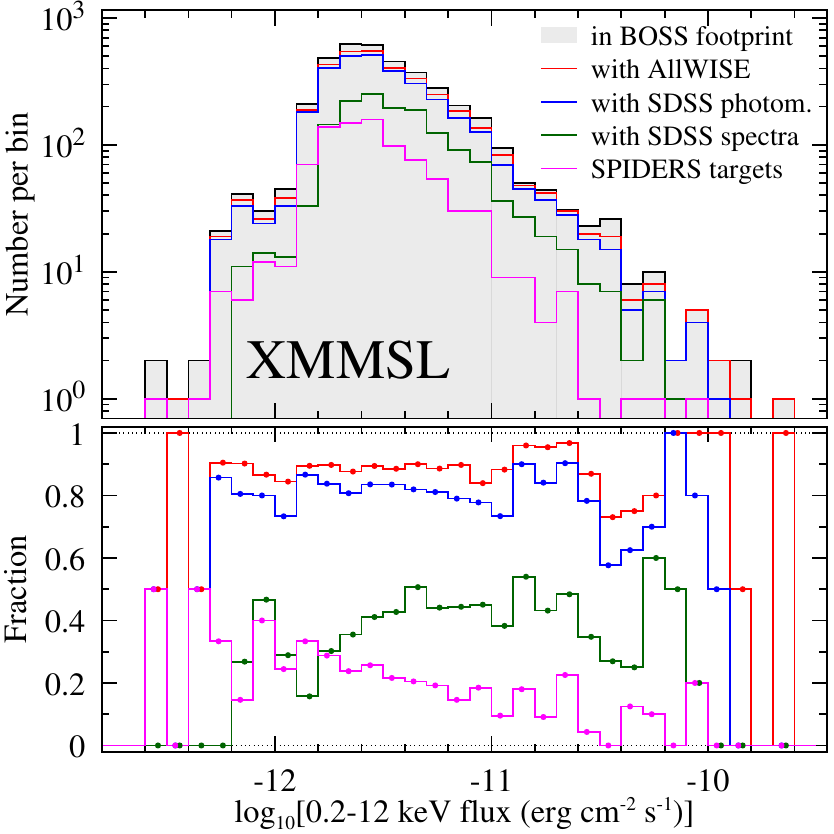}
\end{center}
\caption{{\em First panel:} The X-ray flux distribution of \rass\
  sources at various stages of the cross-matching and down-selection
  process described in section~\ref{sec:assoc:rass}.  From top to
  bottom the curves show the X-ray flux distributions for the
  following subsets: all \rass\ sources within the \boss\ imaging
  footprint (thick black line), \rassw\ sources (red), \rassws\
  sources (blue), \rasswsz\ sources (green), and \rassagn\ sources
  (magenta).  {\em Second panel:} The same information as the upper
  panel, but shown as a ratio.  The curves have been divided by the
  total number of \rass\ sources in each flux bin.  {\em Third and
    fourth panels:} The same information shown for the \xmmsl\
  sample.}
\label{fig:xflux_histo}
\end{figure}

\begin{figure}
\begin{center}
\includegraphics[angle=0,width=84mm]{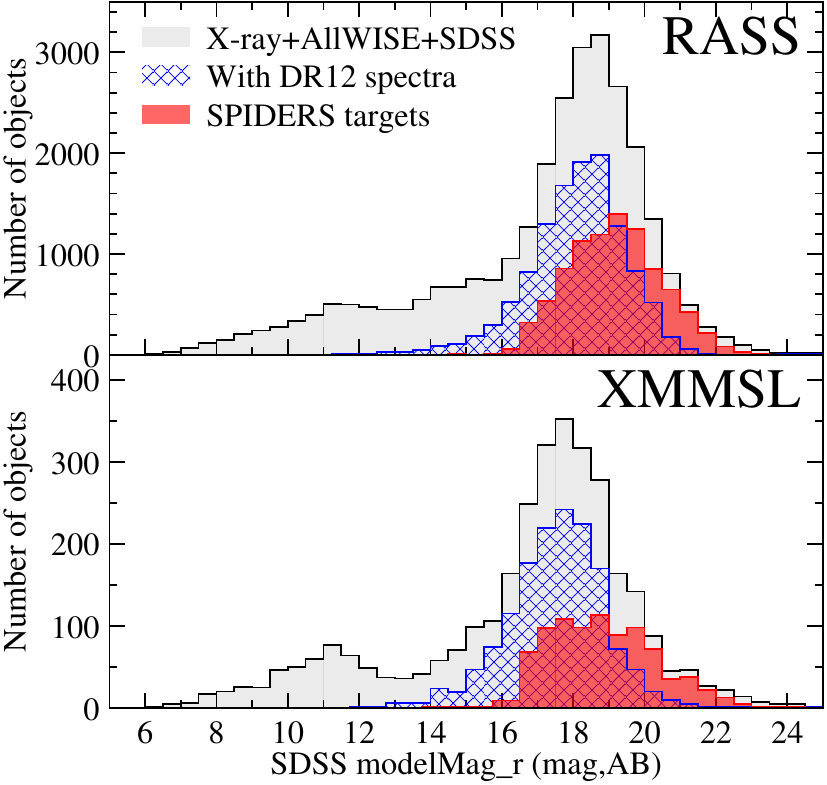}
\end{center}
\caption{The \sdss\ $r$-band magnitude distribution of the optical
  counterparts to X-ray sources (grey histograms), those with
  spectroscopy available in \sdss-DR12 (blue hatched histograms) and
  of the \spiders\ targets to be observed during \sdssiv. (red
  histograms). The upper panel shows \rass\ sources, and the lower
  panel shows \xmmsl\ sources.}
\label{fig:rmag_histo_by_status}
\end{figure}

\begin{figure}
\begin{center}
\includegraphics[angle=0,width=84mm]{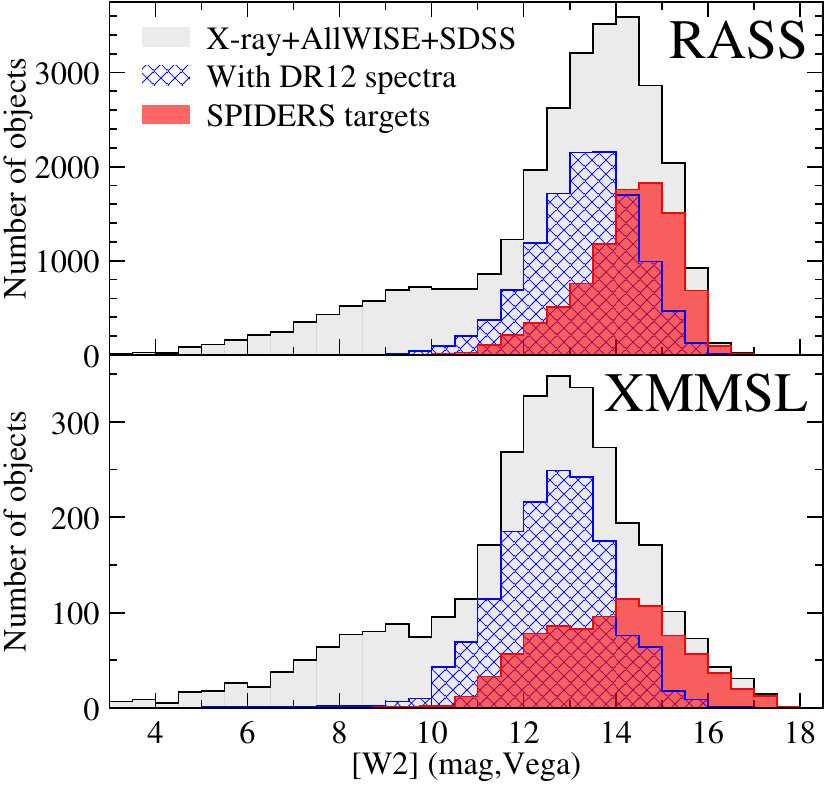}
\end{center}
\caption{The \wt\ magnitude distribution of the \allw\ counterparts of
  X-ray sources (grey histograms), those with spectroscopy available
  in \sdss-DR12 (blue hatched histograms) and of the \spiders\ targets
  to be observed during \sdssiv (red histograms). The upper panel
  shows \rass\ sources, and the lower panel shows \xmmsl\ sources.}
\label{fig:w2mag_histo_by_status}
\end{figure}

The matching and filtering steps described above are summarised in a
flow diagram, see Fig.~\ref{fig:RASS_flow}.  The format of the
catalogue of \rassagn\ targets is described in
Appendix~\ref{sec:RASS_SPIDERS_targets}.  In
Fig.~\ref{fig:RASS_XMMSL_potential_maps} we show the sky distribution
of \rassagn\ targets over the sky. Note that the SGC contains a higher
density of targets than the NGC because only a small part of this
footprint was targeted during \sdss-I/II, and so a smaller fraction of
the \rass\ sources have existing identifications. 
In Figs.~\ref{fig:xflux_histo},
\ref{fig:rmag_histo_by_status} and \ref{fig:w2mag_histo_by_status} we
show the distributions of the \rass\ sample in X-ray flux, $r$-band
model magnitude and \wt\ magnitude at various stages during the
cross-matching and down-selection steps described above.

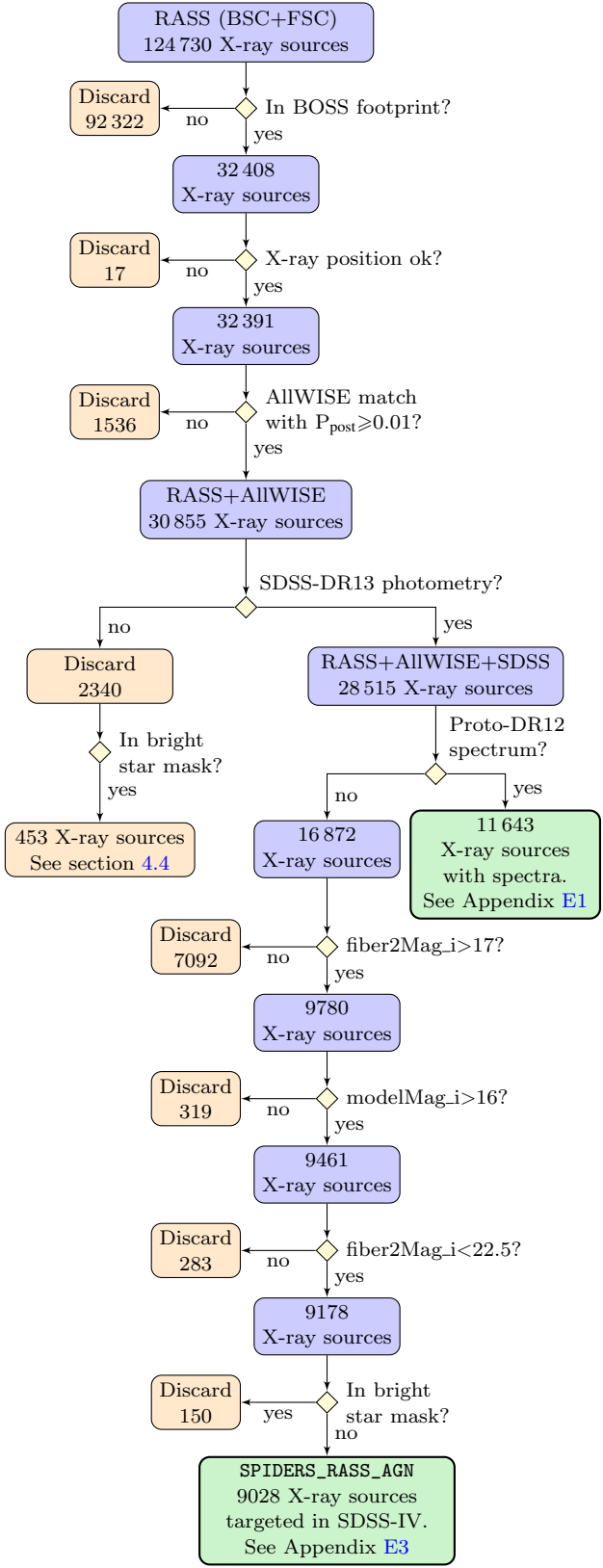
\begin{figure}
\begin{center}
\begin{tikzpicture}[node distance = 1.0cm, auto, >=triangle 60]
    \tikzstyle{every node}=[font=\small]

    \node [block,text width=11em]                                                             (X)            {\rassbf\ 124\,730 X-ray sources };
                                                                                            
    \node [decision, below of=X,label=right:{In \boss\ footprint?}]                           (qXB)          {};
    \node [block,    below of=qXB]                                                            (XB)           {32\,408\\X-ray sources};
    \node [blockd,   left  of=qXB]                                                            (RnB)          {Discard\\92\,322};
                                                       
    \node [decision, below of=XB,label=right:{X-ray position ok?}]                              (qXBP)         {};
    \node [block,    below of=qXBP]                                                           (XBP)          {32\,391\\X-ray sources};
    \node [blockd,   left  of=qXBP]                                                           (XBnP)         {Discard\\17};
                                                                         
    \node [decision, below of=XBP,label={[text width=2.5cm]right:\allw\ match with \post$\ge$0.01?}]         (qXBPW)        {};
    \node [block,    below of=qXBPW,node distance=4.5em,text width=9em]                       (XBPW)         {\rassw\\30\,855 X-ray sources};
    \node [blockd,   left  of=qXBPW]                                                          (XBPnW)        {Discard\\1536};
                                                                                                    
    \node [decision, below of=XBPW,label=above right:{SDSS-DR13 photometry?},node distance=4.5em]   (qXBPWS)       {};
    \node [block,    below right of=qXBPWS,xshift=+5.5em,node distance=4.5em,text width=11em] (XBPWS)        {\rassws\\28\,515 X-ray sources};
    \node [blockd,   below left  of=qXBPWS,xshift=-3.5em,node distance=4.5em,text width=6em]  (XBPWnS)       {Discard\\2340};
                                                                                                            
    \node [decision, below of=XBPWS,label={[text width=2cm]above right:Proto-DR12 spectrum?},node distance=4.5em](qXBPWSZ)      {};
    \node [result,   below right of=qXBPWSZ,yshift=-1.0em,text width=8em]                     (XBPWSZ)       {11\,643\\X-ray sources with spectra.\\See~Appendix~\ref{sec:RASS_DR12_catalogue}};
    \node [block,    below left  of=qXBPWSZ,yshift=-1.0em,xshift=-2.5em]                      (XBPWSnZ)      {16\,872\\X-ray sources};
                                                                                                            
    \node [decision, below of=XBPWSnZ,yshift=-1.0em,label=right:fiber2Mag\_i$>$17?]                          (qXBPWSnZI)    {};
    \node [block,    below of=qXBPWSnZI]                                                      (XBPWSnZI)     {9780\\X-ray sources};
    \node [blockd,   left  of=qXBPWSnZI]                                                      (XBPWSnZnI)    {Discard\\7092};
                                                                                              
    \node [decision, below of=XBPWSnZI,label=right:modelMag\_i$>$16?]                         (qXBPWSnZII)   {};
    \node [block,    below of=qXBPWSnZII]                                                     (XBPWSnZII)    {9461\\X-ray sources};
    \node [blockd,   left  of=qXBPWSnZII]                                                     (XBPWSnZnII)   {Discard\\319};
                                                                                                             
    \node [decision, below of=XBPWSnZII,label=right:fiber2Mag\_i$<$22.5?]                     (qXBPWSnZIIi)  {};
    \node [block,    below of=qXBPWSnZIIi]                                                    (XBPWSnZIIi)   {9178\\X-ray sources};
    \node [blockd,   left  of=qXBPWSnZIIi]                                                    (XBPWSnZnIIi)  {Discard\\283};
                                                                                                 
    \node [decision, below of=XBPWSnZIIi,label={[text width=1.5cm]right:In bright star mask?}]                    (qXBPWSnZIIiB) {};
    \node [result,   below of=qXBPWSnZIIiB,yshift=-0.5em]                                     (XBPWSnZIIiB)  {\rassagn\\9028~X-ray~sources\\targeted in \sdssiv.\\See~Appendix~\ref{sec:RASS_SPIDERS_targets}};

    \node [blockd,   left  of=qXBPWSnZIIiB]                                                   (XBPWSnZnIIiB) {Discard\\150};

    \node [decision, below of=XBPWnS,label={[text width=1.5cm]right:In bright star mask?}]                        (qXBPWnSB)     {};
    \node [blockd,   below of=qXBPWnSB,text width=8em,node distance=4.5em]                    (XBPWnSnB)     {453 X-ray sources\\See section~\ref{sec:xray_no_sdss_photom}};

    \path [line,solid] (X)      --            (qXB);
    \path [line,solid] (qXB)    -- node {yes} (XB);
    \path [line,solid] (qXB)    -- node {no}  (RnB);

    \path [line,solid] (XB)     --            (qXBP);
    \path [line,solid] (qXBP)   -- node {yes} (XBP);
    \path [line,solid] (qXBP)   -- node {no}  (XBnP);

    \path [line,solid] (XBP)    --            (qXBPW);
    \path [line,solid] (qXBPW)  -- node {yes} (XBPW);
    \path [line,solid] (qXBPW)  -- node {no}  (XBPnW);

    \path [line,solid] (XBPW)   --                       (qXBPWS);
    \path [line,solid] (qXBPWS) -| node [pos=0.75] {yes} (XBPWS);
    \path [line,solid] (qXBPWS) -| node [pos=0.75] {no}  (XBPWnS);

    \path [line,solid] (XBPWS)  --                        (qXBPWSZ);
    \path [line,solid] (qXBPWSZ) -| node [pos=0.75] {no}  (XBPWSnZ);
    \path [line,solid] (qXBPWSZ) -| node [pos=0.75] {yes} (XBPWSZ);

    \path [line,solid] (XBPWSnZ) --               (qXBPWSnZI);
    \path [line,solid] (qXBPWSnZI)  -- node {yes} (XBPWSnZI);
    \path [line,solid] (qXBPWSnZI)  -- node {no}  (XBPWSnZnI);

    \path [line,solid] (XBPWSnZI) --               (qXBPWSnZII);
    \path [line,solid] (qXBPWSnZII)  -- node {yes} (XBPWSnZII);
    \path [line,solid] (qXBPWSnZII)  -- node {no}  (XBPWSnZnII);

    \path [line,solid] (XBPWSnZII) --               (qXBPWSnZIIi);
    \path [line,solid] (qXBPWSnZIIi)  -- node {yes} (XBPWSnZIIi);
    \path [line,solid] (qXBPWSnZIIi)  -- node {no}  (XBPWSnZnIIi);

    \path [line,solid] (XBPWSnZIIi) --               (qXBPWSnZIIiB);
    \path [line,solid] (qXBPWSnZIIiB)  -- node {no} (XBPWSnZIIiB);
    \path [line,solid] (qXBPWSnZIIiB)  -- node {yes}  (XBPWSnZnIIiB);

    \path [line,solid] (XBPWnS)   -- (qXBPWnSB);
    \path [line,solid] (qXBPWnSB) -- node {yes} (XBPWnSnB);

\end{tikzpicture}
\end{center}
\caption{A schematic representation of the decision tree which leads to the selection of \rassagn\ targets.
See section~\ref{sec:assoc:rass} for details of the selection steps.}
\label{fig:RASS_flow}
\end{figure}

\begin{figure}
\begin{center}
\includegraphics[angle=0,width=84mm]{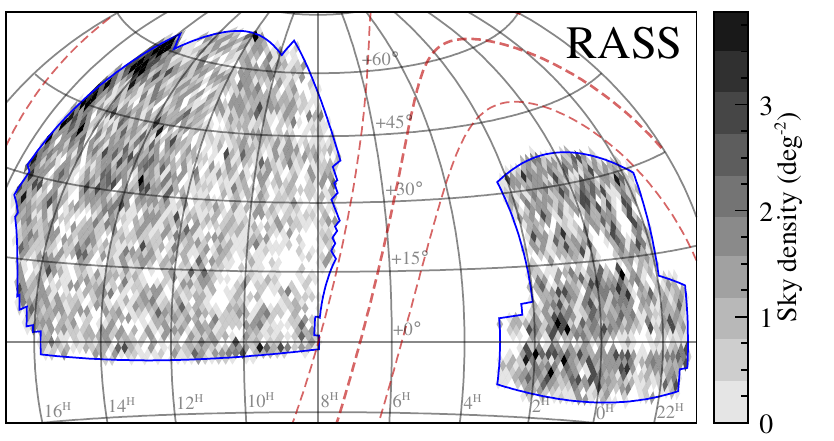}
\end{center}
\vspace{-0.5cm}
\begin{center}
\includegraphics[angle=0,width=84mm]{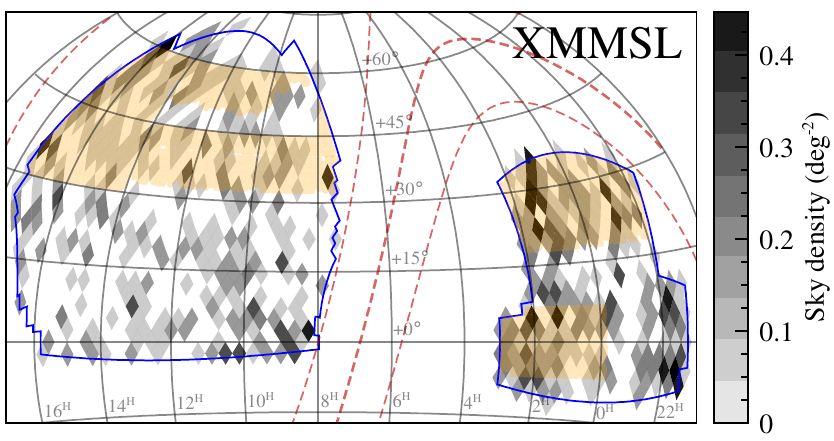}
\end{center}
\caption{{\em Upper panel:} Sky distribution of \rassagn\ sources put
  forward for targeting in \eboss, displayed with a HEALPix
  pixelisation (NSIDE=32, 1 pixel=3.36\,deg$^2$).  {\em Lower panel:}
  Same for \xmmslagn\ sources put forward for targeting in \eboss, but
  shown on a coarser pixel scale (NSIDE=16, 1 pixel=13.4\,deg$^2$).
  The (solid blue) line indicates the perimeter of the \boss\ imaging
  footprint. 
  The Galactic plane is indicated (dashed red lines at $b=0,\pm15$\,deg).
  In the lower panel the (orange) shading indicates the 4010\,\sqdeg\
  area over which \xmmsl\ sources with multiple X-ray detections
  were accidentally excluded during the targeting process (see
  section~\ref{sec:assoc:xmmsl}). The (orange) shaded region is also the sky area for which we compile 
  statistics for the overlap of \spiders-AGN targets with other \ebossts\ target classes 
  (see Appendix~\ref{sec:overlap_eboss}).}
\label{fig:RASS_XMMSL_potential_maps}
\end{figure}

\subsection{Selection of \xmmslagn\ targets}
\label{sec:assoc:xmmsl}

The process of selecting counterparts to the unique \xmmsl\ sources in
the \boss\ footprint followed closely that carried out for \rass\
sources (see section \ref{sec:assoc:rass}).  For each \xmmsl\ source
(all 3843 of which have at least one potential \allw\ counterpart
lying within 1\,arcmin), we used \nway\ to calculate the posterior
probability of it being associated with each of the \allw\
counterparts lying within 1\,arcmin.  We adopted the same priors on
the \wt,\wowt\ plane that were used for the \rass\ sample.  In
  doing this we have ignored the significant differences between the
  X-ray flux distribution of the \xmmsl\ sources and the fluxes of the
  \xmmb\ reference sample (see
  Fig.~\ref{fig:xflux_histo_vs_xmmb}). Using a prior derived from a
  better matched training sample would have been preferable, but we
  chose not to do so because of the very small number of \xmmbw\
  sources above the bright X-ray fluxes probed by the \xmmsl\ survey.
We find that for 3411/3843 (88.8~percent) of \xmmsl\ sources we have a
best matching \allw\ counterpart with posterior probability
\post$\ge$0.01. The 431 \xmmsl\ sources without at least one \allw\
counterpart above this threshold were not considered further.
The \xmmbw\ reference sample suggests that at the
  flux limit of the \xmmsl\ survey, virtually all persistent X-ray
  sources should be detected in the \allw\ survey. However, the
  fraction of \xmmsl\ sources lacking \allw\ counterparts (11~percent)
  is nearly three times larger than the 4~percent of full-band X-ray
  detections in the `Clean' \xmmsl\ catalogue that are
  expected to be entirely spurious (see section~\ref{sec:data:xmmsl}).
  This disparity implies that there is either a residual
  incompleteness in our crossmatching routine, and/or that the \xmmsl\
  catalogue contains a small but significant fraction of transient
  X-ray sources, which may not have persistent mid-IR counterparts
  above the \allw\ detection limits.  We note that if we use a set of
  control X-ray positions (placed 6\,arcmin away from each true X-ray position), and
  rerun the cross-matching algorithm, the fraction of sources without
  any valid \allw\ counterpart increases to 24~percent of cases.  
The
distribution of position differences between the \xmmsl\ sources and
their best matching \allw\ counterparts is shown in
Fig.~\ref{fig:Xray_AllWISE_offset}. The magnitude distributions of the
best matching \allw\ counterparts are shown in
Fig.~\ref{fig:AllWISE_mag_histos}. The `best' \allw' counterpart was
then matched to the \sdss-DR13 photometric catalogue, and 3142/3411
(92.1~percent) of the \xmmslw\ sources have at least one optical
counterpart within 1.5\,arcsec of the \allw\ position (see
Figs.~\ref{fig:AllWISE_SDSS_offset} and \ref{fig:optical_mag_histos}).
We then further down-selected from this list of 3142 optical
counterparts to reach our final list of \xmmslagn\ targets for
observation within \sdssiv.  Matching against the \sdss-proto-DR12
spectral catalogue removes 1411 previously spectroscopically
identified X-ray sources.  The properties of the spectroscopically
identified \xmmsl\ sources are discussed in more detail in
section~\ref{sec:XMMSL_with_spec}.  For the same
  reasons described in section~\ref{sec:assoc:rass} we have removed
from consideration all optical counterparts brighter than the nominal
magnitude limit of \eboss\ (i.e. \texttt{fiber2Mag}\_i$<$17.0, 746
sources), very bright objects (\texttt{modelMag}\_i$<$16.0, 59
sources); very faint targets (\texttt{ fiber2Mag}\_i$>$22.5, 32
sources), and targets which fall within the \boss\ bright star mask
(21 targets).  The final \xmmslagn\ target list contains 873 candidate
targets over the full \boss\ targeting footprint
(0.081\,targets\,\psqdeg).

The matching and filtering steps described above are summarised in a
flow diagram, see Fig.~\ref{fig:XMMSL_flow}.  The format of the
catalogue of \xmmslagn\ targets is described in
Appendix~\ref{sec:XMMSL_SPIDERS_targets}.  In
Fig.~\ref{fig:RASS_XMMSL_potential_maps} we show the sky distribution
of \xmmslagn\ targets over the sky.  In Figs.~\ref{fig:xflux_histo},
\ref{fig:rmag_histo_by_status} and \ref{fig:w2mag_histo_by_status} we
show the distributions of the \xmmsl\ sample in X-ray flux, $r$-band
magnitude and \wt\ magnitude at various stages during the
cross-matching and down-selection steps described above.

\begin{figure}
\begin{center}
\begin{tikzpicture}[node distance = 1.0cm, auto, >=triangle 60]
    \tikzstyle{every node}=[font=\small]

    \node [block,text width=11em]                                                             (X)            {\xmmslver\\ 20\,163 X-ray detections};
                                                                                            
    \node [decision, below of=X,label=right:{In \boss\ footprint?}]                           (qXB)          {};
    \node [block,    below of=qXB,text width=8em]                                             (XB)           {4325\\X-ray detections};
    \node [blockd,   left  of=qXB]                                                            (RnB)          {Discard\\15\,838};
                                                       
    \node [decision, below of=XB,label=right:{Unique X-ray source?}]                            (qXBP)         {};
    \node [block,    below of=qXBP]                                                           (XBP)          {3843\\X-ray sources};
    \node [blockd,   left  of=qXBP]                                                           (XBnP)         {Discard\\483};
                                                                         
    \node [decision, below of=XBP,label={[text width=2.5cm]right:\allw\ match with \post$\ge$0.01?}]           (qXBPW)        {};
    \node [block,    below of=qXBPW,node distance=4.5em,text width=10em]                      (XBPW)         {\xmmslw\\3411 X-ray sources};
    \node [blockd,   left  of=qXBPW]                                                          (XBPnW)        {Discard\\431};
                                                                                                    
    \node [decision, below of=XBPW,label=above right:{SDSS-DR13 photometry?},node distance=4.5em]   (qXBPWS)       {};
    \node [block,    below right of=qXBPWS,xshift=+5.5em,node distance=4.5em,text width=12em] (XBPWS)        {\xmmslws\\3142 X-ray sources};
    \node [blockd,   below left  of=qXBPWS,xshift=-4.0em,node distance=4.5em,text width=6em]  (XBPWnS)       {Discard\\269};
                                                                                                            
    \node [decision, below of=XBPWS,label={[text width=2cm]above right:Proto-DR12 spectrum?},node distance=4.5em](qXBPWSZ)      {};
    \node [result,   below right of=qXBPWSZ,yshift=-1.0em,text width=8em]                     (XBPWSZ)       {1411 X-ray sources with spectra.\\See~Appendix~\ref{sec:XMMSL_DR12_catalogue}};
    \node [block,    below left  of=qXBPWSZ,yshift=-1.0em,xshift=-2.5em]                      (XBPWSnZ)      {1731\\X-ray sources};
                                                                                                            
    \node [decision, below of=XBPWSnZ,yshift=-1.0em,label=right:fiber2Mag\_i$>$17?]            (qXBPWSnZI)    {};
    \node [block,    below of=qXBPWSnZI]                                                      (XBPWSnZI)     {985\\X-ray sources};
    \node [blockd,   left  of=qXBPWSnZI]                                                      (XBPWSnZnI)    {Discard\\746};
                                                                                              
    \node [decision, below of=XBPWSnZI,label=right:modelMag\_i$>$16?]                         (qXBPWSnZII)   {};
    \node [block,    below of=qXBPWSnZII]                                                     (XBPWSnZII)    {926\\X-ray sources};
    \node [blockd,   left  of=qXBPWSnZII]                                                     (XBPWSnZnII)   {Discard\\59};
                                                                                                             
    \node [decision, below of=XBPWSnZII,label=right:fiber2Mag\_i$<$22.5?]                      (qXBPWSnZIIi)  {};
    \node [block,    below of=qXBPWSnZIIi]                                                    (XBPWSnZIIi)   {894\\X-ray sources};
    \node [blockd,   left  of=qXBPWSnZIIi]                                                    (XBPWSnZnIIi)  {Discard\\32};
                                                                                                 
    \node [decision, below of=XBPWSnZIIi,label={[text width=1.5cm]right:In bright star mask?}]                    (qXBPWSnZIIiB) {};
    \node [result,    below of=qXBPWSnZIIiB,yshift=-0.5em]                                    (XBPWSnZIIiB)  {\xmmslagn\\873~X-ray~sources\\targeted in \sdssiv.\\See~Appendix~\ref{sec:XMMSL_SPIDERS_targets}};
    \node [blockd,   left  of=qXBPWSnZIIiB]                                                   (XBPWSnZnIIiB) {Discard\\21};

    \node [decision, below of=XBPWnS,label={[text width=1.5cm]right:In bright star mask?}]                        (qXBPWnSB)     {};
    \node [blockd,   below of=qXBPWnSB,yshift=-0.0em,text width=8em,node distance=4.5em]      (XBPWnSnB)     {67~X-ray sources\\See section~\ref{sec:xray_no_sdss_photom}};

    \path [line,solid] (X)      --            (qXB);
    \path [line,solid] (qXB)    -- node {yes} (XB);
    \path [line,solid] (qXB)    -- node {no}  (RnB);

    \path [line,solid] (XB)     --            (qXBP);
    \path [line,solid] (qXBP)   -- node {yes} (XBP);
    \path [line,solid] (qXBP)   -- node {no}  (XBnP);

    \path [line,solid] (XBP)    --            (qXBPW);
    \path [line,solid] (qXBPW)  -- node {yes} (XBPW);
    \path [line,solid] (qXBPW)  -- node {no}  (XBPnW);

    \path [line,solid] (XBPW)   --                       (qXBPWS);
    \path [line,solid] (qXBPWS) -| node [pos=0.75] {yes} (XBPWS);
    \path [line,solid] (qXBPWS) -| node [pos=0.75] {no}  (XBPWnS);

    \path [line,solid] (XBPWS)  --                        (qXBPWSZ);
    \path [line,solid] (qXBPWSZ) -| node [pos=0.75] {no}  (XBPWSnZ);
    \path [line,solid] (qXBPWSZ) -| node [pos=0.75] {yes} (XBPWSZ);

    \path [line,solid] (XBPWSnZ) --               (qXBPWSnZI);
    \path [line,solid] (qXBPWSnZI)  -- node {yes} (XBPWSnZI);
    \path [line,solid] (qXBPWSnZI)  -- node {no}  (XBPWSnZnI);

    \path [line,solid] (XBPWSnZI) --               (qXBPWSnZII);
    \path [line,solid] (qXBPWSnZII)  -- node {yes} (XBPWSnZII);
    \path [line,solid] (qXBPWSnZII)  -- node {no}  (XBPWSnZnII);

    \path [line,solid] (XBPWSnZII) --               (qXBPWSnZIIi);
    \path [line,solid] (qXBPWSnZIIi)  -- node {yes} (XBPWSnZIIi);
    \path [line,solid] (qXBPWSnZIIi)  -- node {no}  (XBPWSnZnIIi);

    \path [line,solid] (XBPWSnZIIi) --               (qXBPWSnZIIiB);
    \path [line,solid] (qXBPWSnZIIiB)  -- node {no} (XBPWSnZIIiB);
    \path [line,solid] (qXBPWSnZIIiB)  -- node {yes}  (XBPWSnZnIIiB);

    \path [line,solid] (XBPWnS)   -- (qXBPWnSB);
    \path [line,solid] (qXBPWnSB) -- node {yes} (XBPWnSnB);

\end{tikzpicture}
\end{center}
\caption{A schematic representation of the decision tree which leads
  to the selection of \xmmslagn\ targets for observation within \sdssiv. 
  See section~\ref{sec:assoc:xmmsl} for more details.}
\label{fig:XMMSL_flow}
\end{figure}
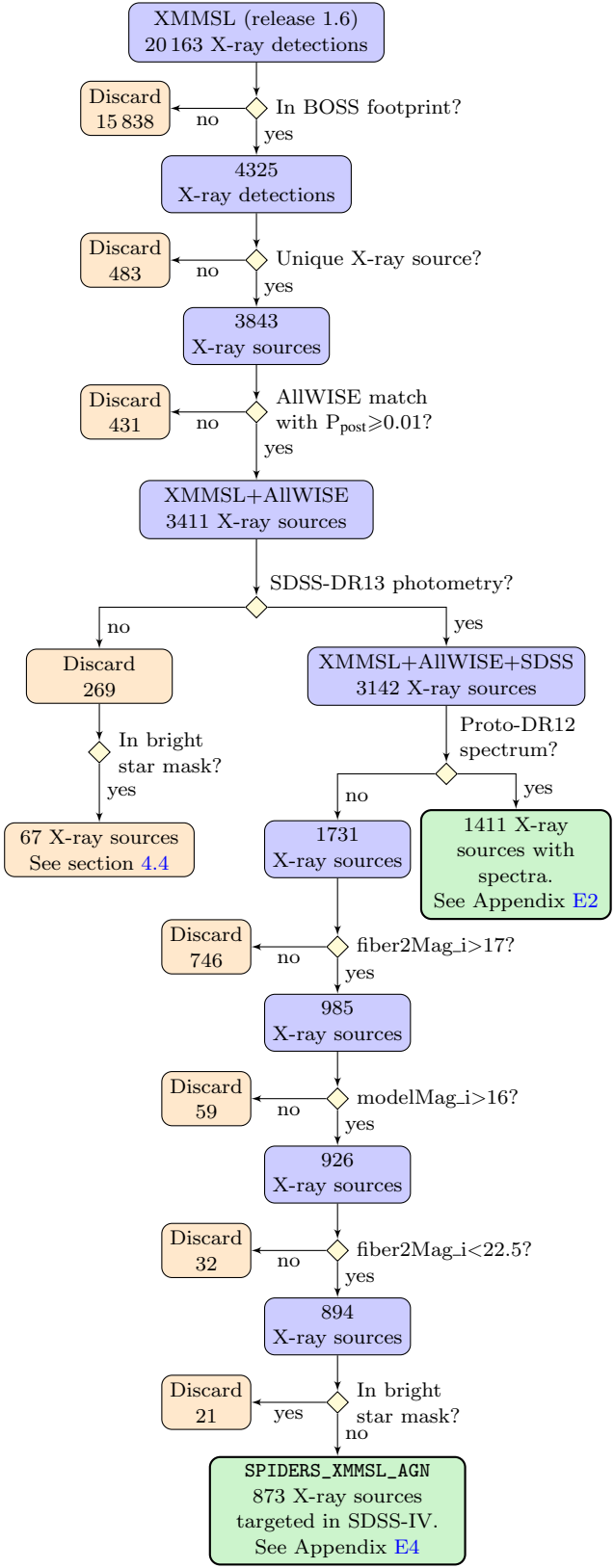

\subsection{X-ray sources with \sdss-DR12 spectra and visual inspections}
\label{sec:VI}
\label{sec:dr12_spectra}

\begin{figure}
\begin{center}
\includegraphics[angle=0,width=84mm]{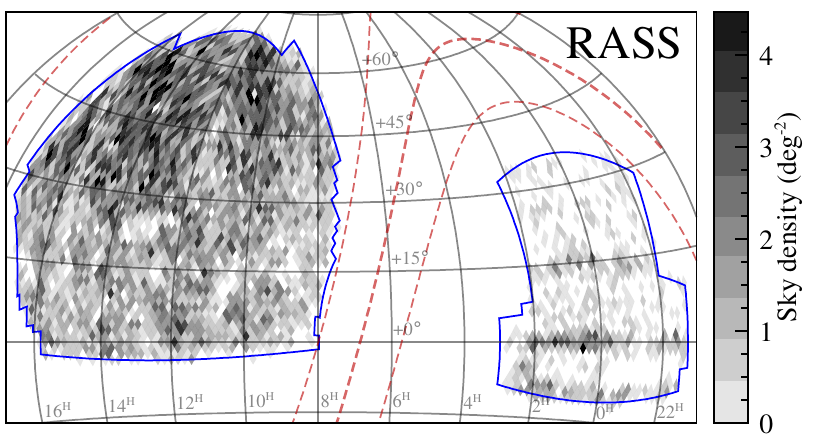}
\end{center}
\vspace{-0.5cm}
\begin{center}
\includegraphics[angle=0,width=84mm]{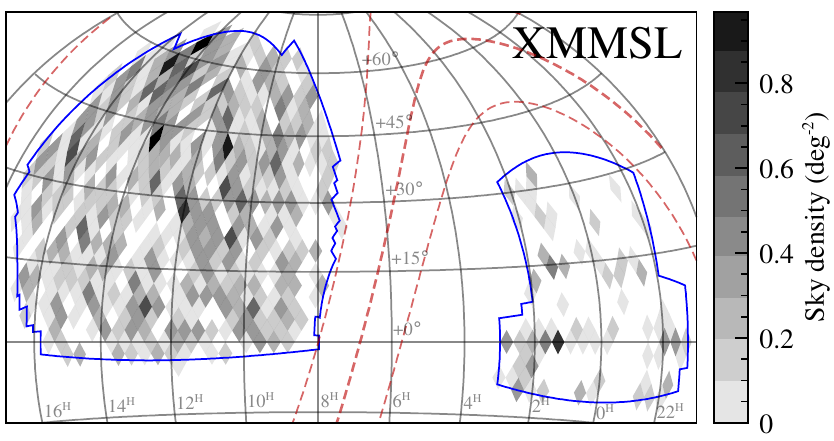}
\end{center}
\caption{{\em Upper panel:} 
Sky distribution of \rass\ sources with spectroscopic identifications in \sdss-DR12, 
displayed with a HEALPix pixelisation (NSIDE=32, 1 pixel=3.36\,deg$^2$). 
{\em Lower panel:} Same for \xmmsl\ sources, 
but shown on a coarser pixel scale (NSIDE=16, 1 pixel=13.4\,deg$^2$).
The (solid blue) line 
indicates the perimeter of the \boss\ imaging footprint.
The Galactic plane is indicated (dashed red lines at $b=0,\pm15$\,deg).
}
\label{fig:RASS_XMMSL_withspec_maps}
\end{figure}

As noted in sections \ref{sec:assoc:rass} and \ref{sec:assoc:xmmsl},
our selection of targets for \spiders\ used a prototype
version of the \sdss-DR12 spectroscopic sample to remove from
consideration any objects having existing spectroscopy.  In order to
form a clean sample, we have repeated the cross-matching
process of \xrayws\ objects to the
official \sdss-DR12 catalogue. Using a simple 1\,arcsec radial search
around the \sdss-DR13 photometric position, we find that \sdss-DR12
contains spectra for 11\,913 \rass, and 1482
\xmmsl\ sources.  
Fig.~\ref{fig:RASS_XMMSL_withspec_maps} shows the sky distribution of
the X-ray sources with spectroscopic identifications in
\sdss-DR12. Note the banded structure in the SGC region, which is
imprinted by the narrow sky coverage of \sdss\ observations prior to
the start of the \boss\ project.

Although the \boss\ redshift fitting algorithm \citep{Bolton12} has
been shown to be accurate and robust for the galaxies and QSOs that
have been targeted in previous iterations of the \sdss\ project, the
routine has not yet been validated for X-ray selected AGN which
dominate the \spiders\ AGN samples. Therefore we have carried out a
programme of visual inspections (VI) in order to check the accuracy of
the pipeline redshifts and spectral classes for \spiders\ targets.

We first collated the existing visual inspection information from
several prior works, namely the DR12Q quasar catalogue
(\citealt[][]{Paris17}; we consider all inspections from the super-set
and supplementary catalogues having \texttt{Z\_PERSON\_CONF}=3), the
DR7Q quasar catalogue \citep{Schneider10}, the visual inspections (of
BLAGN, NLAGN and BL~Lac counterparts to \rass\ sources) from
\citetalias{Anderson07}, and the visually inspected Blazar/BL~Lac catalogue
of \citet{Plotkin10}.  Of the 11913 \rass\ sources with \sdss-DR12
spectra, secure visually inspected redshifts were already available
for 8848 sources: 1875 from DR12Q, 161 from \citet{Plotkin10}, 6196
from \citetalias{Anderson07}, and 4981 from DR7Q.  Likewise, for the
1482 \xmmsl\ sources with \sdss-DR12 spectra, VI information was
already available for 1046 sources: 214 from DR12Q, 53 from
\citet{Plotkin10}, 546 from \citetalias{Anderson07}, and 596 from
DR7Q.  
We have
visually inspected all of the remaining spectra (3325 unique spectra,
associated with 3065 \rass\ and 436 \xmmsl\ sources).  
Following our experience with the BOSS/XMM-XXL sample \citep{Menzel16}, we divided the
spectra into higher- and lower-risk categories.  Higher-risk spectra
were those which met any of the following criteria: a pipeline
redshift outside the range $0.01 \leq z \leq 1.0$, pipeline redshift
uncertainty greater than 0.01, any redshift fitting warning flag 
(\texttt{ZWARNING}$>0$), an $r$-band model magnitude fainter than
20.0, or median per-pixel SNR outside the interval [2:50].  
Please see \citet{Bolton12} for
 a definition of the \texttt{ZWARNING} flag.
There were 569 spectra within the higher-risk
category, which were examined on average by three visual inspectors.
The remaining 2756 lower-risk spectra were each examined by at least
one visual inspector.

\subsubsection{Visual inspection tools and consolidation}
\label{sec:VI_tools}
Visual inspection was performed using the web-browser-based
\speccy\footnote{\url{https://gitlab.rzg.mpg.de/tdwelly/speccy}} tool,
developed (by us) to enable rapid inspection of large numbers of \spiders\ spectra.
The \speccy\ tool
presents users with a plot of flux
versus wavelength for a single observed \sdss\ spectrum,
together with a number of functions to aid the determination of these
parameters, including zooming/panning to regions of interest, and
box-car smoothing of the observed spectrum.  The
user may overlay the observed spectrum with: the observer-frame wavelengths
of common emission/absorption lines, a template spectrum (by default
the best-fitting template found by the \eboss\ pipeline is shown), a
scaled version of the sky background spectrum, the statistical error
spectrum, or the residual (data--model) spectrum.  The user can adjust the displayed
redshift of the template and emission/absorption lines. The user submits
the following information per spectrum: (i) a `visual' redshift
measurement; (ii) a redshift confidence flag (3=highly secure,
2=uncertain, 1=poor/unusable, 0=insufficient data), (iii) a
classification (we used only six classes: QSO (including BLAGN), Broad
Absorption Line (BAL) QSO, Galaxy (including NLAGN), Blazar, Star,
None), and (iv) freehand comments for problematic/unusual spectra.
 
We collected more
than 5000 visual inspections from a cadre of twelve inspectors
(assembled from within the coauthor list of this paper), and we then
consolidated cases where we had multiple inspections per spectrum.  We
defined a standard decision tree based on the relative agreement of
the multiple inspectors in three categories: redshift, redshift
confidence and classification. In nearly all cases inspections from
multiple inspectors are consistent, and
re-inspection/manual-reconciliation (carried out by A.~Merloni) was
deemed necessary in only a small fraction of cases, mostly due to
discrepant redshift assignments. We give a single `best' redshift and spectral
classification, plus a merged confidence flag for each inspected
spectrum. Where visual inspection information was available from
multiple works, we have used the following order of precedence i) our
own visual inspections, ii) DR12Q, iii) \citet{Plotkin10}, iv)
\citetalias{Anderson07}, and finally v) DR7Q.

In a small number of cases (93)
the \sdss-DR12 pipeline redshift estimates disagree by
more than 1~percent
($\frac{|z_{\mathrm{VI}}-z_{\mathrm{DR12}}|}{1+z_{\mathrm{VI}}} > 0.01$) 
from those determined through visual inspection. These
pipeline failures tend to be `catastrophic', and in 70~percent of
cases they over-estimate the true redshifts.  As might be expected, a
large fraction (45/93, 48~percent) of the pipeline redshift failures
are related to sources visually classified as Blazar/BL\,Lac,
i.e. having a strong and relatively featureless continuum. Despite
having higher than average SNR spectra, the visual inspection process
determined high confidence redshifts only for 13/45 (29~percent) of
the Blazar/BL~Lacs.
Another important failure mode, accounting for $\sim$18~percent of
redshift failures, occurs when the pipeline
misidentifies the \mgii\ emission line as \lyalpha.  For most of the redshift failures (57/93,
61~percent) the pipeline \texttt{ZWARNING} flag is set to a non-zero
value, indicating that the algorithm itself has identified a problem
with the spectrum and/or the redshift fitting process.

\section{Assessing the fidelity of our target selection method}
\label{sec:verify}
We have undertaken a number of independent tests designed to assess
the fidelity of the steps which we have used to select \spiders-AGN
targets for observation in \sdssiv.  These tests include i) checking the
X-ray$\rightarrow$\allw\ association step for a sub-sample of \rass\
and \xmmsl\ sources which appear in the \xmm, \chandra\ and \swiftx\
serendipitous catalogues, ii) using blank field populations to estimate
the rate of spurious X-ray$\rightarrow$\allw\ associations over the
whole \rass\ and \xmmsl\ samples, iii) evaluating the fraction of spurious
X-ray detections in the \rass\ sample, and iv) quantifying the success rate of
the association of \allw\ with \sdss-DR13 photometric sources.

\subsection{Verification of our X-ray--\allw\ association 
method using a bright X-ray reference sample}
\label{sec:verify:xref}

In order to estimate the reliability of our X-ray--mid-IR--optical
cross-matching technique we require an independent catalogue of bright
X-ray sources that have well determined positions, and that have
similar X-ray fluxes to the \rass\ and \xmmsl\ samples.  We have
formed a reference sample by selecting a set of well measured bright
X-ray sources from the \xmmiii\ catalogue \citep{Rosen16}, the
\chandra\ Source Catalogue \citep[\CSC;][]{Evans10} and the \swift\
X-ray Telescope Point Source catalogue \citep[\sxps;][]{Evans14}.  In
the following we describe how we built this reference
catalogue, (selected with 
a particular emphasis on astrometric accuracy, which in practice requires 
high signal to noise detections), and how we have used it to measure the reliability of the
cross-matching process.

\noindent
\textbf{\CSC\ sources:}
Starting from the full \CSC\ (v1.1)
catalogue\footnote{\url{http://cxc.cfa.harvard.edu/csc/}} we selected
the sub-sample of 1818 sources that are bright (0.2--2\,keV flux
$>5\times 10^{-14}$\,\cgs), high-quality (detection significance $>5$;
error ellipse major axis $<3$\,arcsec), point-like, unconfused,
unsaturated, and that lie within the \boss\ imaging footprint.

\noindent
\textbf{\xmmiii\ sources:}
We re-used the catalogue of 1049 bright high quality point-like
\xmmiii-DR4 sources previously described in section~\ref{sec:data:3xmmb}.

\noindent
\textbf{\sxps\ sources:}
Starting from the full \sxps\
catalogue\footnote{\url{http://www.swift.ac.uk/1SXPS/docs.php}} we
selected the sub-sample of 2142 non-GRB sources that are bright
(0.3--2\,keV flux $>5\times 10^{-14}$\,\cgs), high-quality ($>$40 net counts;
90~percent error radius $<5$\,arcsec; detection flags $<8$), and that
lie within the \boss\ imaging footprint.

The bright sub-samples of the \CSC, \xmmiii\ and \sxps\ reference
samples were merged to form a single catalogue of bright astrometric
reference sources.  Where multiple detections of
  single X-ray sources were found in more than one catalogue
(within a matching radius of 10\,arcsec) we
adopted a simple hierarchical approach,
  prioritizing \chandra\ detections over \xmm\ detections, and \xmm\
over \swift\ (even if in some cases this leads to a small loss
  of positional information).  We have manually tidied the catalogue
to deal with a handful of degenerate cases (two lensed QSOs, a
non-nuclear source in NGC\,4051, and a case where two detections of a
source made in overlapping observations have not been correctly
associated with each other).  The merged (and tidied) reference
catalogue contains 4752 unique X-ray sources (1813 with best detection
in \CSC, 962 from \xmmiii, and 1977 from \sxps), and is presented in
Appendix~\ref{sec:xray_astro_ref}.

We matched the X-ray sources to the nearest counterpart in the \allw\
catalogue \citep{Cutri13}, limiting our search to within a 5\,arcsec
radius of the X-ray position for \CSC\ and \xmmiii\ sources, and
within 10\,arcsec for \sxps\ sources.  We found 4524/4752 of the
merged X-ray reference sample had a matching \allw\ counterpart. 
In only 2.7~percent of these cases was there more than one 
  potential \allw\ counterpart within the search radius (mostly \sxps\ sources).
  The median position difference between X-ray and nearest \allw\
  positions is 0.9\,arcsec, and 90~percent are separated by less than
  3\,arcsec.  For the X-ray sources having detections in
  more than one of the \CSC, \xmmiii\ and \sxps\ reference samples, we
  have verified that these simple criteria consistently select the
  same `best' \allw\ counterpart for each independent X-ray
  detection.
We therefore make the assumption, given the high quality of the
X-ray positional information for this reference sample, that these
X-ray$\rightarrow$\allw\ associations are secure.

\begin{figure}
\includegraphics[angle=0,width=84mm]{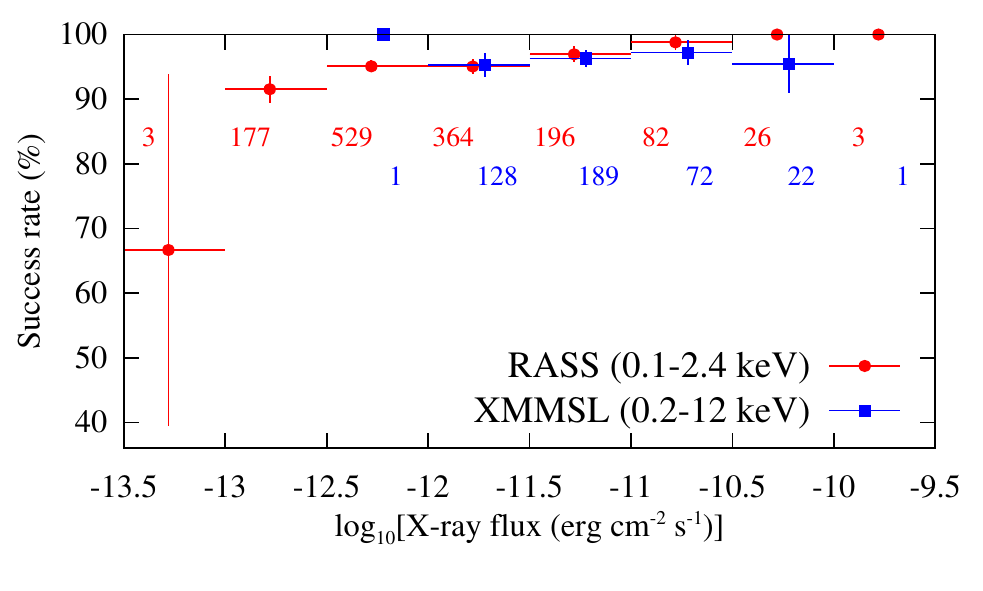}
\vspace{-0.5cm}
\includegraphics[angle=0,width=84mm]{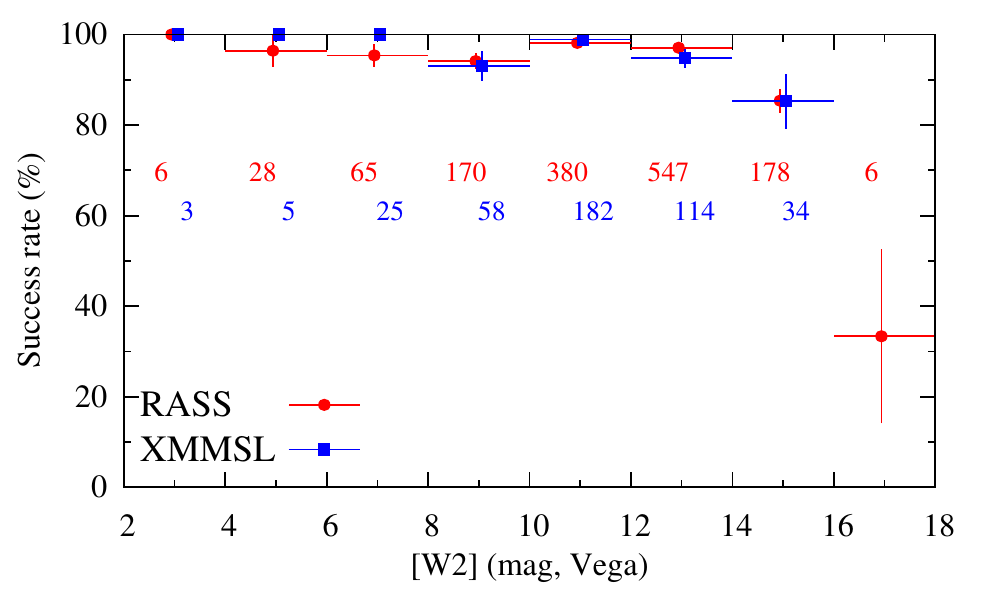}
\caption{The success rate of the Bayesian cross-matching routine,
  defined as the rate at which it selects the correct \allw\
  counterparts for \rass\ and \xmmsl\ sources, as determined from the
  astrometric reference sample.  {\it Upper panel:} Success rate as a
  function of X-ray flux (expressed in the native band-passes for each
  sample).  {\it Lower panel:} Success rate as a function of the \wt\
  magnitude of the correct counterpart.  The number of X-ray sources
  in each bin is indicated.  The vertical error bars show the (na\"{i}ve)
  binomial uncertainty. Markers are slightly offset horizontally for
  clarity.}
\label{fig:Xmatch_success_vs_flux_w2mag}
\end{figure}

Finally we have matched the merged X-ray reference catalogue to
the \rass\ and \xmmsl\ samples described in sections
\ref{sec:assoc:rass} and \ref{sec:assoc:xmmsl} respectively. Using a
search radius of 60\,arcsec (and using X-ray positional information
only), we find 1381 and 421 counterparts to the reference catalogue in
the \rass\ and \xmmsl\ samples respectively.
We calculate the fraction of cases in which, starting with the \rass\ or \xmmsl\
positional information, we have chosen exactly the same \allw\
counterpart that was chosen for the X-ray reference source.  The
success rates for \rass\ and \xmmsl\ sources are comparable, with
1314/1381 = $95.1\pm 0.6$~percent of \rass\ sources successfully
matched and 404/421 = $96.0\pm 1.0$~percent of \xmmsl\ sources (na\"{i}ve 1$\sigma$ binomial
confidence interval).  When only considering \rass\ sources matched to
astrometric reference sources from the \CSC, the success rate is
slightly lower ($92.5\pm 1.7$~percent), which is possibly due to the
lower median X-ray flux of the \CSC\ sources.  There is a slight trend
of decreasing success rate toward fainter X-ray fluxes, see 
Fig.~\ref{fig:Xmatch_success_vs_flux_w2mag}. Except for the very
faintest X-ray flux bin (containing only three \rass\ sources), the
success rate is always $>90$~percent. As might be expected, there is
a stronger trend of decreasing success rate toward fainter \wt\
magnitudes. The success rate in the 14$<$\wt$<$16\,mag range is
\si85~percent for both \rass\ and \xmmsl\ sources.  Approximately
10~percent of the X-ray samples have counterparts in this magnitude
range.

\subsection{Assessment of the rate of incorrect 
X-ray--IR associations using blank field samples}
\label{sec:verify:spurious_rate}
\begin{figure}
\begin{center}
\includegraphics[angle=0,width=84mm]{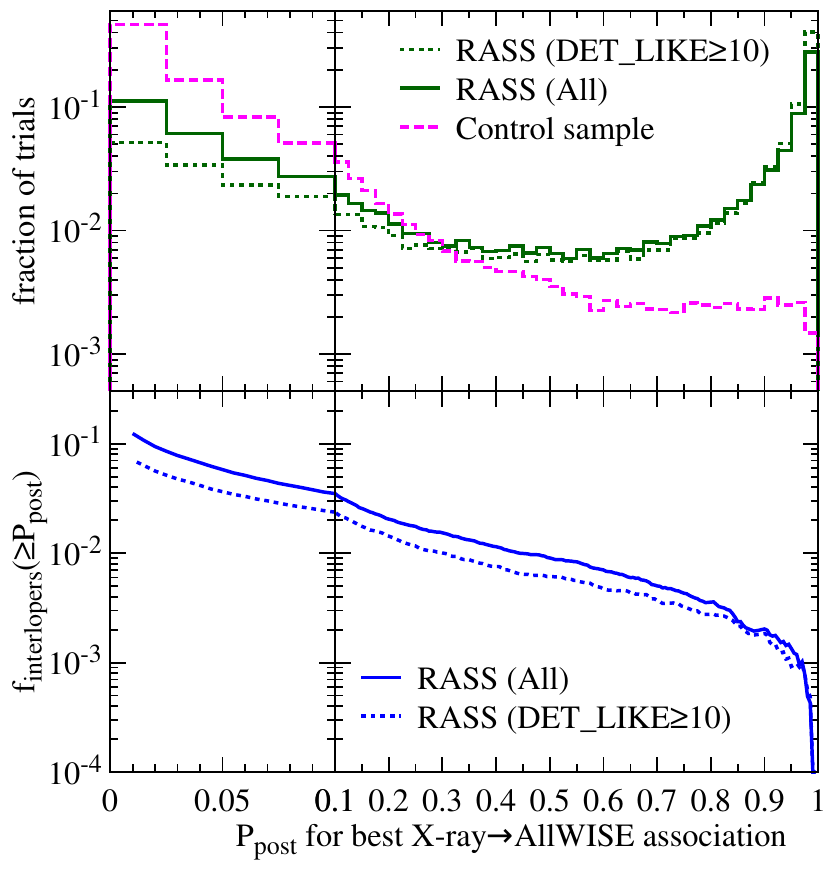}
\end{center}
\caption{{\em Upper Panel:} The distributions of \post\
  for the most probable \allw\ associations for each \rass\ detection
  (green solid histogram), and for a control sample of nearby
  locations (magenta dashed histogram, offset by +0.1\,deg in
  Declination from the \rass\ positions). 
  {\em Lower Panel:} The expected rate of interlopers in the \rassw\
  sample above a given minimum threshold in \post.
  $f_\mathrm{interlopers}(\ge$\post) is the cumulative fraction of
  \rass\ sources for which the most probable X-ray$\rightarrow$\allw\
  association at the offset position has a higher \post\ than the most
  probable association at the actual \rass\ position, and represents
  an estimate of the fraction of spurious associations in the
  sample. In each panel, the dotted curves show the equivalent
  distributions for just the subset of \rass\ sources having
  high-confidence X-ray detections (\textsc{DET\_LIKE}$\ge$10). 
  Note the change in x-axis scale at \post=0.1.}
\label{fig:Pdist}
\end{figure}

The Bayesian cross-matching algorithm implemented in \nway\ reports
the posterior probability (\post) of any two objects being associated,
and hence for each X-ray source gives a ranked list of potential
counterparts.  However, the Bayesian matching algorithm does not tell
us about the rate of false positives, so we carry out an empirical
check to determine the rate of spurious interlopers as a function of
posterior probability (\post).  We have retrieved a `control'
catalogue of \allw\ sources within 1\,arcmin radius of the positions
of the \rass\ sources after the X-ray positions have been offset by
+0.1\,deg in Declination (using locations near the real \rass\
positions is preferred over randomly distributed positions in order to
ensure that the offset catalogue samples approximately the same
distribution of Galactic latitudes and \allw\ exposure depth as the
\rass\ sources).  For each offset position we follow the method
described in section~\ref{sec:post_recipe} to select the \allw\ source
having the largest posterior probability of being associated with the
offset \rass\ position.  The distribution of \post\ for the \rass\ and
offset samples are shown in Fig.~\ref{fig:Pdist} (upper panel).  We
count up the cases where the \post\ for the best association at the
offset position is greater than the \post\ for the best association at
the actual \rass\ position.  The lower panel of Fig.~\ref{fig:Pdist}
shows the rate of such cases, and allows us to make an estimate of the
fraction of the \rassw\ associations that are spurious\footnote{
We did not exclude the 0.4~percent of control positions which
happen to lie close to neighbouring \rass\ sources, and so
slightly overestimate the spurious fraction.}.  For the purposes
of choosing \spiders\ targets for spectroscopic follow up in \sdssiv\
we have made a cut on the posterior probability at the nominally
rather low level of \post$\ge$0.01. However, the cumulative curve in
the lower panel of Fig.~\ref{fig:Pdist} demonstrates that adopting
even this low probability threshold, the total fraction of spurious
associations within the \rassw\ sample is at a reasonably low level
(12.4~percent of the total, equivalent to $\sim$3800 spurious
associations within the \rassw\ sample).  We repeated this exercise
for the subset of \rass\ sources with higher X-ray detection
likelihood (\textsc{DET\_LIKE}$\ge$10), i.e. a sub-sample that should
be only slightly contaminated with spurious X-ray detections.  For
this sub-sample, there is a smaller fraction of associations having
low \post, and hence a lower overall rate of spurious associations
(7~percent for \post$\ge$0.01). This suggests that at least part of
the lowest probability \rassw\ associations are due to spurious X-ray
detections, as we discuss in more detail below.  If a very pure sample
is a priority, then filtering the \rassw\ sample on the basis of a
minimum value of \post\ is possible. For example, in order to achieve
a spurious association fraction $<$3~percent, then one should apply a
cut of \post$\ge$0.123 to the full \rassw\ sample (\post$\ge$0.071 for
the subset of \rass\ sources with \textsc{DET\_LIKE}$\ge$10).

\subsection{An evaluation of spurious X-ray detections within the \rass\ sample}
\label{sec:verify:RASS_spurious}

Another source of uncertainty in our targeting procedure is the
incidence of spurious detections in the \rass\ catalogue. As mentioned
above, the \rassf\ contains sources detected down to a relatively low
confidence level (detection likelihood, \texttt{DET\_LIKE}$>$6.5).
\citet{Boller16} carried out simulations of the detection procedure
used to produce the \iirxs\ catalogue.  They found that the fraction
of spurious detections, averaged over the whole sky and above a
detection likelihood (named \texttt{EXI\_LIKE} in the \iirxs\ catalogue) 
threshold of 6.5, could be as high as $\sim$30~percent.
The spurious fraction is strongly dependent on the detection
likelihood threshold adopted, and drops to $\sim$2~percent at
\texttt{DET\_LIKE}$>$10.  Unfortunately, equivalent simulations are
not available for \irxs, so we are unable to make a direct estimate of
the spurious fraction in the \irxs\ catalogue.  However, since both
\irxs\ and \iirxs\ are based on almost the same X-ray data-sets, and
share many of the same source detection procedures, we assume that the
spurious fraction in our \irxs\ sample is comparable to that in the
\iirxs\ catalogue. 

\begin{figure}
\begin{center}
\includegraphics[angle=0,width=84mm]{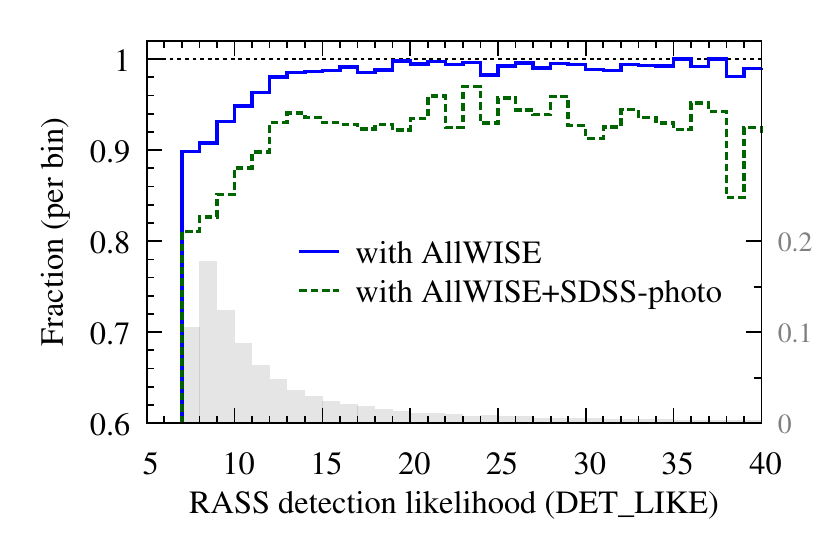}
\end{center}
\caption{Fraction of \rass\ sources that are associated with longer
  wavelength counterparts as a function of X-ray detection likelihood
  \texttt{DET\_LIKE}. The solid blue line shows the fraction of \rass\
  sources having a best \allw\ counterpart with
  \post$\ge$0.01, the green dashed line shows the fraction
  with both \allw\ and \sdss-DR13 photometric counterparts. The shaded
  grey histogram shows the relative fraction of the \rass\ sample falling in 
  each bin of \texttt{DET\_LIKE} (see right hand scale).}
\label{fig:RASS_DET_LIKE_dist}
\end{figure}

We investigate the rate of contamination by spurious sources further
by examining the properties of the \rass\ sample as a function of
\texttt{DET\_LIKE}.  Fig.~\ref{fig:RASS_DET_LIKE_dist} shows the
distribution in X-ray detection likelihood for the \rass\ sample, and
the sub-samples with \allw\ and \sdss-DR13 photometric counterparts.
Clearly, the vast majority of \rass\ sources not
matched to any \allw\ counterpart have low detection likelihoods, and
we should expect about half of them to be spurious detections in the
\rass\ catalogue. The fraction of \rassw\ sources lacking optical
photometric counterparts appears to be relatively independent of the
X-ray detection likelihood.

\begin{figure}
\begin{center}
\includegraphics[angle=0,width=84mm]{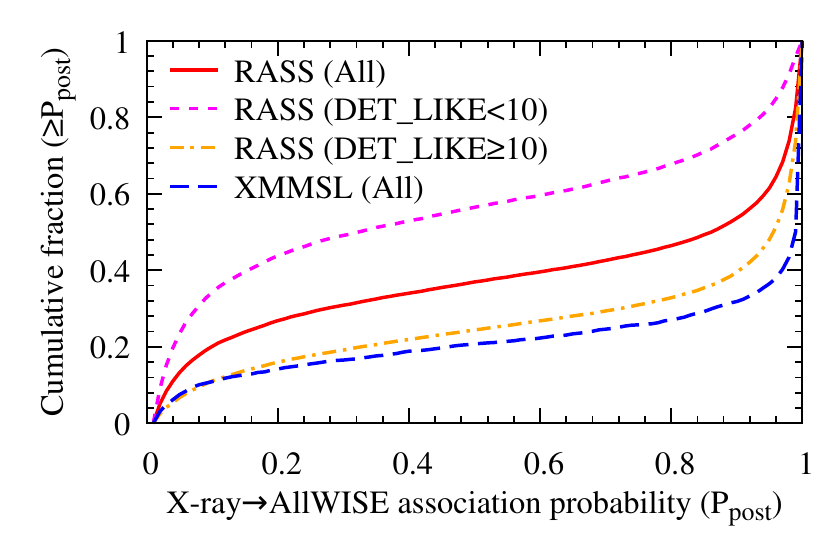}
\end{center}
\caption{Cumulative distribution of the posterior association
  probability \post\ for X-ray$\rightarrow$\allw\
  matches. We show the distributions for the full \rassw\ sample (red
  solid curve), as well as for the high (orange dot-dashed curve) and
  low (magenta short-dashed curve) detection likelihood subsets (split
  at a detection likelihood of \texttt{DET\_LIKE}=10).  For comparison,
  we also show the distribution for the full \xmmslw\ sample (blue
  long-dashed curve).}
\label{fig:post_dist}
\end{figure}

We can also verify the reliability of the source identifications, by
looking at the posterior probability \post\ distribution of those
\rass\ sources which {\it are} matched to \allw\
counterparts. Fig.~\ref{fig:post_dist} shows the cumulative
(normalized) posterior probability distribution for \rassw\ sources,
and the breakup of the same distribution as a function of X-ray
detection likelihood. Clearly, the population of sources with lower
X-ray detection likelihoods have a much lower average posterior
probability of being associated to their \wise\ counterparts.  It can
also be seen that that for high detection likelihood \rass\ sources
(\texttt{DET\_LIKE}$\ge$10) the cumulative distribution of \post\ is
very similar to that for \xmmsl\ sources, which are expected to have
only a $\sim4$~percent contamination by spurious X-ray detections
\citep[][see also the online
documentation\footnote{\url{http://www.cosmos.esa.int/web/xmm-newton/xmmsl1d-ug}}]{Saxton08}.

\subsection{An evaluation of the method used to 
associate \allw\ sources with optical counterparts}
\label{sec:verify:xray_no_sdss_photom}
\label{sec:xray_no_sdss_photom}
The \nway\ tool \salvato\ can compute association probabilities for sets of
targets located in two or more catalogues (e.g. X-ray, IR and
optical), with any number of priors defined for each possible
combination of wave-bands (also see the recent work by \citealt{Pineau17}).
However, in order to avoid imprinting
complex biases into our target selection for \spiders-AGN, we decided
to instead use the rather simple method of choosing the brightest
$r$-band counterpart in the \sdss-DR13 photometric database lying
within 1.5\,arcsec of the \allw\ position.  For a non-negligible
fraction ($\sim$8~percent) of the \xrayw\ sources, no optical counterpart is
found within this small search radius.
Fig.~\ref{fig:RASS_DET_LIKE_dist} demonstrates that the fraction of
\xrayw\ sources lacking optical counterparts does not strongly
depend on X-ray detection likelihood (which roughly scales with
X-ray brightness).

\begin{figure}
\begin{center}
\includegraphics[angle=0,width=84mm]{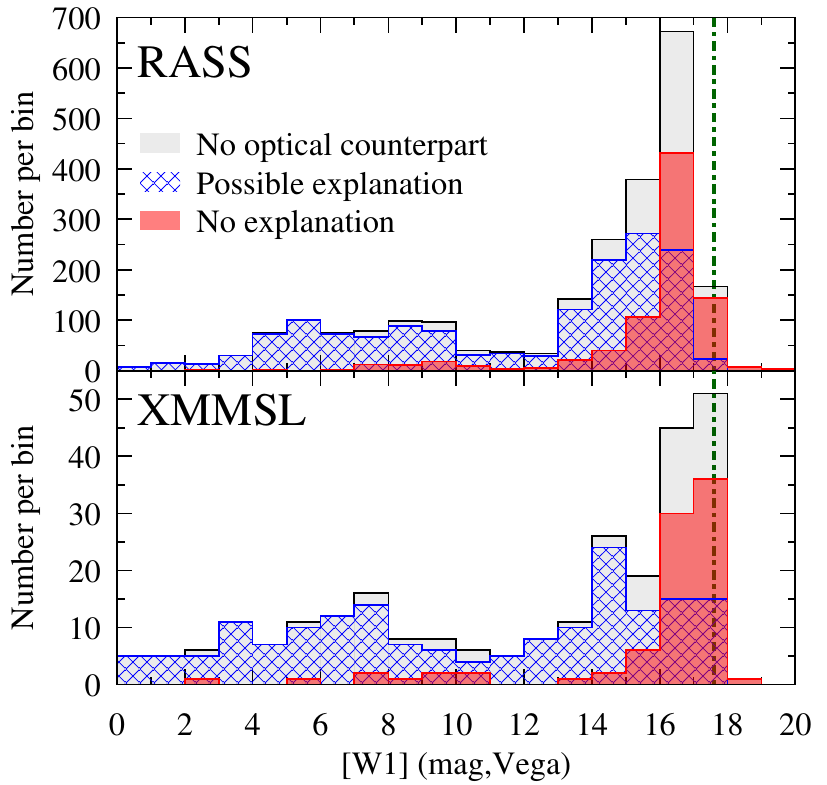}
\end{center}
\caption{Distribution of \allw\ \wo\ magnitude for the \rass\ (upper
  panel) and \xmmsl\ (lower panel) sources which have \allw\
  counterparts but for which we do not find \sdss-DR13 photometric
  counterparts within 1.5\,arcsec of the \allw\ coordinates (grey shaded histogram). 
  The hashed blue regions indicate the subset of sources for which we have a
  possible explanation (they either lie within the \sdss\ bright star mask,
  a bad \sdss\ imaging field or inside the disk of a large bright
  galaxy, have at least one \allw\ warning flag set, or have an
  \sdss-DR13 photometric counterpart within 5\,arcsec of the \allw\
  position).  The red shaded histogram indicates the remaining sources where we
  have no immediate explanation for the lack of an optical counterpart.
  The vertical line (green dot-dashed) shows the nominal \wo\
  5$\sigma$ point source detection limit.  }
\label{fig:AllWISE_noSDSSv5b_mag_histo}
\end{figure}

In Fig.~\ref{fig:AllWISE_noSDSSv5b_mag_histo} we show the \wo\
magnitude distribution of X-ray-\allw\ sources lacking optical
counterparts. We use the \wo\ band here because at the faint end of
the distribution (\wo$>$15\,mag), more than 90~percent of objects
have their highest SNR detection in this band. 
Some of the cases at the bright end of the \wo\ distribution can be attributed to very bright stars, for which
the imaging was heavily saturated, leading to catalogue
incompleteness and/or degraded IR/optical positions.  Indeed,
453 of the \rass\ and 67 of the \xmmsl\ sources that lack optical counterparts lie within the \boss-DR10 bright star rejection
mask.
For 64 \rass\ and 12 \xmmsl\ sources, the
\allw\ positions fall within the optical extent of bright galaxies
\citep{deVaucouleurs92}. In these cases we might reasonably expect
the \allw\ and \sdss\ positions to differ.  A few additional cases (32 \rass\
and 6 \xmmsl\ sources) fall within \sdss\ fields having bad
photometry\footnote{\url{http://data.sdss3.org/sas/dr10/boss/lss/reject_mask/badfield_mask_postprocess_pixs8.ply}}.
For 17~percent of the \rass\ sources lacking
optical counterparts (23~percent for \xmmsl), the
\allw\ \texttt{cc\_flags} column is non-zero in either the \wo\ or \wt\ bands
\citep[indicating potential contaminated or spurious
detections;][]{Cutri13}, much higher than the rate of flagged
detections ($\sim$6~percent) in the \xrayw\ samples for which we
\textit{do} find \sdss\ counterparts.

After filtering out all these cases we are still left with 1689
\rass\ and 173 \xmmsl\ sources lacking
optical counterparts.  The most obvious remaining explanation is that
our adopted search radius was too small.  Indeed, if we expand the
search radius to 5\,arcsec around the \allw\ position then we find
counterparts to a further 833 \rass\
and 81 \xmmsl\ sources.  This leaves 856
\rass\ and 92 \xmmsl\ sources
($\sim$3~percent of the \xrayw\ parent samples) where we cannot
immediately explain the lack of an optical counterpart. Apart from a few very bright objects (likely to be high
proper motion stars or stars which saturate the \sdss\ imaging),
nearly all of the unexplained cases lie at the faint end of the \allw\
magnitude distribution (see
Fig.~\ref{fig:AllWISE_noSDSSv5b_mag_histo}). It is possible that some
fraction of these faint objects could have such red optical--IR
colours that they are undetected in the \sdss\ imaging.  We will carry
out a more thorough analysis of such objects in a future \spiders-AGN
data release, including an inspection of deeper imaging.

\section{Results}
\label{sec:results}
The main goal of this work is to provide a complete account of the
process by which we chose candidate X-ray AGN targets for observation
in the shallower tiers of the \sdss-IV/\spiders\ spectroscopic
survey. This serves the dual purpose of enabling further scientific
investigation by documenting exactly the selection process, as well as
providing large and homogeneous catalogues of bright X-ray sources with
reliable multi-wavelength identification (in a well defined
statistical sense) and spectroscopic information. In this section, we
outline the salient features of these catalogues, and we provide a
outlook of possible applications of the \spiders-AGN samples.

We describe the catalogue format for spectroscopically
identified \rass\ and \xmmsl\ sources in
Appendices~\ref{sec:RASS_DR12_catalogue} and
\ref{sec:XMMSL_DR12_catalogue}.

\subsection{Properties of X-ray sources with existing spectra in \sdss-DR12}
\label{section:X-ray_with_spec}
\label{sec:RASS_with_spec}
\label{sec:XMMSL_with_spec}

Of the X-ray sources with \sdss-DR12 spectra, there are 11788/11913
\rass\ and 1469/1482 \xmmsl\ sources having counterparts with 
visual inspection confidence levels of 3 or equivalent (i.e. secure
redshifts), and pipeline redshifts that agree well with the VI
redshifts (i.e. $|z_{\mathrm{pipe}}-z_{\mathrm{VI}}| < 0.01 [1+z_{\mathrm{VI}}]$),
or that are 
identified as Blazar/BL\,Lac by visual inspection, and have 
a visual inspection confidence level of 2 or greater.

\subsubsection{X-ray luminosity distribution}

For each of the \rass\ and
\xmmsl\ sources with secure redshifts we calculate the rest-frame X-ray luminosities as follows. 
We use a K-correction
term that is appropriate for the same simple power-law spectral models
that were assumed when converting from observed count rate to
flux. Specifically, we compute,
\begin{equation}
L_{[E'_{\mathrm{min}}:E'_{\mathrm{max}}]} = 
4\,\pi\, d^2_{\mathrm{L}}(z)\, K_{\mathrm{corr}}\, F_{[E_{\mathrm{min}}:E_{\mathrm{max}}]}, 
\end{equation}
where $L_{[E'_{\mathrm{min}}:E'_{\mathrm{max}}]}$ is the luminosity in
the rest-frame $[{E'_{\mathrm{min}}:E'_{\mathrm{max}}}]$ energy
interval, $d_{\mathrm{L}}(z)$ is the luminosity distance based on the
visually inspected spectroscopic redshift (see section~\ref{sec:VI}),
and $F_{[E_{\mathrm{min}}:E_{\mathrm{max}}]}$ is the Galactic
absorption corrected flux in the observed
$[{E_{\mathrm{min}}:E_{\mathrm{max}}}]$ energy band.  The K-correction
term, including bandpass conversion, is given by
\begin{equation}
\begin{split}
K_{\mathrm{corr}}(E'_{\mathrm{min}},E'_{\mathrm{max}},E_{\mathrm{min}},E_{\mathrm{max}},\Gamma,z) \qquad \qquad \qquad \qquad \qquad \\  
\qquad \qquad \qquad \qquad \qquad = (1+z)^{\Gamma-2} \frac{E_{\mathrm{max}}^{'2-\Gamma} - E_{\mathrm{min}}^{'2-\Gamma}}{E_{\mathrm{max}}^{2-\Gamma} - E_{\mathrm{min}}^{2-\Gamma}},
\end{split}
\end{equation}
where $\Gamma$ is the assumed spectral index ($\Gamma=2.4$ for \rass\ sources, 
and $\Gamma=1.7$ for \xmmsl\ sources).

\begin{figure}
\begin{center}
\includegraphics[angle=0,width=84mm]{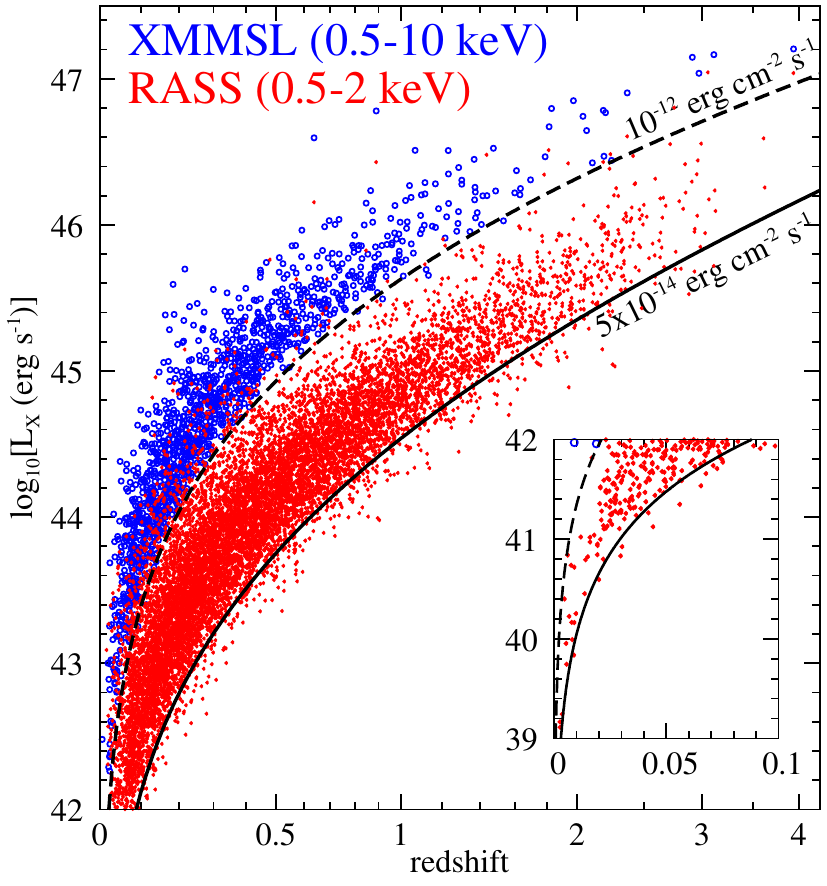}
\includegraphics[angle=0,width=84mm]{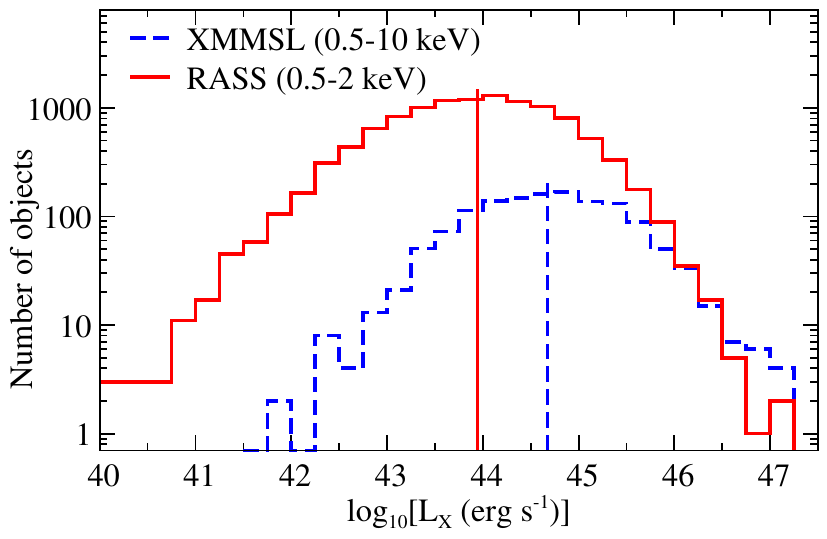}
\end{center}
\caption{\textit{Upper panel:} Distribution of spectroscopically
  identified \rass\ and \xmmsl\ sources with high confidence redshift
  measurements in the redshift -- X-ray-luminosity plane.  X-ray
  luminosities ($\mathrm{L}_\mathrm{X}$) are given in the 0.5--2\,keV
  band for \RASS\ sources (red dots) and in the 0.5--10\,keV band
  for \xmmsl\ sources (blue open circles). The luminosities corresponding to
  two fiducial flux limits are indicated with labelled black curves.
  An inset shows the few X-ray sources falling below the lower
  luminosity limit of the main plot.  \textit{Lower panel:} A
  histogram showing the X-ray luminosity distribution of the high
  confidence extragalactic \rass\ and \xmmsl\ samples.  Vertical lines
  show the median X-ray luminosities of each sample.  }
\label{fig:z_Lx}
\end{figure}

For \rass\ sources we calculate the 0.5--2\,keV luminosity from the
Galactic absorption corrected 0.1--2.4\,keV flux, assuming
$\Gamma=2.4$. Similarly, for the \xmmsl\ sources we assume
$\Gamma=1.7$ to calculate the 0.5--10\,keV luminosity from the
Galactic absorption corrected 0.2--12\,keV flux.  The distribution of
the spectroscopically identified \rass\ and \xmmsl\ sources in the
redshift--X-ray luminosity plane is shown in Fig.~\ref{fig:z_Lx}.
The median X-ray luminosities of the spectroscopically identified \rass\ and \xmmsl\ samples are
$10^{43.9}$ and $10^{44.7}$ \ergs\ respectively. Both samples contain significant 
numbers of sources (1177 for \rass\ and 473 for \xmmsl) at 
extremely high X-ray luminosities ($>10^{45}$\,\ergs).

In Figs.~\ref{fig:class_vs_Lx} and \ref{fig:class_vs_z}, we illustrate the partition of the X-ray
samples by spectral class as a function of X-ray luminosity and redshift respectively.

\subsubsection{Spectral classifications}

\setlength{\tabcolsep}{5pt}
\begin{table}
  \caption{The frequency of spectroscopic classifications within 
    the samples of \rass\ and \xmmsl\ targets with spectra in \sdss-DR12.
    These statistics only include the objects for which we either have 
    i) high confidence visual inspections and pipeline redshifts which 
    agree with the visual inspection redshifts, or ii) 
    where visual inspection has classified the object as Blazar/BL\,Lac 
    with reasonable confidence. Except for Blazar/BL\,Lacs, the 
    \texttt{CLASS} and \texttt{SUBCLASS} parameters are taken directly 
    from the \boss\ pipeline outputs, described in full by \citet{Bolton12}.}
\label{tab:withspec_classes}
\begin{tabular}{@{}llrr}
\hline
\texttt{CLASS} &  \texttt{SUBCLASS}  & N\sub{RASS} & N\sub{XMMSL} \\
\hline
Blazar/BL\,Lac &                               &      259 &       70 \\
\hline
\multirow{9}{*}{\texttt{GALAXY}} & No subclass &     1639 &      123 \\
         & \texttt{AGN                  }      &      190 &       31 \\
         & \texttt{AGN BROADLINE        }      &       50 &        8 \\
         & \texttt{BROADLINE            }      &       98 &       10 \\
         & \texttt{STARBURST            }      &       77 &        5 \\
         & \texttt{STARBURST BROADLINE  }      &        8 &        4 \\
         & \texttt{STARFORMING          }      &      297 &       25 \\
         & \texttt{STARFORMING BROADLINE}      &       18 &        4 \\
\cline{3-4}                                                            
         & \rule{0pt}{3ex}Sub-total            &     2518 &      258 \\
\hline 
\multirow{9}{*}{\texttt{QSO}} &   No subclass  &      246 &        8 \\
         & \texttt{AGN                  }      &       18 &        1 \\
         & \texttt{AGN BROADLINE        }      &      104 &       27 \\
         & \texttt{BROADLINE            }      &     6305 &      731 \\
         & \texttt{STARBURST            }      &       19 &        1 \\
         & \texttt{STARBURST BROADLINE  }      &     2133 &      391 \\
         & \texttt{STARFORMING          }      &        7 &        0 \\
         & \texttt{STARFORMING BROADLINE}      &       93 &       13 \\
\cline{3-4}                                                            
         & \rule{0pt}{3ex}Sub-total            &     9012 &     1189 \\
\hline 
\multirow{11}{*}{\texttt{STAR}} & \texttt{CV}  &       27 &       11 \\
         & \texttt{F5}                         &       10 &        0 \\
         & \texttt{F9}                         &       22 &        1 \\
         & \texttt{K1}                         &       12 &        0 \\
         & \texttt{K3}                         &       10 &        0 \\
         & \texttt{K7}                         &       13 &        1 \\
         & \texttt{M2}                         &       11 &        0 \\
         & \texttt{M3}                         &       12 &        0 \\
         & \texttt{M4}                         &       33 &        0 \\
         & Other subclass                      &       77 &        4 \\
\cline{3-4} 
         & \rule{0pt}{3ex}Sub-total            &      258 &       22 \\
\hline 
Total    &                                     &    11788 &     1469 \\
\hline     
\end{tabular}
\end{table}
\setlength{\tabcolsep}{6pt}

In Table \ref{tab:withspec_classes} we give a breakdown of the \boss\
pipeline spectral classifications (or visual inspection classification
for Blazar/BL\,Lac) for the counterparts to the X-ray sources with
confident spectral classifications.  As expected, the vast majority
(79~percent for \rass, and 84~percent for \xmmsl) of these objects
have optical spectra classified as \texttt{QSO}, or as \texttt{GALAXY}
but with some signature of AGN emission (i.e. \texttt{SUBCLASS}
=\texttt{AGN}, \texttt{AGN BROADLINE}, \texttt{BROADLINE},
\texttt{STARBURST BROADLINE}, or \texttt{STARFORMING
  BROADLINE}). Blazar/BL\,Lacs identified through visual inspections
constitute 2 and 5~percent of the \rass\ and \xmmsl\ samples
respectively.  Another 2--3~percent of the spectra are classified as
\texttt{GALAXY} with signs of ongoing star-formation activity
(\texttt{SUBCLASS} =\texttt{STARBURST} and \texttt{STARFORMING}). It
is feasible that AGN emission features in these star-forming galaxies
(SFG) could be overwhelmed by the emission lines powered by star
formation. Alternatively some or all of the X-ray emission could be
powered by star-formation activity.  Only a small fraction
(1--2~percent) of the X-ray sources are classified as
\texttt{STAR}. However, virtually all of the X-ray sources having
optically bright but unobserved counterparts (unobserved because they
are brighter than the \sdss\ spectroscopic limit), are also likely to
be Galactic stars.

A significant minority of the X-ray sources (14~percent for \rass, and
8~percent for \xmmsl) have spectra classified by the \boss\ pipeline
as \texttt{GALAXY}, with no subclass, i.e. optically quiescent
galaxies \citep[commonly referred to as `X-ray bright, optically normal
galaxies', XBONGs][]{Comastri02}. Many examples of XBONGs have been identified in
previous X-ray surveys \citep[e.g.][and references
within]{Barger01,Georgantopoulos05,Page06,Civano07,Trump09,Menzel16}
with the optical dullness of such galaxies most often attributed to
either dilution of a `normal' AGN signature by the host galaxy light,
or to objects with intrinsically weak AGN emission signatures
\citep{Trump09}.  Alternatively, when the X-ray and optical
measurements are well-separated in time, then strong luminosity
variations could potentially be another reason for the lack of a
clearly observed optical AGN signature.  It is also possible for weak
AGN signatures to remain hidden within the noise of low SNR and/or low
resolution spectra; as typically obtained by spectroscopic follow-up
programs where the primary goal is just to obtain redshift
measurements for as many X-ray sources as possible.

\begin{figure}
\begin{center}
\includegraphics[angle=0,width=84mm]{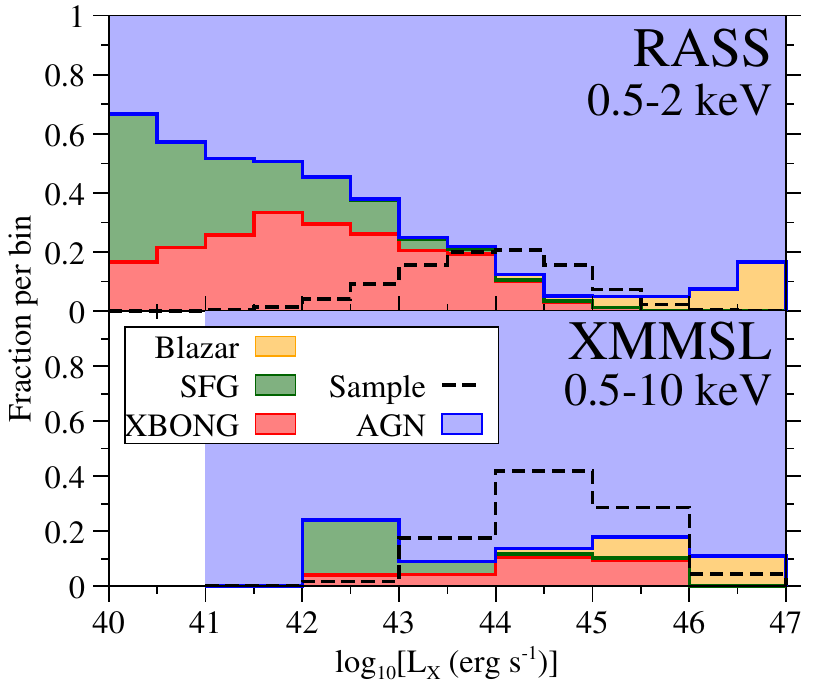}
\end{center}
\caption{The relative frequency of spectroscopic classifications
  of \rass\ (upper panel) and \xmmsl\ sources (lower panel) in bins of
  X-ray luminosity.  In each bin the colours indicate the fraction of objects 
  which are classed in five broad categories: AGN (including QSOs,
  narrow line AGN and any objects exhibiting broad emission lines,
  blue shaded region), Blazars (including objects classed as BL~Lac,
  yellow), SFGs (including objects classed as star-forming or starburst
  galaxies, green), Galactic stars (magenta), and XBONG (objects
  classed as Galaxy with no subclass, red).  The relative number of
  objects in the parent sample falling in each bin is indicated by the
  black dashed histogram.}
\label{fig:class_vs_Lx}
\end{figure}

In previous studies, XBONGs have typically been found in much smaller
quantities than seen in our two X-ray samples (c.f. rates of
2.5~percent in the XMM-COSMOS sample; \citealt{Cappelluti09,Trump09},
and 4~percent in the XMM-XXL survey; \citealt{Menzel16}). We would also
not expect XBONGs to appear in the high X-ray luminosity part of our
samples (see Fig.~\ref{fig:class_vs_Lx}), where the mismatch between
X-ray and optical properties is most marked. In order to explain these
apparent discrepancies we have investigated the \spiders\ XBONG sample in
more detail.

\begin{figure}
\begin{center}
\includegraphics[angle=0,width=84mm]{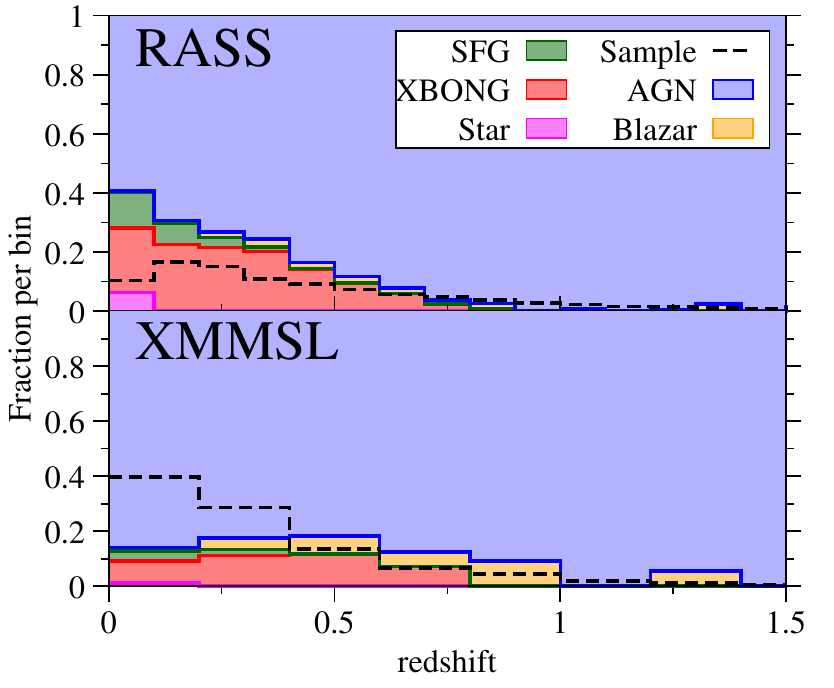}
\end{center}
\caption{The relative frequency of spectroscopic classifications of  
\rass\ (upper panel) and \xmmsl\ sources (lower panel) in bins of redshift. 
Same colour scheme as Fig.~\ref{fig:class_vs_Lx}.}
\label{fig:class_vs_z}
\end{figure}

One potential explanation for the lack of clear spectroscopic
signatures of AGN activity appearing in the spectra of these objects
is that such features (e.g. broad \halpha, \hbeta; strong \oiii) are
redshifted out of the observed wavelength range. However we can see
from Fig.~\ref{fig:class_vs_z} that the \spiders\ XBONGs all lie at low
redshifts.  Indeed, most ($>$75~percent) of the apparent XBONGs lie at
$z<0.4$, where the \halpha\ line is still within the red limits of the
\sdss\ and \boss\ spectrographs \citep[920 and 1040~nm respectively;][]{Smee13}; all of
the apparent XBONGS lie at redshifts where the \hbeta\ and \oiii\
lines are comfortably within the wavelength range of the observed spectra. 

\begin{figure}
\begin{center}
\includegraphics[angle=0,width=84mm]{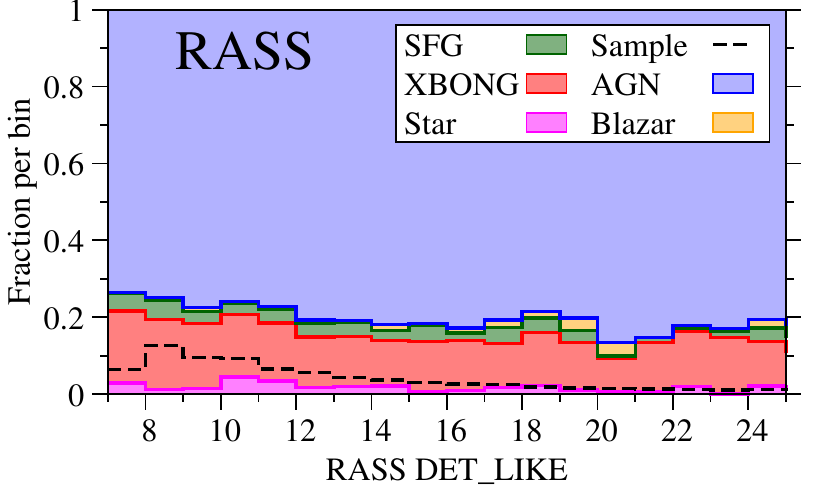}
\end{center}
\caption{The relative frequency of spectroscopic classifications
  of \rass\ sources in bins of the X-ray detection
  likelihood. Same colour scheme as Fig.~\ref{fig:class_vs_Lx}.}
\label{fig:class_vs_RASS_DET_LIKE}
\end{figure}

\begin{figure}
\begin{center}
\includegraphics[angle=0,width=84mm]{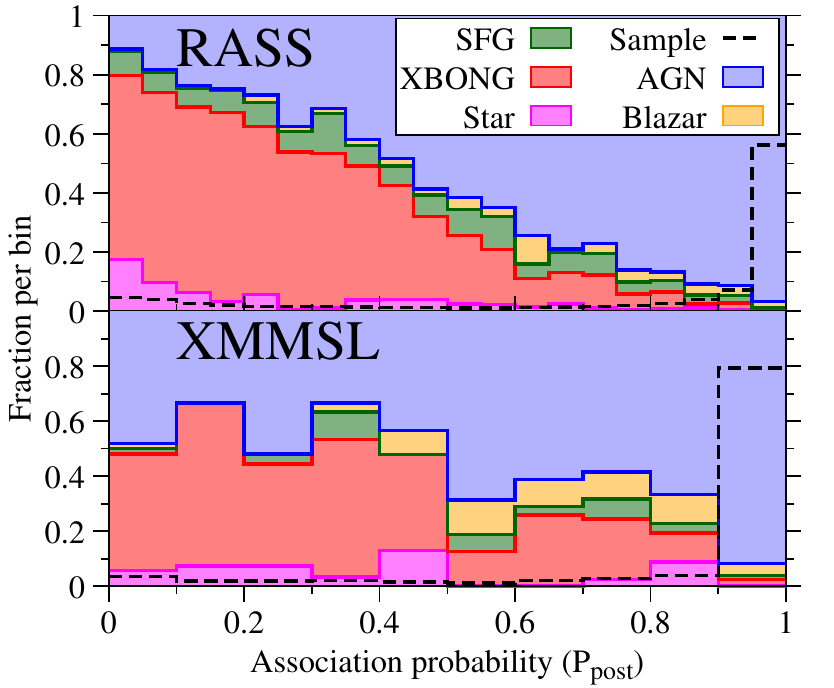}
\end{center}
\caption{The relative frequency of spectroscopic classifications
  of \rass\ (upper panel) and \xmmsl\ sources (lower panel) in bins of
  the X-ray$\rightarrow$\allw\ association posterior probability
  (\post). Same colour scheme as Fig.~\ref{fig:class_vs_Lx}.  }
\label{fig:class_vs_post}
\end{figure}

Fig.~\ref{fig:class_vs_RASS_DET_LIKE} shows that there is no obvious
trend for the fraction of quiescent galaxies to increase at low X-ray
detection likelihoods, so we cannot immediately attribute the apparent
excess of quiescent galaxies to spurious X-ray detections.  However,
Fig.~\ref{fig:class_vs_post} illustrates that these quiescent galaxies
form a significant fraction of the X-ray$\rightarrow$\allw\
associations with small values of \post, and so could
potentially be contaminated by a higher than average rate of incorrect
associations.  Note that the \boss\ pipeline can assign an empty
subclass to galaxies having only weak emission lines, including
objects with weak evidence of AGN emission (such features were 
not uniformly searched for as part of the visual inspection process). 
Therefore, not all of the
\spiders-AGN sources associated with quiescent galaxies will be
completely passive. 

In order to keep our input X-ray samples as complete as possible we
have not filtered the \rass\ and \xmmsl\ catalogues to exclude objects
with significantly extended X-ray emission; in any case the
significant detection of source extent for marginally resolved objects
close to the detection limit is extremely difficult.  Therefore, it is
possible, in some fraction of cases, that we have falsely attributed
X-ray emission to AGN activity, whereas it is actually emission from
the intra-cluster medium of a galaxy cluster.  In order to test this
hypothesis we examined the rate at which the \rass\ and \xmmsl\
sources that are associated with spectroscopically quiescent galaxies
also appear within the optically-selected `red-sequence Matched-filter
Probabilistic Percolation' (redMaPPer) catalogue of candidate cluster
member galaxies presented by \citet[][]{Rykoff14}.  We cross-matched
the redMaPPer catalogue to the optical counterparts of \rass\ and
\xmmsl\ sources having \sdss-DR12 spectra, using the optical positions
and a matching radius of 1\,arcsec. We only consider redMaPPer
galaxies that have cluster membership probabilities of at least
50~percent (and which lie in clusters of `richness'~$\ge$20).  We find
that a significant fraction (\si35~percent) of the XBONGs are matched
to redMaPPer cluster member galaxies (623/1639 and 38/123 of the XBONG
counterparts to \rass\ and \xmmsl\ sources, respectively).  As a
comparison, the frequency with which redMaPPer galaxies are matched to
the non-XBONG counterparts to \rass\ and \xmmsl\ sources is much
lower, of order 1~percent. It is therefore likely that some, but not
all of the apparent XBONGs in our spectroscopic samples are actually
due to X-ray detections of galaxy clusters.  We note that only
  2~percent of the \spidersagn\ targets put forward for observation
  within \ebossts\ are also selected as galaxy cluster targets by
  \citet{Clerc16}. In Appendix~\ref{sec:overlap_eboss} we present a
  full breakdown of the overlap of \spiders-AGN targets with other
  \ebossts\ target classes.

\section{Discussion}
\label{sec:discussion}

\subsection{Comparison with Anderson et al. (2007)}

Our spectroscopically identified sample of \rass\ sources described in
section~\ref{sec:RASS_with_spec} is 70~percent larger that that
presented by \citetalias{Anderson07}. This is due both to the smaller
spectroscopic survey footprint that was available to
\citetalias{Anderson07} (5740~\sqdeg\ versus the 10\,788~\sqdeg\
considered here), and also due to the increased sky density of
spectroscopic identifications in our sample. In the common 
area,
our catalogue contains 32~percent more \rass\ sources with spectra than
the catalogue of \citetalias{Anderson07}.

We find that 6444 of the \rass\ sources with unambiguous
  spectroscopic counterparts presented by \citetalias{Anderson07} also
  appear in our catalogue of \rass\ sources with \allw+\sdss-DR13
  photometric counterparts.  For 95.6~percent of these X-ray sources we choose exactly the
same optical counterpart (within 0.5~arcsec) as chosen by
\citetalias{Anderson07}. For most (207/281) of the
disagreements, we choose a counterpart that is closer to the \rass\
position than that chosen by \citetalias{Anderson07}.  This is not
unexpected since \citetalias{Anderson07} have considered all potential
spectroscopic counterparts lying anywhere within a 1\,arcmin radius
circle centred on the X-ray position (essentially a flat prior),
whereas the Bayesian cross-matching algorithm smoothly down-weights
potential counterparts lying far from the nominal X-ray position.
The two samples have similar bulk properties, e.g. the median $r$-band
magnitude is 18.2 for our spectroscopically identified \rass\ sample
compared to 18.3 for the subset which also appear in
\citetalias{Anderson07}.  Similarly, for extragalactic sources, the
median X-ray luminosities are $10^{43.9}$ and $10^{44.0}$\,\ergs\ (0.5--2\,keV), and
the median redshifts are 0.36 and 0.41 for our spectroscopically
identified \rass\ sample and the \citetalias{Anderson07} subset
respectively.

\subsection{Predictions for the \spiders-AGN\ Tier-0 program}
The combination of existing (\sdss-DR12) redshift measurements and the
forthcoming \sdssiv/\spiders\ data-set will allow us to build a highly
complete sample of X-ray selected AGN. We can make a confident prediction
for the characteristics of the \rass\ and \xmmsl\ samples at the end
of the \spiders\ program. 

Scaling down from the full \boss\ footprint (10\,778\,\sqdeg), we
expect \si22500 \rass\ sources and \si2670 \xmmsl\ sources to lie
within the 7500\,\sqdeg\ predicted to be covered by the final \eboss\
survey footprint. Including both the existing \sdss-DR12\ spectra and
spectra expected to be collected during \sdssiv, we estimate that
\si63~percent of \rass\ and \xmmsl\ sources in the \eboss\ footprint
will have spectroscopic redshift information available at the
completion of the \spiders\ observations. This does not take into
account the large fraction (around 25~percent) of \rass\ and
\xmmsl\ sources which are associated with optical counterparts (mostly
stars) brighter than the bright limit of the \boss\ spectrograph
(\texttt{fiber2Mag\_i}$<$17).  Discounting these optically very bright
sources, the spectroscopic coverage rises to 88~percent.  The success
rate for reliably measuring redshifts/classifications from \sdss\
spectra of counterparts to \rass\ and \xmmsl\ sources is very high,
even for the faint end of the population (\si97~percent), Therefore we
estimate that, at least for \texttt{fiber2Mag\_i}$\ge$17, the fraction
of the \rass\ and \xmmsl\ samples with reliable spectroscopic redshift
measurements will be \si85~percent by the end of the survey.

In summary, the combination of our novel source identification
method, and systematic spectroscopic follow-up with \sdss\ will
result, after completion of the \spiders-AGN Tier-0 program, in highly
complete samples of X-ray selected AGN, probing in a unique way, the
portion of the luminosity-redshift plane populated by the most
luminous accreting black holes.

\section{Conclusions}
\label{sec:conclusions}

We have presented here identification of \rass\ and \xmmsl\ point-like
sources with longer wavelength counterparts in \sdss\ and \wise\
catalogues with a novel Bayesian cross-matching algorithm that allows
priors from multiple catalogues to be considered. In particular, we
have shown how the location of sources in the \wise\ \wt,\wowt\
colour-magnitude diagram provides a very efficient way to identify
correctly (with 88~percent purity overall, and 93~percent for the
subset having high confidence X-ray detections) the mid-IR
counterparts to the bright X-ray sources detected in the \rass\ and
\xmmsl\ surveys.

We have then used our new identifications to inform the targeting for
the ongoing \spiders\ survey, a sub-programme of the \sdssiv\ project, 
and we present the list of \spiders-AGN targets submitted over the
entire \boss\ imaging footprint. In addition, we present samples of
11\,913 \rassw\ and 1482 \xmmsl\ sources that we have associated
with spectra published in the \sdss-DR12 catalogue, already one of the
largest uniformly selected spectroscopic samples of X-ray AGN ever compiled.

Based on such samples, and on the progress of the ongoing \ebossts\
survey, we estimate that, at the end of the \sdssiv\ program,
combining spectra from all generations of \sdss\ surveys, we will
compile a highly complete (\si85~percent) sample of good-quality
optical spectra for more than 15\,000 bright X-ray AGN selected both
in the soft (\rosat) and broad (\xmm) X-ray bands, over an area of
$\sim$7500\,\sqdeg.

\section*{Acknowledgements}
We are very grateful to the referee, F.~Carrera, for his detailed comments
and suggestions which have substantially improved this work.

Funding for SDSS-III has been provided by the Alfred P. Sloan
Foundation, the Participating Institutions, the National Science
Foundation, and the U.S. Department of Energy Office of Science. The
SDSS-III web site is \url{www.sdss3.org}.  SDSS-III is managed by
the Astrophysical Research Consortium for the Participating
Institutions of the SDSS-III Collaboration including the University of
Arizona, the Brazilian Participation Group, Brookhaven National
Laboratory, Carnegie Mellon University, University of Florida, the
French Participation Group, the German Participation Group, Harvard
University, the Instituto de Astrofisica de Canarias, the Michigan
State/Notre Dame/JINA Participation Group, Johns Hopkins University,
Lawrence Berkeley National Laboratory, Max Planck Institute for
Astrophysics, Max Planck Institute for Extraterrestrial Physics, New
Mexico State University, New York University, Ohio State University,
Pennsylvania State University, University of Portsmouth, Princeton
University, the Spanish Participation Group, University of Tokyo,
University of Utah, Vanderbilt University, University of Virginia,
University of Washington, and Yale University.

Funding for the Sloan Digital Sky Survey IV has been provided by the
Alfred P. Sloan Foundation, the U.S. Department of Energy Office of
Science, and the Participating Institutions.  SDSS-IV acknowledges
support and resources from the Center for High-Performance Computing
at the University of Utah. The SDSS web site is \url{www.sdss.org}.  
SDSS is managed by the Astrophysical Research Consortium for the
Participating Institutions of the SDSS Collaboration including the
Brazilian Participation Group, the Carnegie Institution for Science,
Carnegie Mellon University, the Chilean Participation Group, the
French Participation Group, Harvard-Smithsonian Center for
Astrophysics, Instituto de Astrof{\'i}sica de Canarias, The Johns Hopkins
University, Kavli Institute for the Physics and Mathematics of the
Universe (IPMU)/University of Tokyo, Lawrence Berkeley National
Laboratory, Leibniz Institut f{\"u}r Astrophysik Potsdam (AIP),
Max-Planck-Institut f{\"u}r Astronomie (MPIA Heidelberg),
Max-Planck-Institut f{\"u}r Astrophysik (MPA Garching),
Max-Planck-Institut f{\"u}r extraterrestrische Physik (MPE), National
Astronomical Observatories of China, New Mexico State University, New
York University, University of Notre Dame, Observat{\'o}rio Nacional/MCTI, 
The Ohio State University, Pennsylvania State University,
Shanghai Astronomical Observatory, United Kingdom Participation Group,
Universidad Nacional Aut{\'o}noma de M{\'e}xico, University of Arizona,
University of Colorado Boulder, University of Oxford, University of
Portsmouth, University of Utah, University of Virginia, University of
Washington, University of Wisconsin, Vanderbilt University, and Yale
University.

This publication makes use of data products from the Wide-field
Infrared Survey Explorer, which is a joint project of the University
of California, Los Angeles, and the Jet Propulsion
Laboratory/California Institute of Technology, and \neowise, which is a
project of the Jet Propulsion Laboratory/California Institute of
Technology. \wise\ and \neowise\ are funded by the National Aeronautics
and Space Administration.  This research has made use of the NASA/IPAC
Infrared Science Archive, which is operated by the Jet Propulsion
Laboratory, California Institute of Technology, under contract with
the National Aeronautics and Space Administration.  This research has
made use of `Aladin sky atlas' developed at CDS, Strasbourg
Observatory, France \citep{Bonnarel00,Boch14}.  This research has made
use of data obtained from the \chandra\ Source Catalog, provided by the
\chandra\ X-ray Center (CXC) as part of the \chandra\ Data Archive.  This
research has made extensive use of the {\sc stilts} toolset
\citep{Taylor06}.  
MB acknowledges support from the FP7 Career
Integration Grant `eEASy: supermasssive black holes through cosmic
time: from current surveys to eROSITA-Euclid Synergies' (CIG~321913).

\bibliographystyle{mnras}
\bibliography{SPIDERS_AGN_targeting}

\appendix
\section{Estimating X-ray fluxes}
\label{sec:Xray_fluxes}
\subsection{Estimating X-ray fluxes for \rass\ sources}
\label{sec:RASS_ECF_etc}

A full X-ray spectral analysis of each \rass\ source is not feasible.
Therefore, in order to convert the observed instrumental \rosat\ count
rates (which are collected over the 0.1--2.4\,keV range) into physical
fluxes we  assume a simple spectral model of a
power-law absorbed by the full Galactic column of neutral material.
This simplifying assumption,
valid for the extragalactic AGN in which we are most interested, does
however result in 
overestimated unabsorbed fluxes, for any Galactic sources that lie in
front of a significant fraction of the total Galactic column.

\begin{figure}
\begin{center}
\includegraphics[angle=0,width=84mm]{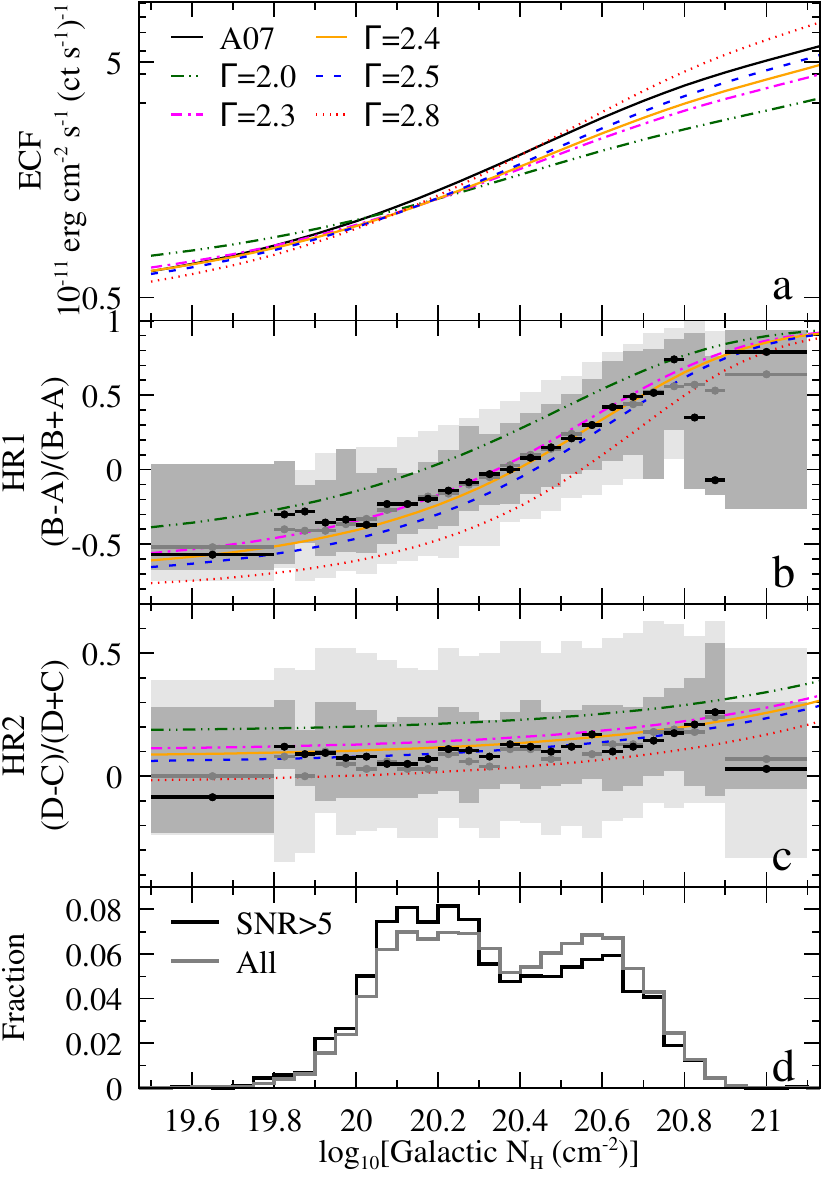}
\end{center}
\caption{\textit{Panel a:} Energy conversion factors (ECF) 
  for the \rosat\ PSPC-C, instrument assuming a simple
  spectral model of a power law absorbed by the Galactic column.  Also
  shown is the ECF assumed by \citetalias{Anderson07}.  \textit{Panel b:}
  Distribution of observed X-ray hardness ratio HR1 (see text for
  description) for \rass\ sources in the \boss\ imaging footprint as a
  function of Galactic column. The grey points with horizontal bars
  and the light grey shaded areas show the median and inter-quartile
  range for all the \rass\ sources found in each bin of Galactic
  \nh. The black points with horizontal bars and darker shaded area
  show the same but for just those \rass\ sources detected with
  SNR$\ge$5.  The curves show the values expected for the absorbed
  power law spectral model using the same colour-code indicated in
  panel \textit{a}.  \textit{Panel c:} Same as panel \textit{b}, but for the X-ray
  hardness ratio measure `HR2'.  \textit{Panel d:} Distribution of
  Galactic column density for \rass\ sources in the \boss\ footprint
  (grey histogram).  Also shown is the distribution for those
  \rass\ sources detected with SNR$\ge$5. }
\label{fig:RASS_ECF_etc}
\end{figure}

We estimate the Galactic column density in the direction of 
each \rass\ source using the map of Galactic
\nh\ provided by the NASA LAMBDA
team\footnote{\url{http://lambda.gsfc.nasa.gov/product/foreground/combnh_map.cfm}},
which in the region of interest (Dec$>$-30\,deg) is based on data from the Leiden/Dwingeloo HI
Survey \citep{Hartmann97}. The \rass\ sources in our sample 
have a range of Galactic 
column density in the range $10^{19.6}\le$\nh$\le 10^{21.1}$\,cm$^{-2}$, with 
median \nh=$10^{20.36}$\,cm$^{-2}$, see the bottom panel of Fig.~\ref{fig:RASS_ECF_etc}.

To predict the instrumental count rates expected
per unit flux for the above mentioned spectral model we use the \xspec\
tool combined with the \rosat\ PSPC-C (Position Sensitive
Proportional Counter-C) on-axis response
matrix\footnote{\url{ftp://legacy.gsfc.nasa.gov/caldb/data/rosat/pspc/cpf/matrices/pspcc_gain1_256.rmf}},
which is appropriate for \rass\ observations made prior to 1991~Jan~25.  
We calculated the count rates and model fluxes over a  grid covering the range $10^{19.5} \le$\nh$\le
10^{22}$\,cm$^{-2}$, and for spectral slopes in the range $1.0 \le
\Gamma \le 3.0$.  From the \xspec\ outputs we then calculate the multiplicative
energy conversion factors (ECF), which convert from observed count
rate to energy flux, and also the ratios of count rates expected
between different detector energy ranges (i.e. hardness ratios).
\citet{Voges99,Voges00} provide two hardness ratios for the \rass\
sources; \texttt{HR1} is calculated between energy bands `A' (PSPC-C channels
11--41, $\sim$0.1--0.4\,keV) and `B' (channels 52--201,
$\sim$0.5--2\,keV), and \texttt{HR2} is calculated between energy bands `C'
(channels 52--90, $\sim$0.5--0.9\,keV) and `D' (channels 91--201,
$\sim$0.9--2\,keV). The hardness ratios are computed as $HR =
(H-S)/(H+S)$, where $H$ and $S$ are the vignetting-corrected count
rates in the harder and softer bands, respectively.

We compare the predicted model hardness ratios (HR1, HR2) with the
observed distributions, binned in small steps of Galactic \nh, see
Fig.~\ref{fig:RASS_ECF_etc}.  The \rass\ sources have a wide scatter
of hardness ratios, presumably the result of some intrinsic range of
spectral shapes within the source population that is additionally
broadened by measurement errors. The smaller scatter in \texttt{HR1} and \texttt{HR2}
for \rass\ sources with SNR$>$5 indicates that at least some of this
scatter is likely to be due to measurement errors.  However, a trend
of increasing median (and inter-quartile range) of the hardness ratios with
Galactic \nh\ can be clearly seen
for both the full source sample and for the SNR$>$5 subset.  Over the
majority of the \nh\ range the median of the \texttt{HR2} distribution
is well traced by a spectral model having $\Gamma=2.5$.  At Galactic
columns smaller than \nh$\sim 10^{20.5}$\,cm$^{-2}$ the distribution
of \rass\ sources in \texttt{HR1} appears to favour a slightly harder
spectrum with $\Gamma\sim2.3$, hinting at mild spectral curvature over
the 0.1--2\,keV energy range.
Therefore, in order to convert the \rass\ instrumental count rates and errors
to fluxes, we have adopted the intermediate spectral index of
$\Gamma=2.4$.  

Note that \citetalias{Anderson07} assumed a slightly softer slope of
$\Gamma=2.5$ to convert from \rass\ count rates to fluxes.  In the top panel of 
Fig.~\ref{fig:RASS_ECF_etc} we show the ECF derived using the
\citetalias{Anderson07} recipe in comparison to the ECFs derived using
\xspec. The \citetalias{Anderson07} ECF is always equal to or larger than the ECF
adopted here, with the ratio increasing from $\sim 1$ at Galactic \nh=$10^{19.5}$\,cm$^{-2}$ 
to $\sim 1.2$ at \nh=$10^{21}$\,cm$^{-2}$, and then declining at larger values of \nh.
At the median Galactic \nh\ of the \rass\ sources
($10^{20.36}$\,cm$^{-2}$), the ECF of \citetalias{Anderson07} is 12~percent
larger than that used in our work.

\subsection{Estimating unabsorbed X-ray fluxes for \xmmb\ sources}
\label{sec:3XMM_ECF_etc}

The flux estimates supplied in the \xmmiii\ catalogue were calculated
assuming a simple power-law spectrum with photon-index $\Gamma = 1.7$
absorbed by a fixed neutral column density of \nh=$2\times
10^{20}$\,cm$^{-2}$ \citep{Mateos09}.  These fluxes have not been
corrected for Galactic absorption.  

We have adopted the following method to estimate the unabsorbed fluxes
of the \xmmiii\ sources in the 0.1--2.4\,keV band to allow direct
comparison with the fluxes of \rass\ sources.  We first make the
assumption that the \xmmiii\ catalogue fluxes represent a good
estimate of the absorbed energy flux in each of the relatively narrow
energy bands over which they were calculated: 0.2--0.5\,keV (band 1),
0.5--1\,keV (band 2) and 1--2\,keV (band 3).  We then use a look-up
table to correct the flux in each of these three energy bands by an
amount appropriate for the Galactic column density local to the
direction of each source \citep{Hartmann97}. The look-up table was calculated using
\xspec\ and assumes an intrinsic spectral model of a power-law having
photon index $\Gamma = 2.4$.
The unabsorbed flux in the 0.2--0.5\,keV band is then extrapolated to
0.1--0.5\,keV and the 1--2\,keV band is extrapolated to 1--2.4\,keV
using correction factors appropriate for a spectral model of a
power-law having photon index $\Gamma = 2.4$.  Specifically, for a
power-law this bandwidth correction factor is given by
$F_{[E'_{\mathrm{min}}:E'_{\mathrm{max}}]}/F_{[E_{\mathrm{min}}:E_{\mathrm{max}}]} = 
(E_{\mathrm{max}}^{'2-\Gamma} - E_{\mathrm{min}}^{'2-\Gamma})/
(E_{\mathrm{max}}^{2-\Gamma} - E_{\mathrm{min}}^{2-\Gamma})$, where
[$E'_{\mathrm{min}}:E'_{\mathrm{max}}$] and [$E_{\mathrm{min}}:E_{\mathrm{max}}$] give the energy bounds of
the new and original energy bands respectively.  Finally, our estimate of
the unabsorbed 0.1--2.4\,keV flux is given by the unweighted sum of the
unabsorbed flux estimates in the 0.1--0.5, 0.5--1 and 1--2.4\,keV
bands.

\section{Analysis of \xmmb\ sources without counterparts in \allw}
\label{sec:xmmb_ir_faint}
In order to determine if the few \xmmb\ sources lacking mid-IR
counterparts represent a population of X-ray bright but IR faint
emitters, we have visually inspected the \xmm, \sdss\ and \allw\
images at the X-ray locations of each of the 49/1049 \xmmb\ sources
that do not have at least one \allw\ counterpart within 3\,arcsec. 
The reasons for this can be categorised as follows. 
i) In
19/49 cases, the lack of a counterpart in the \allw\ catalogue is due
to pairs or groups of astrophysical objects being blended in the
$\sim$6\,arcsec resolution \allw\ images. In many of these cases the
correct counterpart to the X-ray source (as determined from inspection
of higher angular resolution \sdss\ optical images) is lost in the
wings of the PSF of a bright star.  ii) In 15/49 cases, the X-ray source
lies very close to the edge of the \xmm\ field of view, and hence the
X-ray positions are less well constrained than the nominal positional
errors in the catalogue would suggest. For most (10/15) of these high
off-axis-angle sources, a potential \allw\ counterpart lies within
5\,arcsec of the X-ray position, and for the remainder an \allw\
counterpart lies within 15\,arcsec. iii) In 9/49 cases, the X-ray position
lies within the optical disk of a nearby well resolved galaxy. iv) In 3/49
cases, the X-ray source can be associated with a high proper motion
star, which has moved more than 3\,arcsec from the X-ray position in
the time that passed between the \xmm\ observation and the \allw\
observations. v) Finally, in just 3/49 cases, the \xmmb\ source can be associated with
a faint optical counterpart that in the mid-IR is below the detection
limit of the \allw\ survey.  One of these (3XMM\,J003027.4+045139) is
the pulsar PSR\,J0030+0451 \citep{Bogdanov09}, one
(3XMM\,J133935.6-003132) is associated with a faint counterpart
($i'$=20.9$\pm$0.1, SDSS\,J133935.59-003132.7) in the \sdss\ imaging,
and one (3XMM\,J220221.4+015330) has no optical counterpart visible in
the \sdss\ imaging.  Fortuitously, the latter source lies within the
footprint of the CHFTLS-Wide survey, and we find that there is a
single catalogued detection within 3\,arcsec of the \xmmb\ position:
an extremely faint object with $i$=24.06$\pm$0.16
\citep[CFHTLS\,1411\_033310;][]{Hudelot12Cat}.  We conclude that mid-IR-faint 
objects do not form a large fraction of the X-ray source
population at the bright X-ray fluxes probed by the \xmmb\ sample.

\section{SDSS targeting efficiency and overlap 
of \spiders-AGN targets with other \ebossts\ target types}
\label{sec:overlap_eboss}
Within the \ebossts\ project, a heterogeneous mix of targets from a
variety of classes share the focal plane during each observation.
Designing a near-optimal set of \sdss\ plug-plates based on these
target lists (`tiling' in the \sdss\ parlance) is a non trivial task,
described in full by \citet{Dawson16}. In summary, the method is to
first assign each target category into one of several ranked collision
groups. Each group is tiled in sequence, taking account of the results
of previous tiling rounds. Within each collision group a priority
ranking scheme is used to determine which target receives a fibre in
cases of collisions.  \spiders\ targets are considered within the
first round of fibre assignments, and as they form the numerically
smallest target type and have the highest scientific requirement for
 completeness, they are given the highest priorities within
that round. Therefore, \spiders\ targets can potentially miss out on
being assigned a fibre only if they are within 62\,arcsec of another
\spiders\ target (or if they are flagged as belonging to one of the
\eboss\ cosmology samples, see below).  Within \spiders\ we adopt an
internal target priority ranking, again following a scheme wherein rarer
populations are tiled with higher priority. The \spiders\
targets are ranked as follows (from highest to lowest priority): i)
Brightest Cluster Galaxies from the \spiders-Clusters survey, ii)
\xmmslagn\ targets, iii) \rassagn\ targets, and iv) \spiders-Clusters
member galaxies.  We note that, in order to avoid imprinting a bias in the
selection function of \eboss\ main samples, any \spiders\ targets that
are also selected by one of the \eboss\ cosmology target selection
algorithms (see below) are tiled with the same priority as targets
from those other programs (i.e. at a lower priority than for unique
\spiders\ targets).

\begin{table}
  \caption{The numbers of \rassagn\ and \xmmslagn\ targets 
    that are associated with other \ebossts\ target types, calculated 
    for the first 4009.6\,\sqdeg\ of \eboss. 
    The `ET1' and `ET2' columns give the index of the bits that 
    identify this target class in the  \texttt{EBOSS\_TARGET1} and \texttt{EBOSS\_TARGET2} 
    bitmasks respectively.  The N\sub{RASS} and N\sub{XMMSL} 
    columns give the respective numbers of \rassagn\ and \xmmslagn\ targets that 
    overlap with the given target class. Values in bold are the self matches.}
\label{tab:target_overlap}
\begin{tabular}{@{}lllcc}
\hline
\eboss\ Target Flag           &  ET1   & ET2    & N\sub{RASS} & N\sub{XMMSL} \\
\hline
\texttt{LRG1\_WISE}           &   1    & n/a    &     9       &    1      \\  
\texttt{QSO1\_VAR\_S82}       &   9    & n/a    &     6       &    0      \\
\texttt{QSO1\_EBOSS\_CORE}    &  10    & n/a    &   705       &   47      \\  
\texttt{QSO1\_PTF}            &  11    & n/a    &   270       &   15      \\
\texttt{QSO1\_EBOSS\_FIRST}   &  14    & n/a    &    96       &   11      \\
\rule{0pt}{3ex}\noindent
\texttt{TDSS\_TARGET}         &  30    &  any   &   467       &   48      \\
\texttt{TDSS\_FES\_DE}        &  30    &  21    &    15       &    6      \\
\texttt{TDSS\_FES\_NQHISN}    &  30    &  23    &     3       &    0      \\
\texttt{TDSS\_FES\_VARBAL}    &  30    &  25    &     1       &    0      \\
\texttt{TDSS\_B}              &  30    &  26    &   161       &   18      \\
\texttt{TDSS\_FES\_HYPQSO}    &  30    &  27    &    11       &    2      \\
\texttt{TDSS\_FES\_HYPSTAR}   &  30    &  28    &     5       &    1      \\
\texttt{TDSS\_CP}             &  30    &  31    &   274       &   22      \\
\rule{0pt}{3ex}\noindent
\texttt{\rassagn}             &  31    &  0     & {\bf 4057}  &  120      \\
\texttt{\xmmslagn}            &  31    &  4     &   120       & {\bf 376} \\
\texttt{\rassclus}            &  31    &  1     &    93       &   12      \\
\texttt{\xclassclus}          &  31    &  5     &     3       &    3      \\
\hline
\rule{0pt}{3ex}\noindent
Any non-\spiders\ target      &        &        &    1030     &  82       \\
Unique targets                &        &        &    2852     &  200      \\
\rule{0pt}{3ex}\noindent
Total                         &        &        &    4057     &  376      \\
\hline
\end{tabular}
\end{table}

At the time of writing (early 2017), a substantial fraction of the
tiling process for the \ebossts\ survey has been carried out, covering
4010\,\sqdeg\ of sky (the tiling `chunks' named internally as
\texttt{eboss1--5}, \texttt{eboss9}, and \texttt{eboss16}).  In this
initial sky area, indicated in
Fig.~\ref{fig:RASS_XMMSL_potential_maps}, we find that 3971/4057 of
\rassagn\ targets and 369/376 of \xmmslagn\ targets are assigned a fibre;
an impressive targeting efficiency of 98~percent.

Within \ebossts\ there are a number of independent routes through
which an object can be selected to receive a fibre, and so some
targets are included in more than one target selection scheme.  In
Table~\ref{tab:target_overlap} we list the \ebossts\ target categories
which overlap with the \rassagn\ and \xmmslagn\ targets in the first
4010\,\sqdeg\ of tiled \eboss\ sky. We find that 25~percent of
\rassagn\ targets and 22~percent of \xmmslagn\ targets overlap with at
least one other non-\spiders\ target class.  We can see that for
\rassagn\ targets, the most frequent overlap (17~percent) is with the
core \eboss\ QSO target sample (\texttt{QSO1\_EBOSS\_CORE}), which is
targeting optical+mid-IR selected objects in the redshift range
$0.9<z<2.2$ \citep{Myers15}. There is also some overlap with the
\tdss\ project (12~percent of \rassagn\ targets), see
\citet{Morganson15} for more details.  For \xmmslagn, the most
frequent overlap, as would be expected, is with \rassagn\ targets
(32~percent), although there is also a significant overlap with the
\eboss\ QSOs and \tdss\ targets.  Note that the entries for \xmmsl\
targets in Table~\ref{tab:target_overlap} should be considered
indicative only, as for this sky area we used a version of the
\xmmslagn\ catalogue which mistakenly excluded sources having multiple
X-ray detections (see section~\ref{sec:data:xmmsl}).

\section{\sequels: \spidersagn\ targets in the \eboss\ pilot survey}
\label{sec:SEQUELS}
The \textbf{S}loan \textbf{E}xtended \textbf{QU}asar, \textbf{E}LG and
\textbf{L}RG \textbf{S}urvey (\sequels), was a pilot programme
designed to demonstrate the target selection for the main \ebossts\
survey \citep{Alam15,Dawson16}.  The spectroscopic observations for
\sequels\ were carried out in the MJD range [56660:57166] and include
a total of 117 good quality plates, covering a sky area of
471.9\,\sqdeg\ within the NGC. A summary of the target selection
processes for all categories of \sequels\ target types is given in
section A.3 of \citet{Alam15}, here we provide a detailed description
of the process by which \spiders-AGN targets were selected for
\sequels. The latter differs substantially from the two-step
\rass$\rightarrow$\allw$\rightarrow$\sdss\ method by which
\spiders-AGN targets were selected for observation in the main part of
\sdssiv\ (as described in section~\ref{sec:assoc:rass}). Note that the
\xmmsl\ catalogue was not considered when choosing targets for
\sequels.

The total time available for \sequels\ observations was not well known
at the time of selecting targets (2013), and so all contributing teams
were asked to provide sufficient targets to cover an optimistically
large footprint; defined by the 813\,\sqdeg\ of the \boss\ imaging
footprint lying within the bounds 120$\le$R.A.$\le$210\,deg and
+45$\le$Dec$\le$+60\,deg. There are 3049 \rassbf\ sources lying
inside this region, 3042 of which have at least one optical
counterpart within 1\,arcmin in the \sdss-DR8 photometric catalogue
\citep{Aihara11}.  We removed from further consideration any \rass\
sources with X-ray positions lying within 30\,arcsec radius of any
bright star in the Tycho-II catalogue \citep[][276 X-ray
sources]{Hog00}, or any object in the AGN catalogue of \citet[][307
X-ray sources]{Veron-Cetty10}, or any object having a pipeline
classification of \texttt{QSO} in the \sdss-DR11 spectroscopic
catalogue \citep[][1174 X-ray sources]{Alam15}.  After these filtering
steps we are left with 1563 X-ray sources.

\begin{figure}
\begin{center}
\includegraphics[angle=0,width=84mm]{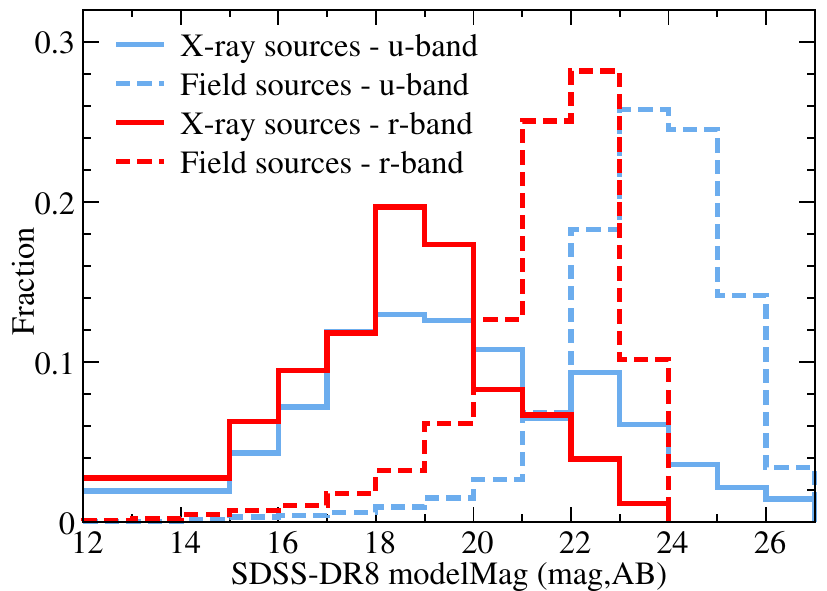}
\end{center}
\caption{A graphical representation of the Bayesian priors that were
  used when choosing optical counterparts to \rass\ sources for the
  \sequels\ survey.  For each filter considered ($u$: light blue
  lines, $r$: red lines), we show the magnitude distribution of the
  X-ray reference sample (solid) and the field population
  (dashed). The ratio of the reference curve to the field curve gives
  the prior for any potential candidate.}
\label{fig:SEQUELS_priors_ur}
\end{figure}

We used a prototype version of \nway\ \salvato\ to choose the most
probable optical counterpart for each remaining \rass\ source. We
instructed \nway\ to take account of the X-ray positions and their
uncertainties, the positions and model $u$,~$r$ magnitudes of the
candidate optical counterparts, plus pre-defined priors on the
expected distribution of counterparts in $u$ and $r$.  To derive the
priors for the \sequels\ \spiders-AGN target selection, we first
computed the $u$- and $r$-band magnitude distributions of a sample
of well-understood, X-ray bright \xmm-detected sources matched to the
\sdss-DR7 catalogue taken from \citet{Georgakakis11}, and then divided
these distributions by the $u$- and $r$-band magnitude distributions
of all \sdss-DR8 photometric objects in the sky area
considered\footnote{Before reaching the decision to use $u$- and
  $r$-band magnitude priors, we first experimented with a number of
  alternative combinations of priors, both in magnitude and colour.
  However, none of the tested combinations performed as well as the
  $u$- and $r$-band magnitude priors finally adopted here.}. Our
adopted priors for \sequels\ are illustrated in
Fig.~\ref{fig:SEQUELS_priors_ur}.  After running \nway, which produces
a ranked list of potential counterparts for each remaining X-ray
source, we further down-selected the sample to form a list of
potential spectroscopic targets.  For each \rass\ source we considered
only the `best' optical counterpart (the one having the highest
posterior probability as determined by \nway).  Many (408/1563) of the
best optical counterparts already had spectroscopic
identifications/redshifts in the \sdss-DR11 spectroscopic catalogue
\citep[][80 with \texttt{CLASS=QSO}; 77 with \texttt{CLASS=STAR}; 240
with \texttt{CLASS=GALAXY}]{Alam15} and so were removed from further
consideration.  Of the remaining 1155 X-ray sources, we retained only
the 630 having counterparts inside the magnitude range accessible by
$\sim$1\,hour exposures with the \boss\ spectrograph ($17\le r \le
22$\,mag; 455 were too bright, and 30 were too faint).  Finally, we
discarded three of the remaining \rass\ sources which had poorly
determined X-ray positions and uncertainties.  The filtering steps
left 627 optical counterparts to be put forward for targeting within
the \sequels\ program. A catalogue of these targets is presented in
Appendix~\ref{sec:RASS_SEQUELS_targets}.

At the plate design stage (after collation of all categories of
\sequels\ targets), 599/627 of our targets were allocated a fibre
within at least one of the 222 plates designed for \sequels.  There
were 307 \rass-selected AGN targets of which lie within the 117 plates
that were actually manufactured and observed before the conclusion of the
\sequels\ programme.  The \sdss\ spectra associated with these targets
have the \texttt{SPIDERS\_RASS\_AGN} flag set within the
\texttt{EBOSS\_TARGET0} bitmask in the \sdss-DR13 spectroscopic
database \citep{Albareti16}.

\section{Detailed description of \spiders-AGN catalogues}
\label{sec:catalogues}

In this section we describe the format and content of the catalogues
released as part of this paper.  All of these catalogues are supplied
in the Flexible Image Transport System (FITS) file format
\citep{Pence10} and are distributed via the MPE X-ray surveys website
(\url{http://www.mpe.mpg.de/XraySurveys/SPIDERS/SPIDERS_AGN}).  The
\spiders-AGN target catalogues are also hosted (albeit, in a slightly
different format to that described below) on the \sdss-DR13 website
(\url{https://data.sdss.org/sas/dr13/eboss/spiders/target}), and their
content is described on the corresponding `data model' webpage
(\url{https://data.sdss.org/datamodel/files/SPIDERS_TARGET}).  For
each column of each catalogue we give the column name, the FITS format
code, and a description of the column's meaning, including units where
appropriate.

\subsection{Catalogue of \rass\ sources with spectroscopy in \sdss-DR12}
\label{sec:RASS_DR12_catalogue}
See section~\ref{sec:dr12_spectra} for a description of how this catalogue was created.
The \rass+\sdss-DR12 catalogue contains 11913 entries, with the following columns:
\begin{description}[font=$\bullet$~\scshape\bfseries]
\item[RASS\_*                      ] X-ray source properties derived from the \rass\ source catalogues \citep{Voges99,Voges00}
\item[RASS\_ID                (16A)] \rass\ source identifier
\item[RASS\_BF\_FLAG           (1A)] Flag indicating if this X-ray source is taken from the \rass-BSC (`B') or \rass-FSC (`F') 
\item[RASS\_RA                 (1D)] Right Ascension of X-ray source (deg, J2000)
\item[RASS\_DEC                (1D)] Declination of X-ray source (deg, J2000)
\item[RASS\_RADEC\_ERR         (1E)] 1~$\sigma$ uncertainty on X-ray position (arcsec)
\item[RASS\_SRC\_RATE          (1E)] Observed X-ray count rate in the 0.1--2.4\,keV band (\cps)
\item[RASS\_SRC\_RATE\_ERR     (1E)] Statistical uncertainty on X-ray count rate (\cps)
\item[RASS\_BACK\_RATE         (1E)] Estimated background count rate at source location (\cpspa)
\item[RASS\_TEXP               (1E)] Effective exposure time at source location (s)
\item[RASS\_HR1                (1E)] First hardness ratio measure 
\item[RASS\_HR1\_ERR           (1E)] Uncertainty on first hardness ratio measure 
\item[RASS\_HR2                (1E)] Second hardness ratio measure 
\item[RASS\_HR2\_ERR           (1E)] Uncertainty on second hardness ratio measure 
\item[RASS\_EXT\_LIKE          (1E)] Likelihood of X-ray source being extended
\item[RASS\_DET\_LIKE          (1E)] Likelihood of X-ray source detection
\item[RASS\_ECF\_GAMMA2p4      (1E)] Multiplicative energy conversion factor used to convert from observed count rate to unabsorbed flux assuming a power-law spectrum with slope $\Gamma$=2.4 (\ecfunitsb) 
\item[RASS\_SRC\_FLUX          (1E)] Unabsorbed flux in the 0.1--2.4\,keV band corrected for full Galactic column (\cgs) 
\item[RASS\_SRC\_FLUX\_ERR     (1E)] Statistical uncertainty on unabsorbed flux (\cgs) 
\item[RASS\_GAL\_LONG          (1D)] Galactic longitude of X-ray source (deg)
\item[RASS\_GAL\_LAT           (1D)] Galactic latitude of X-ray source (deg)
\item[RASS\_LOGGALNH           (1E)] Logarithm of Galactic column density in direction of X-ray source (\logcm)  
\item[RASS\_ALLW\_DIST         (1D)] Distance between X-ray position and position of best \allw\ counterpart (arcsec) 
\item[RASS\_ALLW\_BIAS         (1D)] Bayesian prior (or bias) factor, $\Pi(\mathbf{x})$, of the best \allw\ counterpart, derived from its location in \wt-\wowt\ parameter space 
\item[RASS\_ALLW\_BFPOST       (1D)] Bayesian posterior probability of the best X-ray$\rightarrow$\allw\ association, before considering bias factor 
\item[RASS\_ALLW\_POST         (1D)] Bayesian posterior probability of the best X-ray$\rightarrow$\allw\ association, including the bias factor 
\item[ALLW\_*                      ] Properties of the best \allw\ counterpart to the X-ray source \citep[all taken from][]{Cutri13}
\item[ALLW\_ra                 (1D)] The Right Ascension of the \allw\ source (deg, J2000)
\item[ALLW\_dec                (1D)] The Declination of the \allw\ source (deg, J2000)
\item[ALLW\_sigra              (1E)] The component of the positional uncertainty of the \allw\ source parallel to Right Ascension (arcsec)
\item[ALLW\_sigdec             (1E)] The component of the positional uncertainty of the \allw\ source parallel to Declination (arcsec)
\item[ALLW\_w1mpro             (1E)] The \wo\ magnitude of the \allw\ source (mag, Vega)
\item[ALLW\_w1sigmpro          (1E)] The \wo\ magnitude uncertainty of the \allw\ source (mag, Vega)
\item[ALLW\_w2mpro             (1E)] The \wt\ magnitude of the \allw\ source (mag, Vega)
\item[ALLW\_w2sigmpro          (1E)] The \wt\ magnitude uncertainty of the \allw\ source (mag, Vega)
\item[ALLW\_w3mpro             (1E)] The \wthree\ magnitude of the \allw\ source (mag, Vega)
\item[ALLW\_w3sigmpro          (1E)] The \wthree\ magnitude uncertainty of the \allw\ source (mag, Vega)
\item[ALLW\_w4mpro             (1E)] The \wfour\ magnitude of the \allw\ source (mag, Vega)
\item[ALLW\_w4sigmpro          (1E)] The \wfour\ magnitude uncertainty of the \allw\ source (mag, Vega)
\item[ALLW\_W1\_W2             (1D)] The \wowt\ colour of the \allw\ source (mag, Vega)
\item[ALLW\_cc\_flags          (4A)] A set of `contamination and confusion' quality flags for the \allw\ source
\item[ALLW\_r\_2mass           (1E)] The distance from the \allw\ source to the nearest 2MASS counterpart (arcsec)
\item[ALLW\_pa\_2mass          (1E)] The position angle of the nearest 2MASS counterpart from the \allw\ source  (degrees)
\item[ALLW\_n\_2mass           (1J)] The number of 2MASS counterparts within 3\,arcsec of the \allw\ source
\item[ALLW\_j\_m\_2mass        (1E)] The $J$ band magnitude of the nearest 2MASS counterpart to the \allw\ source (mag, Vega)
\item[ALLW\_j\_msig\_2mass     (1E)] The $J$ band magnitude uncertainty of the nearest 2MASS counterpart (mag, Vega)
\item[ALLW\_h\_m\_2mass        (1E)] The $H$ band magnitude of the nearest 2MASS counterpart (mag, Vega)
\item[ALLW\_h\_msig\_2mass     (1E)] The $H$ band magnitude uncertainty of the nearest 2MASS counterpart (mag, Vega)
\item[ALLW\_k\_m\_2mass        (1E)] The $K_s$ band magnitude of the 2MASS counterpart (mag, Vega)
\item[ALLW\_k\_msig\_2mass     (1E)] The $K_s$ band magnitude uncertainty of the nearest 2MASS counterpart (mag, Vega)
\item[SDSS\_*                      ] Properties of the best \sdss-DR13 photometric counterpart to the \allw\ source \citep[taken from][]{Albareti16}
\item[SDSS\_RUN                (1J)] Element of the standard five-part \sdss\ photometric source identification descriptor 
\item[SDSS\_RERUN              (3A)] see above
\item[SDSS\_CAMCOL             (1J)] see above
\item[SDSS\_FIELD              (1J)] see above
\item[SDSS\_ID                 (1J)] see above
\item[SDSS\_RA                 (1D)] The Right Ascension of the \sdss-DR13 photometric counterpart (deg, J2000)
\item[SDSS\_DEC                (1D)] The Declination of the \sdss-DR13 photometric counterpart (deg, J2000)
\item[SDSS\_MODELMAG\_u        (1E)] The `model magnitude' of the optical counterpart to the X-ray source in the $u$ band, assuming a 2\,arcsec diameter fibre (mag, AB)
\item[SDSS\_MODELMAG\_g        (1E)] as above, but for the $g$ band
\item[SDSS\_MODELMAG\_r        (1E)] as above, but for the $r$ band
\item[SDSS\_MODELMAG\_i        (1E)] as above, but for the $i$ band
\item[SDSS\_MODELMAG\_z        (1E)] as above, but for the $z$ band
\item[SDSS\_FIBER2MAG\_u       (1E)] The `fiber magnitude' of the optical counterpart to the X-ray source in the $u$ band, assuming a 2\,arcsec diameter fibre (mag, AB)
\item[SDSS\_FIBER2MAG\_g       (1E)] as above, but for the $g$ band
\item[SDSS\_FIBER2MAG\_r       (1E)] as above, but for the $r$ band
\item[SDSS\_FIBER2MAG\_i       (1E)] as above, but for the $i$ band
\item[SDSS\_FIBER2MAG\_z       (1E)] as above, but for the $z$ band
\item[DR12\_*                      ] Properties extracted from the \sdss-DR12 spectroscopic catalogue \citep[][]{Alam15}
\item[DR12\_SURVEY             (6A)] The \sdss\ survey in which this spectrum was obtained
\item[DR12\_PLATE              (1J)] The \sdss\ spectroscopic plate index 
\item[DR12\_MJD                (1J)] The date on which the last spectroscopic data for the plate were obtained (MJD, days) 
\item[DR12\_FIBERID            (1J)] The index of the spectroscopic fibre through which this spectrum was obtained 
\item[DR12\_PLUG\_RA           (1D)] The Right Ascension of the spectroscopic fibre position (deg, J2000)
\item[DR12\_PLUG\_DEC          (1D)] The Declination of the spectroscopic fibre position (deg, J2000)
\item[DIST\_PHOT\_PLUG         (1E)] The distance between the \sdss-DR13 photometric catalogue position and the spectroscopic fibre position (arcsec)  
\item[DR12\_Z                  (1E)] The best fitting redshift solution computed by the \sdss-DR12 spectroscopic pipeline 
\item[DR12\_Z\_ERR             (1E)] Uncertainty on the best redshift solution computed by the \sdss-DR12 spectroscopic pipeline 
\item[DR12\_ZWARNING           (1J)] Flag that is set to $>0$ when the pipeline has encountered problems during the spectral reduction and redshift fitting process 
\item[DR12\_SN\_MEDIAN\_ALL    (1E)] The median SNR (per pixel) of the spectrum 
\item[DR12\_RCHI2              (1E)] The reduced $\chi^2$ of the best redshift solution
\item[DR12\_CLASS              (6A)] The broad spectral classification computed by the \sdss-DR12 spectroscopic pipeline 
\item[DR12\_SUBCLASS          (21A)] The detailed spectral classification computed by the \sdss-DR12 spectroscopic pipeline 
\item[DR12\_NSPECOBS           (1J)] Number of \sdss-DR12 spectra that are associated with this object
\item[DR12\_RUN2D              (6A)] Version code for the lower level \sdss\ spectral reduction pipeline used to process this spectrum
\item[DR12\_RUN1D              (6A)] Version code for the higher \sdss\ spectral reduction software used to process this spectrum
\item[DR7Q\_*                      ] Visual inspection information for this spectrum extracted from the \sdss-DR7Q quasar catalogue \citep[][]{Schneider10}
\item[DR7Q\_MEMBER             (1B)] Flag, set to 1 if a visual inspection for this spectrum was part of the \sdss-DR7Q catalogue
\item[DR7Q\_Z\_VI              (1E)] Visual inspection redshift from \sdss-DR7Q
\item[DR12Q\_*                     ] Visual inspection information for this spectrum extracted from the \sdss-DR12Q quasar catalogue \citep[][]{Paris17}
\item[DR12Q\_MEMBER            (1B)] Flag, set to 1 if a visual inspection for this spectrum was part of the \sdss-DR12Q catalogue
\item[DR12Q\_Z\_VI             (1E)] Visual inspection redshift from \sdss-DR12Q
\item[DR12Q\_CLASS\_PERSON     (1J)] Visual inspection spectral classification from \sdss-DR12Q
\item[DR12Q\_Z\_CONF\_PERSON   (1J)] Visual inspection redshift confidence from \sdss-DR12Q
\item[DR12Q\_ORIGIN            (8A)] Whether this spectrum appeared in the \sdss-DR12Q main catalogue, the superset catalogue or the supplementary bad spectrum catalogue 
\item[AO7\_*                       ] Visual inspection information for this spectrum extracted from \citet[][]{Anderson07}
\item[A07\_MEMBER              (1B)] Flag, set to 1 if a visual inspection for this spectrum was part of the \citetalias[][]{Anderson07} catalogue
\item[A07\_Z                   (1E)] Visual inspection redshift from \citetalias[][]{Anderson07}
\item[A07\_CLASS               (6A)] Visual inspection spectral classification from \citetalias[][]{Anderson07}
\item[A07\_FXcor               (1E)] X-ray flux estimate (0.1--2.4\,keV) from \citetalias[][]{Anderson07} (\cgs)
\item[A07\_logLX               (1E)] X-ray luminosity estimate (0.1--2.4\,keV) from \citetalias[][]{Anderson07} (\logergs)
\item[P10\_*                       ] Visual inspection information for this spectrum extracted from \citet[][]{Plotkin10}
\item[P10\_MEMBER              (1B)] Flag, set to 1 if a visual inspection for this spectrum was part of the \citet[][]{Plotkin10} catalogue
\item[P10\_Z                   (1E)] Visual inspection redshift from \citet[][]{Plotkin10}
\item[P10\_ZSP                 (1A)] Visual inspection spectral classification from \citet[][]{Plotkin10}
\item[P10\_CONF                (1A)] Original visual inspection confidence measure from \citet[][]{Plotkin10}
\item[P10\_Z\_CONF             (1J)] Derived visual inspection confidence measure from \citet[][]{Plotkin10}
\item[D17\_*                       ] Visual inspection information for this spectrum produced by the \spiders\ team
\item[D17\_MEMBER              (1B)] Flag, set to 1 if at least one visual inspection was carried out by the \spiders\ team
\item[D17\_NINSPECTORS         (1J)] Number of \spiders\ visual inspectors who examined the spectrum
\item[D17\_Z                   (1E)] Redshift derived from the visual inspections carried out by the \spiders\ team
\item[D17\_Z\_CONF             (1J)] Redshift confidence derived from the visual inspections carried out by the \spiders\ team
\item[D17\_CLASS               (8A)] Spectral class derived from the visual inspections carried out by the \spiders\ team
\item[D17\_RECONCILED          (1J)] Flag, set to 1 if manual reconciliation of the \spiders\ team visual inspections was carried out 
\item[VI\_BEST                 (7A)] The origin of the `best' visual inspection (one of DR7Q, DR12Q, AO7, P10, or SPIDERS)  
\item[Z\_BEST                  (1E)] The best visual inspection redshift
\item[Z\_CONF\_BEST            (1J)] The confidence of the best visual inspection redshift 
\item[Z\_DIFF                  (1E)] The difference between the \sdss-DR12 pipeline redshift and the best visual inspection redshift
\item[CLASS\_BEST              (8A)] The best visual inspection spectral classification
\item[DL\_BEST                 (1E)] The luminosity distance derived from the best visual inspection redshift (Mpc)
\item[RASS\_LOGLX              (1E)] Estimate of the X-ray luminosity (corrected for Galactic absorption) in the rest-frame 0.1--2.4\,keV band (\logergs)
\item[RASS\_LOGLX\_ERR         (1E)] Statistical uncertainty on the X-ray luminosity (\logergs)
\item[RASS\_LOGLX\_05\_2       (1E)] Estimate of the X-ray luminosity (corrected for Galactic absorption) in the rest-frame 0.5--2\,keV band (\logergs)
\end{description}

\subsection{Catalogue of \xmmsl\ sources with spectroscopy in \sdss-DR12}
\label{sec:XMMSL_DR12_catalogue}
See section~\ref{sec:dr12_spectra} for a description of how this catalogue was created.
The \rass+\sdss-DR12 catalogue contains 1482 entries, with the following columns:
\begin{description}[font=$\bullet$~\scshape\bfseries]
\item[XMMSL\_*                     ] X-ray source properties derived from the \xmmsl\ source catalogue \citep{Saxton08}
\item[XMMSL\_UNIQUE\_SRCNAME  (40A)] \xmmsl\ source identifier
\item[XMMSL\_OBSID            (10A)] \xmm\ observation identifier of the slew in which the source was detected
\item[XMMSL\_RA                (1D)] Right Ascension of X-ray source (deg, J2000)
\item[XMMSL\_DEC               (1D)] Declination of X-ray source (deg, J2000)
\item[XMMSL\_RADEC\_ERR        (1D)] 1~$\sigma$ uncertainty on X-ray position (arcsec)
\item[XMMSL\_DATE\_OBS        (19A)] Date of start of \xmm\ slew (YYYY-MM-DDThh:mm:ss) 
\item[XMMSL\_DATE\_END        (19A)] Date of end of \xmm\ slew (YYYY-MM-DDThh:mm:ss) 
\item[XMMSL\_SCTS\_FULL        (1E)] Background-subtracted source counts in the 0.2--12\,keV energy band (cts)
\item[XMMSL\_SCTS\_FULL\_ERR   (1E)] Statistical uncertainty on above (cts)
\item[XMMSL\_SCTS\_SOFT        (1E)] Same as above but for the 0.2--2\,keV band 
\item[XMMSL\_SCTS\_SOFT\_ERR   (1E)] Same as above but for the 0.2--2\,keV band 
\item[XMMSL\_SCTS\_HARD        (1E)] Same as above but for the 2--12\,keV band 
\item[XMMSL\_SCTS\_HARD\_ERR   (1E)] Same as above but for the 2--12\,keV band 
\item[XMMSL\_BG\_MAP\_FULL     (1E)] Estimated number of background counts within source extraction aperture in the 0.2--12\,keV band (cts)
\item[XMMSL\_BG\_MAP\_HARD     (1E)] Same as above but for the 0.2--2\,keV band 
\item[XMMSL\_BG\_MAP\_SOFT     (1E)] Same as above but for the 2--12\,keV band 
\item[XMMSL\_EXP\_MAP\_FULL    (1E)] Effective exposure time in the 0.2--12\,keV band (s)
\item[XMMSL\_EXP\_MAP\_SOFT    (1E)] Same as above but for the 0.2--2\,keV band 
\item[XMMSL\_EXP\_MAP\_HARD    (1E)] Same as above but for the 2--12\,keV band
\item[XMMSL\_RATE\_FULL        (1E)] Background-subtracted and vignetting-corrected count rate in the  0.2--12\,keV energy band (\cps)
\item[XMMSL\_RATE\_FULL\_ERR   (1E)] Statistical uncertainty on above (\cps)
\item[XMMSL\_RATE\_SOFT        (1E)] Same as above but for the 0.2--2\,keV band 
\item[XMMSL\_RATE\_SOFT\_ERR   (1E)] Same as above but for the 0.2--2\,keV band 
\item[XMMSL\_RATE\_HARD        (1E)] Same as above but for the 2--12\,keV band
\item[XMMSL\_RATE\_HARD\_ERR   (1E)] Same as above but for the 2--12\,keV band
\item[XMMSL\_FLUX\_FULL        (1E)] Estimated X-ray flux (corrected for full Galactic column) in the 0.2--12\,keV energy band (\cgs)
\item[XMMSL\_FLUX\_FULL\_ERR   (1E)] Statistical uncertainty on above (\cgs)
\item[XMMSL\_FLUX\_SOFT        (1E)] Same as above but for the 0.2--2\,keV band 
\item[XMMSL\_FLUX\_SOFT\_ERR   (1E)] Same as above but for the 0.2--2\,keV band 
\item[XMMSL\_FLUX\_HARD        (1E)] Same as above but for the 2--12\,keV band
\item[XMMSL\_FLUX\_HARD\_ERR   (1E)] Same as above but for the 2--12\,keV band
\item[XMMSL\_HR1               (1E)] Hardness ratio compute between the 0.2--2\,keV (soft) and 2--12\,keV (hard) energy bands
\item[XMMSL\_HR1\_ERR          (1E)] Statistical uncertainty on above
\item[XMMSL\_RASS\_DIST        (1E)] Distance to nearest RASS source (arcsec)
\item[XMMSL\_GAL\_LONG         (1D)] Galactic longitude of X-ray source (deg)
\item[XMMSL\_GAL\_LAT          (1D)] Galactic latitude of X-ray source (deg)
\item[XMMSL\_LOGGALNH          (1E)] Logarithm of Galactic column density in direction of X-ray source (\logcm)  
\item[XMMSL\_ALLW\_DIST        (1D)] Distance between X-ray position and position of best \allw\ counterpart (arcsec)
\item[XMMSL\_ALLW\_BIAS        (1D)] Bayesian prior (or bias) factor, $\Pi(\mathbf{x})$, of the best \allw\ counterpart, derived from its location in \wt-\wowt\ parameter space
\item[XMMSL\_ALLW\_BFPOST      (1D)] Bayesian posterior probability of the best X-ray$\rightarrow$\allw\ association, before considering bias factor 
\item[XMMSL\_ALLW\_POST        (1D)] Bayesian posterior probability of the best X-ray$\rightarrow$\allw\ association, including the bias factor 
\item[ALLW\_ra                 (1D)] See section~\ref{sec:RASS_DR12_catalogue}
\item[ALLW\_dec                (1D)] See section~\ref{sec:RASS_DR12_catalogue}
\item[ALLW\_sigra              (1E)] See section~\ref{sec:RASS_DR12_catalogue}
\item[ALLW\_sigdec             (1E)] See section~\ref{sec:RASS_DR12_catalogue}
\item[ALLW\_w1mpro             (1E)] See section~\ref{sec:RASS_DR12_catalogue}
\item[ALLW\_w1sigmpro          (1E)] See section~\ref{sec:RASS_DR12_catalogue}
\item[ALLW\_w2mpro             (1E)] See section~\ref{sec:RASS_DR12_catalogue}
\item[ALLW\_w2sigmpro          (1E)] See section~\ref{sec:RASS_DR12_catalogue}
\item[ALLW\_w3mpro             (1E)] See section~\ref{sec:RASS_DR12_catalogue}
\item[ALLW\_w3sigmpro          (1E)] See section~\ref{sec:RASS_DR12_catalogue}
\item[ALLW\_w4mpro             (1E)] See section~\ref{sec:RASS_DR12_catalogue}
\item[ALLW\_w4sigmpro          (1E)] See section~\ref{sec:RASS_DR12_catalogue}
\item[ALLW\_W1\_W2             (1D)] See section~\ref{sec:RASS_DR12_catalogue}
\item[ALLW\_cc\_flags          (4A)] See section~\ref{sec:RASS_DR12_catalogue}
\item[ALLW\_r\_2mass           (1E)] See section~\ref{sec:RASS_DR12_catalogue}
\item[ALLW\_pa\_2mass          (1E)] See section~\ref{sec:RASS_DR12_catalogue}
\item[ALLW\_n\_2mass           (1J)] See section~\ref{sec:RASS_DR12_catalogue}
\item[ALLW\_j\_m\_2mass        (1E)] See section~\ref{sec:RASS_DR12_catalogue}
\item[ALLW\_j\_msig\_2mass     (1E)] See section~\ref{sec:RASS_DR12_catalogue}
\item[ALLW\_h\_m\_2mass        (1E)] See section~\ref{sec:RASS_DR12_catalogue}
\item[ALLW\_h\_msig\_2mass     (1E)] See section~\ref{sec:RASS_DR12_catalogue}
\item[ALLW\_k\_m\_2mass        (1E)] See section~\ref{sec:RASS_DR12_catalogue}
\item[ALLW\_k\_msig\_2mass     (1E)] See section~\ref{sec:RASS_DR12_catalogue}
\item[SDSS\_RUN                (1J)] See section~\ref{sec:RASS_DR12_catalogue}
\item[SDSS\_RERUN              (3A)] See section~\ref{sec:RASS_DR12_catalogue}
\item[SDSS\_CAMCOL             (1J)] See section~\ref{sec:RASS_DR12_catalogue}
\item[SDSS\_FIELD              (1J)] See section~\ref{sec:RASS_DR12_catalogue}
\item[SDSS\_ID                 (1J)] See section~\ref{sec:RASS_DR12_catalogue}
\item[SDSS\_RA                 (1D)] See section~\ref{sec:RASS_DR12_catalogue}
\item[SDSS\_DEC                (1D)] See section~\ref{sec:RASS_DR12_catalogue}
\item[SDSS\_MODELMAG\_u        (1E)] See section~\ref{sec:RASS_DR12_catalogue}
\item[SDSS\_MODELMAG\_g        (1E)] See section~\ref{sec:RASS_DR12_catalogue}
\item[SDSS\_MODELMAG\_r        (1E)] See section~\ref{sec:RASS_DR12_catalogue}
\item[SDSS\_MODELMAG\_i        (1E)] See section~\ref{sec:RASS_DR12_catalogue}
\item[SDSS\_MODELMAG\_z        (1E)] See section~\ref{sec:RASS_DR12_catalogue}
\item[SDSS\_FIBER2MAG\_u       (1E)] See section~\ref{sec:RASS_DR12_catalogue}
\item[SDSS\_FIBER2MAG\_g       (1E)] See section~\ref{sec:RASS_DR12_catalogue}
\item[SDSS\_FIBER2MAG\_r       (1E)] See section~\ref{sec:RASS_DR12_catalogue}
\item[SDSS\_FIBER2MAG\_i       (1E)] See section~\ref{sec:RASS_DR12_catalogue}
\item[SDSS\_FIBER2MAG\_z       (1E)] See section~\ref{sec:RASS_DR12_catalogue}
\item[DR12\_SURVEY             (6A)] See section~\ref{sec:RASS_DR12_catalogue}
\item[DR12\_PLATE              (1J)] See section~\ref{sec:RASS_DR12_catalogue}
\item[DR12\_MJD                (1J)] See section~\ref{sec:RASS_DR12_catalogue}
\item[DR12\_FIBERID            (1J)] See section~\ref{sec:RASS_DR12_catalogue}
\item[DR12\_PLUG\_RA           (1D)] See section~\ref{sec:RASS_DR12_catalogue}
\item[DR12\_PLUG\_DEC          (1D)] See section~\ref{sec:RASS_DR12_catalogue}
\item[DIST\_PHOT\_PLUG         (1E)] See section~\ref{sec:RASS_DR12_catalogue}
\item[DR12\_Z                  (1E)] See section~\ref{sec:RASS_DR12_catalogue}
\item[DR12\_Z\_ERR             (1E)] See section~\ref{sec:RASS_DR12_catalogue}
\item[DR12\_ZWARNING           (1J)] See section~\ref{sec:RASS_DR12_catalogue}
\item[DR12\_SN\_MEDIAN\_ALL    (1E)] See section~\ref{sec:RASS_DR12_catalogue}
\item[DR12\_RCHI2              (1E)] See section~\ref{sec:RASS_DR12_catalogue}
\item[DR12\_CLASS              (6A)] See section~\ref{sec:RASS_DR12_catalogue}
\item[DR12\_SUBCLASS          (21A)] See section~\ref{sec:RASS_DR12_catalogue}
\item[DR12\_NSPECOBS           (1J)] See section~\ref{sec:RASS_DR12_catalogue}
\item[DR12\_RUN2D              (6A)] See section~\ref{sec:RASS_DR12_catalogue}
\item[DR12\_RUN1D              (6A)] See section~\ref{sec:RASS_DR12_catalogue}
\item[DR7Q\_MEMBER             (1B)] See section~\ref{sec:RASS_DR12_catalogue}
\item[DR7Q\_Z\_VI              (1E)] See section~\ref{sec:RASS_DR12_catalogue}
\item[DR12Q\_MEMBER            (1B)] See section~\ref{sec:RASS_DR12_catalogue}
\item[DR12Q\_Z\_VI             (1E)] See section~\ref{sec:RASS_DR12_catalogue}
\item[DR12Q\_CLASS\_PERSON     (1J)] See section~\ref{sec:RASS_DR12_catalogue}
\item[DR12Q\_Z\_CONF\_PERSON   (1J)] See section~\ref{sec:RASS_DR12_catalogue}
\item[DR12Q\_ORIGIN            (8A)] See section~\ref{sec:RASS_DR12_catalogue}
\item[A07\_MEMBER              (1B)] See section~\ref{sec:RASS_DR12_catalogue}
\item[A07\_Z                   (1E)] See section~\ref{sec:RASS_DR12_catalogue}
\item[A07\_CLASS               (6A)] See section~\ref{sec:RASS_DR12_catalogue}
\item[A07\_FXcor               (1E)] See section~\ref{sec:RASS_DR12_catalogue}
\item[A07\_logLX               (1E)] See section~\ref{sec:RASS_DR12_catalogue}
\item[P10\_MEMBER              (1B)] See section~\ref{sec:RASS_DR12_catalogue}
\item[P10\_Z                   (1E)] See section~\ref{sec:RASS_DR12_catalogue}
\item[P10\_ZSP                 (1A)] See section~\ref{sec:RASS_DR12_catalogue}
\item[P10\_CONF                (1A)] See section~\ref{sec:RASS_DR12_catalogue}
\item[P10\_Z\_CONF             (1J)] See section~\ref{sec:RASS_DR12_catalogue}
\item[D17\_MEMBER              (1B)] See section~\ref{sec:RASS_DR12_catalogue}
\item[D17\_NINSPECTORS         (1J)] See section~\ref{sec:RASS_DR12_catalogue}
\item[D17\_Z                   (1E)] See section~\ref{sec:RASS_DR12_catalogue}
\item[D17\_Z\_CONF             (1J)] See section~\ref{sec:RASS_DR12_catalogue}
\item[D17\_CLASS               (8A)] See section~\ref{sec:RASS_DR12_catalogue}
\item[D17\_RECONCILED          (1J)] See section~\ref{sec:RASS_DR12_catalogue}
\item[VI\_BEST                 (7A)] See section~\ref{sec:RASS_DR12_catalogue}
\item[Z\_BEST                  (1E)] See section~\ref{sec:RASS_DR12_catalogue}
\item[Z\_CONF\_BEST            (1J)] See section~\ref{sec:RASS_DR12_catalogue}
\item[Z\_DIFF                  (1E)] See section~\ref{sec:RASS_DR12_catalogue}
\item[CLASS\_BEST              (8A)] See section~\ref{sec:RASS_DR12_catalogue}
\item[DL\_BEST                 (1E)] See section~\ref{sec:RASS_DR12_catalogue}
\item[XMMSL\_LOGLX\_FULL        (1E)] Estimate of the X-ray luminosity (corrected for Galactic absorption) in the rest-frame 0.2--12\,keV band (\logergs)
\item[XMMSL\_LOGLX\_FULL\_ERR   (1E)] Statistical uncertainty on the X-ray luminosity in the rest-frame 0.2--12\,keV band (\logergs)
\item[XMMSL\_LOGLX\_SOFT        (1E)] As above, but for the rest-frame 0.2--2\,keV band
\item[XMMSL\_LOGLX\_SOFT\_ERR   (1E)] As above, but for the rest-frame 0.2--2\,keV band
\item[XMMSL\_LOGLX\_HARD        (1E)] As above, but for the rest-frame 2--12\,keV band
\item[XMMSL\_LOGLX\_HARD\_ERR   (1E)] As above, but for the rest-frame 2--12\,keV band
\item[XMMSL\_LOGLX\_05\_2       (1E)] As above, but for the rest-frame 0.5--2\,keV band
\item[XMMSL\_LOGLX\_05\_2\_ERR  (1E)] As above, but for the rest-frame 0.5--2\,keV band
\item[XMMSL\_LOGLX\_05\_10      (1E)] As above, but for the rest-frame 0.5--10\,keV band
\item[XMMSL\_LOGLX\_05\_10\_ERR (1E)] As above, but for the rest-frame 0.5--10\,keV band
\end{description}

\subsection{Catalogue of \rass\ sources (\rassagn) to be targeted in \ebossts}
\label{sec:RASS_SPIDERS_targets}
See section~\ref{sec:assoc:rass} for a description of how this catalogue was created.
The \rassagn\ catalogue contains 9028 entries, with the following columns:

\begin{description}[font=$\bullet$~\scshape\bfseries]
\item[RA                       (1D)] The Right Ascension of the \sdss-DR13 photometric counterpart to the X-ray source \citep[deg, J2000;][]{Albareti16}
\item[DEC                      (1D)] The Declination of the \sdss-DR13 photometric counterpart to the X-ray source \citep[deg, J2000;][]{Albareti16}
\item[FIBER2MAG                (5E)] The `fiber magnitude' of the optical counterpart to the X-ray source in the $ugriz$ bands, assuming a 2\,arcsec diameter fibre \citep[mag, AB;][]{Albareti16}
\item[RUN                      (1J)] Element of the standard five-part \sdss-DR13 source identification descriptor for the optical counterpart to the X-ray source \citep{Albareti16}
\item[RERUN                    (3A)] see above
\item[CAMCOL                   (1J)] see above
\item[FIELD                    (1J)] see above
\item[ID                       (1J)] see above
\item[PRIORITY                 (1J)] Priority assigned to this target for the purposes of the \ebossts\ tiling process (0=highest priority, 99=lowest priority)
\item[XRAY\_SRC\_NAME         (16A)] Source identifier from the \rass\ catalogue \citep{Voges99,Voges00}
\item[XRAY\_RA                 (1D)] The Right Ascension of the X-ray source \citep[deg, J2000;][]{Voges99,Voges00}
\item[XRAY\_DEC                (1D)] The Declination of the X-ray source \citep[deg, J2000;][]{Voges99,Voges00}
\item[XRAY\_RADEC\_ERR         (1E)] The positional uncertainty of the X-ray source \citep[arcsec;][]{Voges99,Voges00}
\item[XRAY\_FLUX               (1D)] The X-ray flux of the source (\cgs), corrected for Galactic absorption, in the 0.1--2.4\,keV band (see section \ref{sec:RASS_ECF_etc}) 
\item[XRAY\_DET\_ML            (1D)] The detection likelihood for the X-ray source \citep{Voges99,Voges00}
\item[BAYES\_POSTERIOR\_PROB   (1E)] The Bayesian posterior probability (\post) of the association between the X-ray and \allw\ source.
\item[ALLWISE\_DESIGNATION    (19A)] The source identifier for the \allw\ counterpart to the X-ray source \citep{Cutri13}
\item[ALLWISE\_RA              (1D)] The Right Ascension of the \allw\ counterpart to the X-ray source \citep[deg, J2000;][]{Cutri13}
\item[ALLWISE\_DEC             (1D)] The Declination of the \allw\ counterpart to the X-ray source \citep[deg, J2000;][]{Cutri13}
\item[ALLWISE\_W2MPRO          (1E)] The \wt\ magnitude of the \allw\ counterpart to the X-ray source \citep[mag, Vega;][]{Cutri13}
\item[ALLWISE\_W1\_W2          (1E)] The \wowt\ colour of the \allw\ counterpart to the X-ray source (mag, Vega)
\end{description}

Note that this catalogue also formed part of the \sdss-DR13 data release, and can be
downloaded from
\url{https://data.sdss.org/sas/dr13/eboss/spiders/target/spiderstargetAGN-SPIDERS_RASS_AGN-v2.1.fits}.

\subsection{Catalogue of \xmmsl\ sources (\xmmslagn) to be targeted in \ebossts}
\label{sec:XMMSL_SPIDERS_targets}
See section~\ref{sec:assoc:xmmsl} for a description of how this catalogue was created.
The \xmmslagn\ catalogue contains 873 entries, with the following columns:

\begin{description}[font=$\bullet$~\scshape\bfseries]
\item[RA                       (1D)] See section~\ref{sec:RASS_SPIDERS_targets}
\item[DEC                      (1D)] See section~\ref{sec:RASS_SPIDERS_targets}
\item[FIBER2MAG                (5E)] See section~\ref{sec:RASS_SPIDERS_targets}
\item[RUN                      (1J)] See section~\ref{sec:RASS_SPIDERS_targets}
\item[RERUN                    (3A)] See section~\ref{sec:RASS_SPIDERS_targets}
\item[CAMCOL                   (1J)] See section~\ref{sec:RASS_SPIDERS_targets}
\item[FIELD                    (1J)] See section~\ref{sec:RASS_SPIDERS_targets}
\item[ID                       (1J)] See section~\ref{sec:RASS_SPIDERS_targets}
\item[PRIORITY                 (1J)] See section~\ref{sec:RASS_SPIDERS_targets}
\item[XRAY\_SRC\_NAME         (24A)] Source identifier from the \xmmsl\ catalogue (column \texttt{UNIQUE\_SRCNAME} in official \xmmsl\ catalogue)
\item[XRAY\_OBSID             (10A)] The \xmm\ Observation Identifier of the observation in which this detection was made
\item[XRAY\_RA                 (1D)] The Right Ascension of the X-ray source (deg, J2000)
\item[XRAY\_DEC                (1D)] The Declination of the X-ray source (deg, J2000)
\item[XRAY\_RADEC\_ERR         (1E)] The positional uncertainty of the X-ray source (arcsec)
\item[BAYES\_POSTERIOR\_PROB   (1E)] See section~\ref{sec:RASS_SPIDERS_targets}
\item[ALLWISE\_DESIGNATION    (19A)] See section~\ref{sec:RASS_SPIDERS_targets}
\item[ALLWISE\_RA              (1D)] See section~\ref{sec:RASS_SPIDERS_targets}
\item[ALLWISE\_DEC             (1D)] See section~\ref{sec:RASS_SPIDERS_targets}
\item[ALLWISE\_W2MPRO          (1E)] See section~\ref{sec:RASS_SPIDERS_targets}
\item[ALLWISE\_W1\_W2          (1E)] See section~\ref{sec:RASS_SPIDERS_targets}
\item[MISSING\_IN\_V3p1        (1B)] Set to 1 if this source was excluded in an earlier version of the catalogue (see section~\ref{sec:data:xmmsl} for details) 
\end{description}

Note that an older version of this catalogue (v3.1, containing 819 entries, see caveats in section
\ref{sec:data:xmmsl}) formed part of the \sdss-DR13 data release, and
can be downloaded from
\url{https://data.sdss.org/sas/dr13/eboss/spiders/target/spiderstargetAGN-SPIDERS_XMMSL_AGN-v3.1.fits}. The
revised version of the catalogue presented here, which fixes these problems, is only 
available from the MPE X-ray surveys website: \url{http://www.mpe.mpg.de/XraySurveys/SPIDERS/SPIDERS_AGN}.

\subsection{Catalogue of \rass\ sources targeted in \sequels}
\label{sec:RASS_SEQUELS_targets}
See section~\ref{sec:SEQUELS} for a description of how this catalogue was created.
This catalogue contains 627 entries, with the following columns:

\begin{description}[font=$\bullet$~\scshape\bfseries]
\item[RA                       (1D)] The Right Ascension of the \sdss-DR8 photometric counterpart to the X-ray source \citep[deg, J2000;][]{Aihara11}
\item[DEC                      (1D)] The Declination of the \sdss-DR8 photometric counterpart to the X-ray source \citep[deg, J2000;][]{Aihara11}
\item[FIBER2MAG                (5E)] The `fiber magnitude' of the optical counterpart to the X-ray source in the $ugriz$ bands, assuming a 2\,arcsec diameter fibre \citep[mag, AB;][]{Aihara11}
\item[RUN                      (1J)] Element of the standard five-part \sdss-DR8 source identification descriptor for the optical counterpart to the X-ray source \citep{Aihara11}
\item[RERUN                    (3A)] see above
\item[CAMCOL                   (1J)] see above
\item[FIELD                    (1J)] see above
\item[ID                       (1J)] see above
\item[PRIORITY                 (1J)] See section~\ref{sec:RASS_SPIDERS_targets}
\item[XRAY\_SRC\_NAME         (16A)] See section~\ref{sec:RASS_SPIDERS_targets}
\item[XRAY\_RA                 (1D)] See section~\ref{sec:RASS_SPIDERS_targets}
\item[XRAY\_DEC                (1D)] See section~\ref{sec:RASS_SPIDERS_targets}
\item[XRAY\_RADEC\_ERR         (1E)] See section~\ref{sec:RASS_SPIDERS_targets}
\item[XRAY\_FLUX               (1D)] See section~\ref{sec:RASS_SPIDERS_targets}
\item[XRAY\_DET\_ML            (1D)] See section~\ref{sec:RASS_SPIDERS_targets}
\item[BAYES\_POSTERIOR\_PROB   (1E)] The Bayesian posterior probability (\post) of the association between the X-ray and optical source.
\end{description}

This catalogue formed part of the \sdss-DR13 data release, and so can also be downloaded 
from \url{https://data.sdss.org/sas/dr13/eboss/spiders/target/spiderstargetSequelsAGN-SPIDERS_RASS_AGN-v1.1.fits}.

\subsection{Catalogue of bright X-ray sources used as an astrometric reference sample}
\label{sec:xray_astro_ref}
For completeness, we provide here the astrometric reference catalogue of bright
X-ray sources (from \chandra, \xmm\ and \swift\ serendipitous catalogues) described in~\ref{sec:verify:xref}.
This catalogue contains 4752 entries with the following columns:

\begin{description}[font=$\bullet$~\scshape\bfseries]
\item[XBEST\_CAT               (5A)] A code (one of `CSC', `3XMM' or `SXPS') identifying the catalogue from which the `best' X-ray properties of this source were derived
\item[XBEST\_RA                (1D)] The `best' X-ray determined Right Ascension of this source (deg, J2000)
\item[XBEST\_DEC               (1D)] The `best' X-ray determined Declination of this source (deg, J2000)
\item[XBEST\_FLUX\_SOFT        (1D)] An estimate of the X-ray flux in the `soft' energy band (0.2--2\,keV for \CSC\ and \xmmiii, 0.3-2\,keV for \sxps; all with units \cgs)
\item[XBEST\_FLUX\_FULL        (1D)] An estimate of the X-ray flux in the `full' energy band (0.2--7\,keV for \CSC, 0.2--12\,keV for \xmmiii, and 0.3-10\,keV for \sxps; all with units \cgs)
\item[XBEST\_POSERR            (1D)] The positional uncertainty in the form given by the parent catalogue (error ellipse half-major axis for \CSC, 1\,$\sigma$ error radius for \xmmiii, and 90~percent containment radius for \sxps; all with units of arcsec)  
\item[CSC\_name               (20A)] Name of this source (if and) as it appears in the \CSC\ catalogue \citep{Evans10}
\item[XMM\_NAME               (21A)] Name of this source (if and) as it appears in the \xmmiii\ catalogue \citep{Rosen16}
\item[SXPS\_Name              (22A)] Name of this source (if and) as it appears in the \sxps\ catalogue \citep{Evans14}
\item[XBEST\_NWISE             (1J)] Number of potential \allw\ associations within the search radius (i.e. within 5\,arcsec for \CSC\ and \xmmiii, and within 10\,arcsec for \sxps)
\item[DIST\_XBEST\_ALLW        (1D)] Distance from the best X-ray position to the nearest \allw\ counterpart (arcsec)
\item[ALLW\_designation       (20A)] The source identifier for the nearest \allw\ counterpart to the X-ray source \citep{Cutri13}       
\item[ALLW\_ra                 (1D)] The Right Ascension of the nearest \allw\ counterpart to the X-ray source \citep[deg, J2000;][]{Cutri13}
\item[ALLW\_dec                (1D)] The Declination of the nearest \allw\ counterpart to the X-ray source \citep[deg, J2000;][]{Cutri13}
\item[ALLW\_w1mpro             (1D)] The \wo\ magnitude of the nearest \allw\ counterpart to the X-ray source \citep[mag, Vega;][]{Cutri13}
\item[ALLW\_w1sigmpro          (1D)] The \wo\ magnitude uncertainty of the nearest \allw\ counterpart to the X-ray source \citep[mag, Vega;][]{Cutri13}
\item[ALLW\_w2mpro             (1D)] The \wt\ magnitude of the nearest \allw\ counterpart to the X-ray source \citep[mag, Vega;][]{Cutri13}
\item[ALLW\_w2sigmpro          (1D)] The \wt\ magnitude uncertainty of the nearest \allw\ counterpart to the X-ray source \citep[mag, Vega;][]{Cutri13}
\item[ALLW\_W1\_W2             (1D)] The \wt\ magnitude uncertainty of the nearest \allw\ counterpart to the X-ray source \citep[mag, Vega;][]{Cutri13}
\end{description}

\bsp	
\label{lastpage}
\end{document}